\documentclass[useAMS,usenatbib]{mn2e}
\usepackage{epsfig}

\title[Optical IDV of HBLs]{Quasi$-$simultaneous two band optical variability of the
blazars 1ES 1959+650 and 1ES 2344+514} 

\author[Gaur et al.]
{Haritma Gaur$^{1,2}$\thanks{E-mail: haritma@aries.res.in}, 
Alok C.\ Gupta$^{1,2}$,  
A. Strigachev$^{3}$,
R. Bachev$^{3}$,
E. Semkov$^{3}$, 
\newauthor Paul J. Wiita$^{4}$,
S. Peneva$^{3}$,
S. Boeva$^{3}$,
N. Kacharov$^{3,5}$,
B. Mihov$^{3}$,
E. Ovcharov$^{5}$ 
\\ 
$^{1}$Aryabhatta Research Institute of Observational Sciences (ARIES),
Manora Peak, Nainital -- 263129, India\\
$^{2}$Department of Physics, DDU Gorakhpur University, Gorakhpur - 273 009, India\\
$^{3}$Institute of Astronomy and National Astronomical Observatory,  
Bulgarian Academy of Sciences, 72 Tsarigradsko Shosse Blvd., \\ 1784 Sofia, Bulgaria\\
$^{4}$Department of Physics, The College of New Jersey, P.O.\ Box 7718, Ewing, NJ 08628, USA\\
$^{5}$Department of Astronomy, University of Sofia, 5 James Bourchier, 1164 Sofia, Bulgaria
}

\begin{document}

\date{Accepted ....... Received  ......; in original form ......}

\pagerange{\pageref{firstpage}--\pageref{lastpage}} \pubyear{2010}

\maketitle

\label{firstpage}

\begin{abstract}
We report the results of quasi-simultaneous two filter optical monitoring of two high-energy peaked blazars, 
1ES 1959+650 and 1ES 2344+514, to search for microvariability and short-term variability (STV). We carried out 
optical photometric monitoring of these sources in an alternating sequence of B and R pass-bands,
and have 24 and 19 nights of new data for these two sources, respectively.
No genuine microvariability (intra-night variability) was detected in either of these sources.  This
non-detection of intra-night
variations is in agreement with the conclusions of previous studies 
that high-energy peaked BL Lacs are intrinsically less variable than low-energy peaked BL Lacs
in the optical bands.
We also report the results of STV studies for these two sources between 
 July 2009 and August 2010. Genuine STV is found for the source 1ES 1959+650 but not for 
1ES 2344+514.  We briefly discuss possible reasons for the difference between the intra-night variability
behaviour of high- and low-energy peaked blazars.
\end{abstract}

\section{Introduction}

Blazars constitute an enigmatic subclass of radio-loud active galactic nuclei (AGNs).
Blazars include both BL Lacertae (BL Lac) objects and flat spectrum radio quasars (FSRQs), as both are
characterized by strong and rapid flux variability across the entire electromagnetic spectrum. 
In addition, BL Lacs show largely featureless optical continua.   Blazars exhibit strong polarization from radio
to optical wavelengths and usually have core-dominated radio structures. According to the orientation based 
unified model of radio-loud AGNs, blazar jets usually make an angle $\leq 10^{\circ}$ to our 
line-of-sight (e.g., Urry \& Padovani 1995).   These extreme AGNs provide a natural laboratory to 
study the mechanisms of energy extraction from the central super-massive black holes and the physical 
properties of astrophysical jets. \par
The electromagnetic (EM) radiation from blazars is predominantly 
non-thermal. At lower frequencies (through the UV or X-ray bands) the mechanism is almost certainly 
synchrotron emission, while at higher frequencies the emission mechanism is probably due to 
Inverse-Compton (IC) emission (e.g., Sikora \& Madejski 2001; Krawczynski 2004). The spectral energy 
distributions (SEDs) of blazars have a double-peaked structure (e.g., Giommi, Ansari \& Micol 1995; 
Ghisellini et al.\ 1997; Fossati et al.\ 1998). Based on the location of the first peak of their 
SEDs, $\nu_{\rm peak}$, and blazars are often sub-classified into 
 low energy peaked blazars (LBLs) and high energy peaked blazars (HBLs)  (Padovani \& Giommi 1995).
The first component peaks in the near-infrared/optical 
for LBLs and in the UV or X-rays for HBLs, while the second component usually peaks at GeV energies 
for LBLs and at TeV energies for HBLs.  

Observations of blazars reveal that they are variable at all accessible timescales, from a few tens of 
minutes to years, and even decades, at many frequencies (e.g., Carini \& Miller 1992; Wagner \& Witzel 1995; 
Gupta et al.\ 2004; Villata et al.\ 2007). 
These different timescales allow the variability of blazars to be broadly divided into three 
classes, e.g., intra-day variability (IDV), short-term variability (STV), and long-term variability (LTV). 
Variations in the flux of a source of a couple of hundredths of a magnitude up
to a few tenths of magnitude over a time scale of a day or less is known as IDV
(Wagner \& Witzel 1995) or microvariability or intra-night optical variability. 
Flux changes observed over days to a few months are often considered to be STV, while those taking from 
several months to many years are usually called LTV (e.g., Gupta et al.\ 2004).

For a better understanding of how the properties of HBLs relate to the previously known LBLs,
we need to discriminate among the theoretical models that try to explain them. Optical observations
offer a wealth of information on the variability of blazars. 
Over the past many years, a large amount of  optical variability data for LBLs
have been reported and rapid optical variability has been shown to be common (Rani et al. 2010, 2011 and
references therein).
However, because most of the HBLs were more recently found by X-ray and gamma-ray sky surveys (Perlman 
et al. 2006, Abdo et al. 2010 and refernces therein) 
they have been less extensively monitored, and the properties of the optical variability of HBLs are not yet
clear.
Preliminary observations of  HBLs  suggested that they show statistically lesser amounts
of optical variability  (Heidt \& Wagner 1996, 1998) and polarized light than that of 
LBLs (Stoke et al. 1985, 1989, Jannuzi et al. 1993, 1994, Villata et al. 2000). 
Heidt \& Wagner (1998) noticed these objects likely display different duty cycles and variability
amplitudes from those of  the LBLs.
Romero, Cellone \& Combi (1999) found such differences in the case of  microvariability, as
just one out of three HBLs in their sample showed micro-variations, while eight LBLs out of 12 did 
so.  These differences were attributed to the possible presence of stronger magnetic fields in 
the high-energy peaked BL Lacs which could prevent the formation of small-scale jet 
inhomogeneities in these objects.
However, due to the relative paucity of optical monitoring for HBLs, 
we can not yet definitively say whether and how HBLs are different from LBLs in optical variability
and therefore any additional high quality observations of HBLs are particularly useful. 

Here we report quasi-simultaneous observations of two HBLs, 1ES 1959$+$650
and 1ES 2344$+$514 in two optical bands (B and R) 
for the first time. In addition to looking for any microvariability in these two optical bands,
our observations also give additional information on any microvariations in (B$-$R) colour.
We also have searched for  short-term variability timescales for these HBLs in the optical bands.
We used five telescopes located in India, Bulgaria and Greece to monitor the optical variability 
of 1ES 1959$+$650 and 1ES 2344$+$514  between July 2009 and November 2010. 

The paper is structured as follows.  In Section 2, we discuss the key features of these two
HBLs.  Section 3 describes the observations and data reduction 
procedure.  Section 4 reports our results and we present our discussion and conclusions in Section 5.

\section{Previous observations}

\subsection{1ES 1959$+$650}

One of the best studied member of the HBL subclass is 1ES 1959$+$650, with a redshift, $z=0.046$ (Veron-Cetty \& Veron 2006).
It was first detected at X-rays during the Slew Survey made by the Einstein satellite's 
Imaging Proportional Counter
(Elvis et al.\ 1992). Based on the X-ray/radio versus X-ray/optical colour-color diagram,
the source was classified as a BL Lac object by Schachter et al.\ (1993). The mass of the central
black hole (BH) in 1ES 1959$+$650 has been estimated to be $\sim$1.5 $\times10^{8}$ M$_\odot$
(Falomo et al.\ 2002).

Heidt et al.\ (1999) performed a detailed study of the optical band and found a complex structure
composed of a luminous elliptical galaxy (M$_{R} =-23$) plus a disk and an absorption dust lane. 
Three photometric optical values were found in the literature, showing a
great variation of the source brightness in the optical band. Its brightness  was reported as: V = 16 by
Marcha (1994); V = 12.8, as derived from digitalized plates, using B$-$V = 0.5 (Schachter
et al. 1993); and an approximate magnitude of 13.67 obtained from the Cambridge Automatic
Plate Measurement instrument (Perlman et al.\ 1996). Kurtanidze et al.\ (1999)  reported data taken during 
1997 to 1999, finding brightnesses in the range R = 14.50 to 14.92. Krawczynski et al.\ (2004) carried out 
multi-wavelength observations of strong flares from this source. Villata  et al.\ (2000) observed
the source from February 29, 1996 to May 30, 1997, and recorded a rapid decrease of 0.28 mag in 4 days.
More recently, Ma et al.\ (2010) observed the source between 4 May 2003 to 19 September 2003 but saw no significant optical variations.

The blazar 1ES 1959$+$650 was further detected at X-rays with ROSAT \& BeppoSAX (Beckmann et al.\ 2002).
In 2002 May, the X-ray flux of the source had increased significantly. Both the Whipple (Holder
et al. 2003) and {\it High Energy Gamma Ray Astronomy} (HEGRA) (Aharonian et al.\ 2003) collaborations 
subsequently confirmed a higher Very High Energy (VHE) $\gamma$-ray flux as well.
 Observing the source in 2000, 2001 and early 2002, the HEGRA collaboration yielded a marginal
signal (Horns et al.\ 2002). 

An interesting aspect of the source activity in 2002, was the observations of a so-called ``orphan
flare'' (i.e., a flare of VHE $\gamma$-rays not accompanied by correlated increased activity at other
wavelengths), recorded on June 4 by the Whipple Collaboration (Krawczynski et al.\ 2004; Daniel
et al.\ 2005). The HEGRA collaboration had observed another, albeit less significant, orphan VHE signal
during moonlight 2 days earlier (Tonello \& Kranich 2003).
Both of these flares in VHE $\gamma$-rays, observed in the absence of high activity in X-rays,
are very difficult to reconcile with the standard synchrotron self-Compton model (Ghisellini et al.\ 1998), 
which is routinely very successfully employed to explain VHE $\gamma$-ray production.
So, a hadronic synchrotron mirror model was proposed by B{\"o}ttcher et al.\ (2005) to explain this orphan
TeV flare.
1ES 1959$+$650 was observed several times at TeV energies (Aharonian et al.\ 2003, Albert et al.\ 2006,
 Tagliaferri et al.\ 2008,  and references therein), and unsurprisingly, this source is 
included in the first Fermi catalog of AGNs (Abdo et al.\ 2010).

\subsection{1ES 2344$+$514}

This object, at $z=0.044$, was identified as a BL Lac object (Perlmann et al.\ 1996) and
a 2 keV X-ray flux of 1.14 $\mu$Jy was found while it had an optical brightness of V=15.5
magnitude (with no galaxy subtraction). It was discovered in TeV $\gamma$-rays ($>$350 GeV) with
the Whipple Observatory telescope and thus identified as a TeV BL Lac object (Catanese et al.\ 1998).
Giommi et al.\ (2000) reported  rapid variability
in the X-ray band  on a timescale of  5000 sec.  The very high energy gamma-ray emission from 1ES 2344$+$514 has been observed by the MAGIC
collaboration (Albert et al. 2007)  and Grube (2008) reported that this source showed a nightly variability
in the integrated flux above 300 GeV.

Miller et al.\ (1999) observed 1ES 2344+514  and found a positive detection of optical
microvariability, with a maximum range of about 0.08 magnitude over one night in the V-band.
Measurements in R band from September 1998 to October 2000 by  Kurtanidze and
Nikolashvili (2002) have shown LTV at the level of 0.1 mag; any intraday variability
was confined within 0.05 magnitudes and no evidence of microvariability at scales of hours or
smaller was found. Dai et al.\ (2001) observed this source and reported a possible micro-variation of 0.14
magnitude in the V band within 26 minutes, but the data quality was poor. Xie et al.\ (2002)  
monitored the source in 2000 and found
no significant intraday variation (maximum flickering amplitudes are $\Delta$V $\sim$ 0.18 mag, $\Delta$R
$\sim$ 0.10 mag) but, they didn't find significant intra-day variation because their errors on
IDV light curve were $\sim$ 0.05 mag. During observations in  the R-band in October 2000 no microvariability
was observed (Fan et al.\ 2004).  Ma et al.\ (2010) have claimed an optical
variability amplitude in R band to be 0.69$\pm$0.16 mag 
and timescale to be 4738 sec, similar to the rapid variability
of 5000 sec noted earlier in the X-ray band, though this is presumably a coincidence.

\section{Observations and data reduction}

Observations of these two HBLs were performed using five optical telescopes, 
one in India, one in Greece and three in Bulgaria. All of these telescopes are equipped 
with CCD detectors and Johnson UBV and Cousins RI filters. Details of the telescopes, 
detectors and other parameters related to the observations are given in Table 1.
 A complete log of observations of these two HBLs from those five telescopes are given in Tables 2 and 3.   \par
We carried out optical photometric observations  during the period July
2009 to November 2010. The raw photometric data  were processed by standard methods that we now
briefly describe.  For image processing or pre-processing, we generated a master bias frame 
for each observing night by taking the median of all bias frames. The master bias frame for the 
night was subtracted from all flat and source image frames taken on that night. Then the master flat 
in each pass-band was generated by median combine of all flat frames in each pass-band. Next, 
the normalized master flat for each pass-band was generated.
Each source image frame was flat-fielded by dividing by the normalized master flat in the respective band to 
remove pixel-to-pixel inhomogeneities. Finally, cosmic ray removal was done from all source image
frames. Data pre-processing used the standard routines in Image Reduction and Analysis
Facility\footnote{IRAF is distributed by the National Optical Astronomy Observatories, which are operated
by the Association of Universities for Research in Astronomy, Inc., under cooperative agreement with the
National Science Foundation.} (IRAF)  and ESO-MIDAS\footnote{ESO-MIDAS is the acronym for the European
Southern Observatory Munich Image Data Analysis System which is developed and maintained 
by European Southern Observatory} softwares.  \par
We processed the data using the Dominion Astronomical Observatory Photometry (DAOPHOT II) 
software to perform the circular concentric aperture photometric technique (Stetson 1987, 1992).
For each night we carried out aperture photometry with four different aperture radii, i.e., 
1$\times$FWHM, 2$\times$FWHM, 3$\times$FWHM and 4$\times$FWHM. On comparing the photometric 
results we found that aperture radii of 2$\times$FWHM almost always provided the best S/N ratio, 
so we adopted that aperture for our final results.   \par

For these two HBLs, we observed  three or more local standard stars on the 
same fields. The magnitudes of these standard stars are 
given in Table 4.  We employed two standard stars from each blazar field with magnitudes 
similar to those of the target so as to not induce any significant 
errors due to differences in photon statistics. We used standard stars 4 and 6 for the
blazars 1ES 1959+650 and stars C2 and C3 for the HBL 1ES 2344+514 and  show
their differential instrumental magnitudes for the IDV light curves.  
As the fluxes of the blazars and the standard stars were obtained simultaneously and so at the same 
air mass and with identical instrumental and weather conditions, the flux ratios are considered 
to be very reliable.
Finally, to calibrate the photometry of the blazars, we used the one standard star of each pair that had a
colour closer to that of the blazar;
star 6 and star C2  were the calibrators for 1ES 1959$+$650 and 1ES 2344$+$514, respectively.
 We adopted the magnitudes for these stars listed in Table 4 but did not incorporate the 
errors in their standardized magnitudes when quoting blazar magnitudes.  Typical photometric errors 
for 1ES 1959$+$650 are $\sim 0.01$ mag in each of the BVRI bands, while for 1ES 2344$+$514 the 
errors in the B and V bands are $\sim 0.02$ mag while in the R and I bands it is $\sim 0.01$ mag.

\section{Results}
\subsection{Variability Parameter Detection Techniques}

Romero et al.\ (2002) pointed out how an inappropriate choice of the comparison stars
used for differential photometry could result in spurious fluctuations in the differential 
light curve, and hence claims of spurious variability.  To prevent this, we have selected non-variable
or standard stars that closely match the target's magnitude and to be conservative we have removed  isolated
apparently discrepant points from our analysis. We have  quantified our results by employing up to 
four different statistics (e.g., de Diego 2010). In our analysis, we have used standard stars 4 and 6
as starA and starB for 1ES 1959$+$650 and standard stars C2 and C3 as starA and starB for 
1ES 2344$+$514. 

\subsubsection{C-test}
The variability detection parameter, $C$ was 
introduced by Romero et al.\ (1999), and is defined as the average of $C1$ and $C2$ where
\begin{equation}
C1 = \frac{\sigma(BL - starA)}{\sigma(starA - starB)} \hspace{0.2 cm} {\rm \&} \hspace{0.2 cm} C2 = \frac{\sigma(BL - starB)}{\sigma(starA - starB)}.
\end{equation}
Here $(BL - starA)$ and $(BL - starB)$ are the differential instrumental magnitudes of the blazar and 
standard star A and the blazar and standard star B, respectively, while $\sigma (BL-starA)$, $\sigma 
(BL-starB)$ and $\sigma (starA-starB)$ are the observational 
scatters of the differential instrumental magnitudes of the blazar and star A, the blazar and star 
B and starA and star B, respectively. If $C \geq 2.576$, the nominal confidence level
of a variability detection is $> 99$\%, and we follow much of the previous literature
(Jang \& Miller 1997, Stalin et al. 2000, Gupta et al. 2008)
in considering this to be a positive detection of a variation.  However, this $C$-test is not
a true statistic as it is not appropriately distributed and this criterion is usually too 
conservative (de Diego 2010).
\par

\subsubsection{F-test}

We test our variability results using the standard $F$-test, which is a properly distributed
statistic (de Diego 2010). Given two sample variances such as
s$_Q ^{2}$ for the blazar instrumental light curve measurements and s$_* ^{2}$ for that of the standard star,
then
\begin{equation}
F=\frac {s_Q ^{2}}{s_* ^{2}} .
\end{equation}

The number of degrees of freedom for each sample, $\nu_Q$ and $\nu_*$ will be the same and equal to
the number of measurements N minus 1 ($\nu =$ N $-$ 1). The $F$ value is then compared with the
$F^{(\alpha)}_{\nu_Q,\nu_*}$ critical value, where $\alpha$ is the significance level set for the test.
The smaller the $\alpha$ value, the more improbable the result is produced by chance. If $F$ is larger
than the critical value, the null hypothesis (no variability) is discarded. We have performed
$F$-test at two significance levels (0.1\% and 1\%) which correspond to 3$\sigma$ and 2.6$\sigma$
detections, respectively.

\subsubsection{$\chi^{2}$-test}

We also performed a $\chi^{2}$-test on the data. Given a number of observations of a source 
over a given period of time, the $\chi^{2}$ statistic is expressed by

\begin{equation}
\chi^2 = \sum_{i=1}^N \frac{(V_i - \overline{V})^2}{\sigma_i^2},
\end{equation}
where, $\overline{V}$ is the mean magnitude, and the $i$th observation yields a magnitude $V_i$
with a corresponding standard error $\sigma_i$. 
Here $\sigma_i$ is the expected error,  the error from considering photon noise from the source 
and sky, along with the CCD read-out and all possible non-systematic sources of error. Since such errors are 
often calculated from the usually underestimated values yielded by the IRAF reduction package, 
an error rescaling is necessary. Usually, theoretical errors are smaller than the real
errors by a factor of typically 1.3-1.75 (e.g., Gopal Krishna et al.\ 2003, Bachev et al.\ 2005, Gupta et al.\ 2008). 
Our analysis of the current data indicates this value is, on average, $\sim$1.5. 
So the errors that the program codes 
return should be multiplied by 1.5 to get better estimates of the real errors of the photometry.
This statistic is compared against a critical value $\chi_{\alpha,\nu}^2$ obtained from 
the $\chi^{2}$ probability function, where $\alpha$ is again the significance level and $\nu = N -1$ 
are the degrees of freedom. If $\chi^2 > \chi_{\alpha,\nu}^2$ the test indicates a larger than 
expected scattering of the data points, and hence evidence of variability.

\subsubsection{ANOVA-test}

ANOVA tests are used to compare the means of a number of samples. de Diego et al.\ (1998) used the one-way ANOVA
test to investigate the variability in the light curves of quasars. This method consists of
measuring $k$ groups of $n_{j}$=5; in short cadence observations such as ours, these groups should be ideally separated 
by 20$-$30 minutes. Variance of the means of each group of five observations and the mean for the dispersion
within the groups are computed. Then, the ratio between these variances is calculated and multiplied 
by the number of observations in each group. The number obtained behaves as the $F$-statistic and 
we compare the $F$-value with the critical values $F^{(\alpha)}_{\nu_1,\nu_2}$. 
In this test, if $N$ is the number of observations and $k$ is the number of groups ($k=N/n$), 
the number of degrees of freedom are $\nu_{1}$ = $k-1$ for the groups and $\nu_{1}$ = $N-k$ for the errors;
therefore, $\nu_{1}+\nu_{1} =N-1$, corresponds to the degrees of freedom of the original data set.
For a certain significance
level $\alpha$, if $F$ exceeds the critical value, the null hypothesis will be rejected. 
We have used the inbuilt ANOVA code available in R. 

\subsubsection{Variability amplitude}

Heidt \& Wagner (1996) introduced the variability amplitude, defined as
\begin{equation}
A = \frac{100}{<A>}\times \sqrt{{(A_{max}-A_{min}})^2 - 2\sigma^2}(\%),
\end{equation}  
where $A_{max}$ and $A_{min}$ are the maximum and minimum fluxes in the calibrated LCs of
the blaza, $<A>$ is their mean, and the average measurement error of the blazar LC is $\sigma$.  We use this approach to
quantify any variability.

\subsection{Results for intra-day variability}

\subsubsection{1ES 1959+650}

We observed the source 1ES 1959$+$650 on 21 nights (simultaneously in optical bands, B and R)
and 3 nights in the R band alone from July 2009 to August 2010 for microvariability. 
We have performed the four tests discussed above on these data.  The $C$, $F$, $\chi^{2}$ and ANOVA parameters never exceed the 99\% 
confidence level (Table 5) for  B, R or (of course)
(B-R). So this source was stable with respect to IDV during our observations. The 
differential light curves of the source are given in figure 1 and 2 for these 24 nights.

\subsubsection{1ES 2344+514}

We observed the source 1ES 2344$+$514 on 14 nights (simultaneously in the B and R bands) 
and 5 nights in the R band from August 2009 to July 2010 for microvariability. We have tested for 
IDV using all four tests but no positive results were ever obtained in the B, R or (B-R) bands as 
the $C$, $F$, $\chi^{2}$ and ANOVA results never showed significance levels above 99\% considering
both stars
 (Table 6); although several $F$-test values exceed 0.99 significance with respect to Star 1 
they never do so with respect to Star 2.   Hence, this source was also very stable during each 
individual night of the observation period. The differential light curves of 1ES 2344+514 are
given in figure 3 and 4 for these 14 nights.

\subsection{Results for short-term variability}
 We quantify our results of short-term variability using two statistics: C-- and F-tests.
Since, in ANOVA test, we used to compare the means of a number of samples or groups (of 5
observations each), which should be ideally seperated by 20-30 minutes which is not possible
for our short-term observations.
Also, $\chi^{2}$ test compares the variation of an object with its photometric errors. And
the errors from DAOPHOT are usually underestimated. Because of which, we need to introduce a factor
by which these errors are multiplied. This factor comes from a comparison between the variation 
and the errors of two intrinsically nonvariable stars. 
As mentioned below, for our source 1ES 2344+514, two or perhaps more than two out of three 
standard stars present in the field are slowly varying. So, in this case, the factor is underestimated
and its not just of typically 1.3-1.75. So, $\chi^{2}$ test is not applicable here.
Hence, we are not using ANOVA and $\chi^{2}$ test in short-term variability analysis.

\subsubsection{1ES 1959+650}
We have examined the source 1ES 1959+650 for short-term variability 
during 44 nights (in B, V, R and I bands) between July
2009 and November 2010  (Figure 5).

{\bf B pass-band:}

The short-term light curve of 1ES 1959$+$650 in the B band is displayed in the upper panel
of Figure 5. The maximum variation noticed in the light curve of the source
is 1.23 magnitudes (between its brightest level at 14.99 mag on JD 2455036.40207 and the 
faintest level at 16.22 mag on JD 2455330.40554). The values of $C-$ and $F$-tests support 
the existence of short-term variation of the source in the B-band observations.  
We calculated the short-term variability
amplitude using Equation (3) and found that source has varied $\approx$102 per cent.

{\bf V pass-band:}

The V-band short-term light curve of 1ES 1959$+$650 is shown in the second panel
of Figure 5. The maximum variation seen 
is 1.06 mag (between its brightest level at 14.429 on JD 2455036.40811 and the faintest level
at 15.490 on JD 2455330.40301). The $C-$ and $F$-tests again support the existence of short-term
variation of the source in the V-band observations. We calculated that this source has a 
variability amplitude of $\approx$ 90 per cent.

{\bf R pass band:}

The third panel of Figure 5 gives the R-band short-term light curve of 1ES 1959$+$650. 
The maximum variation noticed in the source is 0.94 mag (between its brightest level at 
14.015 on JD 2455041.32782 and the faintest level
at 14.959 on JD 2455326.41514). The values of the $C-$ and $F$-test calculations 
also support the existence
of STV in our source. The amplitude of variability in the source is $\approx$82  per cent.

{\bf I pass-band:}

The short-term light curve of 1ES 1959$+$650 in the I band is also displayed, in the fourth panel
of Figure 5. The maximum variation noticed in the source
is 0.85 mag (between its brightest level at 13.49 mag on JD 2455036.40052  and the faintest level
at 14.34 mag on JD 2455330.39305).  Again, the  $C-$ and $F$-tests both support the existence of 
short-term variations of the source in these I-band observations. The amplitude of variability 
in the source is $\approx$ 75 per cent.

{\bf (V-R) Colour:}

The (V-R) colour index of 1ES 1959+650 in the (V-R) over this period comprises the bottom panel of
Figure 5. The maximum variation we saw in the source is 0.39  (between its colour index of 0.29 
on JD 2455395.38234 and 0.68  at JD 2455161.26157).  Both the $C-$ and $F-$tests show significant
 (V-R) colour variations in our observations. The amplitude of variability in the source's colour  
is $\approx$ 36 per cent.

\subsubsection{1ES 2344+514}

We have examined the source 1ES 2344$+$514 on 39 nights (in B,V,R and I bands) ranging from August
2009 to November 2010 for short-term variability (shown in figure 6). Because of the variation 
in the standard stars present in the field of this blazar, we noticed variations in 1ES 2344+514 in 
each band. Perhaps two out of three standard stars have slow variations which are propagating in 
the source. Therefore, despite the nominally large variation in the blazar calibrated magnitude, 
neither the $C-$ nor the $F-$test support the existence of short-term variation of the source 
in 1ES 2344+514.    
Also, the similar behaviour of standard stars 
were noticed in Tuorla observatory  observations\footnote {http://users.utu.fi/kani/1m/}

The maximum variation noticed in the short-term differential B-band light curve of 1ES 2344$+$514
is 0.80 mag  (between its brightest level at 2.02 mag on JD 2455065.48144 and the faintest level
at 1.22 mag on JD 2455449.57756) and the variation in the differential instrumental magnitudes
of standard stars is 0.53.  Because the stellar differential magnitudes varied so much, 
none of them could be truly considered to be of fixed brightness. \par

The maximum variation we found in the short-term light curve of this source in V-band
is 0.73 mag (between its brightest level at 15.47 mag on JD 2455449.57230 and the faintest level
at 16.20 mag on JD 2455065.47806). The variation in the differential instrumental magnitudes
of standard stars is 0.45.  \par

The R-band maximum variation  we found was 
 0.65 mag (between its brightest level at 14.72 mag on JD 2455449.56190 and the faintest level
at 15.37 mag on JD 2455070.41122). The variation in the differential instrumental magnitudes
of standard stars is 0.44. \par

The maximum variation noticed in the light curve of the source in I-band
is 0.62 mag (between its brightest level at 13.92 mag on JD 2455449.55730 and the faintest level
at 14.52 mag on JD 2455063.50316). The variation in the differential instrumental magnitudes
of standard stars is 0.55. \par

Finally, the maximum difference in the (V-R) Colour for 1ES 2344$+$514 is 0.22 mag 
(between its colour range 0.62 mag
on JD 2455070.41047 and 0.84 mag at JD 2455065.47656). The variation in the differential 
instrumental magnitudes of standard stars is 0.16.

\section{Discussion and Conclusion}

We have presented our results  of quasi-simultaneous and relatively dense temporal
observations of two HBLs, 1ES 1959$+$650 
and 1ES 2344$+$514, in two optical bands (B and R) during 2009--2010 .  
These are the first extensive observations of this type of these two blazars and 
thus adds to the rather small amount of  data on HBL microvariability and STV.
We have observed the sources 1ES 1959$+$650 and 
1ES 2344$+$514 on 24 and 19 nights respectively, but no significant intra-day variability 
was observed during any night for those two targets.
In addition, we can report the results of short-term variability studies of these sources.
The blazar 1ES 1959$+$650 has shown genuine short-term variability and it can be
characterized by a preliminary fading trend of $\sim$0.6 mag
(from JD 2455036 to 2455161)
and then a brightening trend of $\sim$0.74 (from JD 2455326 to 2455506 in V band).
However, despite nominally substantial magnitude changes,
genuine short-term variability was not found for the source 1ES 2344+514.
We found that most of the apparent variations by the blazar could have arisen from
 slow variations in the ``standard'' stars in the field.  
The results presented here make an important contribution  to the study of these target objects, 
as the such extensive studies have not yet been performed on their intra-night timescales
and the optical behaviour of this class of BL Lacs is still largely unknown.
Our measurements provide quasi-simultaneous measurements in two bands, unlike
most earlier microvariability studies.\par

There are several theoretical models that might be able to explain the observed variability 
over wide time-scales for all bands, with the leading contenders based upon shocks propagating 
down relativistic jets (e.g., Marscher \& Gear 1985; Qian et al.\ 1991; Hughes, Aller \& Aller 1991;
 Marscher, Gear \& Travis 1992; Wagner \& Witzel 1995; Marscher 
1996). In this class of models much of the variability arises when a shock strikes a feature in the jet.
Some of the variability may arise from helical structures, precession, or other geometrical 
effects occurring within the jets (e.g., Camenzind \& Krockenberger 1992; Gopal-Krishna \& 
Wiita 1992).   
These models seem to provide reasonable, albeit generic, representations of the variations observed on timescales
from about a day up through months and probably, years.
Hot spots or disturbances in or above accretion disks surrounding the black holes at the centers of AGNs  
(e.g., Mangalam \& Wiita 1993; Chakrabarti \& Wiita 1993) are likely to play a key role in the variability 
of non-blazar AGNs and might provide seed fluctuations that could be advected into the blazar jet 
and then Doppler amplified in the relativistic jets. \par    
 
Previous observational results have indicated that HBLs show a statistically significantly
lesser amount of 
optical variability then that do LBLs (Stocke et al.\ 1985, 1989, Heidt \& Wagner 1996), and this conclusion
is supported by our results, in that neither of these two HBLs ever showed microvariability.
Sambruna et al.\ (1996) investigated the multifrequency spectral properties of LBLs and HBLs and
mentioned this spectral properties requires a systematic change of intrinsic physical parameters, such as
magnetic field, jet size, and maximum electron energy and this change is in the sense that HBLs have
higher magnetic fields/electron energies and smaller sizes than LBLs.
The difference in the intra-day variability behaviour of HBLs could be 
due to the effect of stronger magnetic fields
and/or electron energies in these objects (Sambruna et al.\ 1996) that could prevent 
or delay the formation of 
features like density inhomogeneities and bends in the bases of the jets by Kelvin-Helmholtz 
 instabilities (Romero et al.\ 1999).
For instance, in the model proposed by Sol et al.\ (1989), extragalactic jets are composed of two different
fluids, a highly relativistic spine and a slower sheath. 
Kelvin-Helmholtz instabilities must arise at the interface between these two fluids with different
bulk velocities. Such instabilities might be responsible for many of the knots and bends observed
in the large scale structures of jets.
Romero (1995) showed that axial magnetic fields prevent the development of Kelvin-Helmholtz 
instabilities in sub-parsec to parsec scale jets if their values exceed the critical value B$_{c}$
given by
\begin{equation}
B_{c}=[4 \pi n m_{e}c^{2}(\gamma^{2}-1)]^{1/2} \gamma^{-1},
\end{equation}
where $n$ is the local electron density, $m_{e}$ is the electron rest mass, and $\gamma$ is the 
central flows bulk Lorentz factor. 
So, if $B > B_{c}$ this class of instabilities will be suppressed.  However, if B $< B_{c}$
the Kelvin-Helmholtz instabilities can produce significant changes in the jet morphology,
and these features could be responsible for rapid variability when the 
relativistic shock waves interact with them.  So if 
HBLs typically possess stronger magnetic fields for given electron densities or lower electron
densities for given field strengths,
 than we suggest that such a reduction in Kelvin-Helmholtz instabilities would
reduce the incidence of microvariability in the optical light curves.

Another possible explanation for the difference between the optical variability behaviour of the
LBL and HBL classes of blazars can dispense with the hypothesis of suppressed Kelvin-Helmholtz instabilities.
We recall that the optical band is near or above the peak of the first hump in the SED for LBLs but 
below that peak for HBLs. Therefore changes in the efficiency of acceleration of, and/or in the
rates at which energy is radiated by, the highest energy
electrons available for synchrotron emission would have a large and rapid effect on
the emitted flux in those bands for LBLs.  However, those variations would have a more modest 
and more retarded effect on optical variability in HBLs, since their optical emission is below the 
peak of the synchrotron emission. 
 Such variations in acceleration efficiency could arise from changes in the local number density of the most energetic electrons or the strengths of the maximum localized magnetic fields.  While magneto-hydrodynamic instabilities might be the origin of such variations, so might
the presence of turbulence in the vicinity of the shock (e.g., Marscher, Gear \& Travis 1992). 
If this is the case, then X-ray variability should be more pronounced for HBLs than for
LBLs, in that the peak of the synchrotron emission lies near the X-ray band for the former class.

\section*{Acknowledgments} 
This research was partially supported by Scientific Research Fund of the
Bulgarian Ministry of Education and Sciences (BIn - 13/09 and DO 02-85) and by 
Indo $-$ Bulgaria bilateral scientific exchange project INT/Bulgaria/B$-$5/08 funded 
by DST, India.
The Skinakas Observatory is a collaborative project of the University of Crete, the 
Foundation for Research and Technology -- Hellas, and the
Max-Planck-Institut f\"ur Extraterrestrische Physik.

\begin{table*}
\caption{ Details of telescopes and instruments}
\textwidth=6.0in
\textheight=9.0in
\vspace*{0.2in}
\noindent
\begin{tabular}{llllll} \hline
Site:        &ARIES Nainital & NAO Rozhen    & NAO Rozhen         & AO Belogradchik     &Skinakas Observatory, Crete   \\\hline
Telescope:   &1.04-m RC Cassegrain           & 2-m Ritchey-Chr\'etien  & 50/70-cm Schmidt  & 60-cm Cassegrain  &1.3-m Modified RC             \\
CCD model:   & Wright 2K CCD                 & PI VersArray:1300B      & FLI PL160803     & FLI PL09000 &Andor DX436-BV-9CQ  \\
Chip size:   & $2048\times2048$ pixels     & $1340\times1300$ pixels & $4096\times4096$ pixels  & $3056\times3056$ pixels & $2048\times2048$ pixels  \\
Pixel size:  &$24\times24$ $\mu$m             & $20\times20$ $\mu$m     & $9\times9$ $\mu$m  & $12\times12$ $\mu$m          & $13.5 \times13.5$ $\mu$m  \\
Scale:       &0.37\arcsec/pixel               &0.258\arcsec/pixel       & 1.079\arcsec/pixel   & 0.330\arcsec/pixel$^{\rm a}$  &0.2825\arcsec/pixel \\
Field:       & $13\arcmin\times13\arcmin$     &$5.76\arcmin\times5.59\arcmin$ & $73.66\arcmin \times 73.66 \arcmin$  & $16.8\arcmin\times16.8\arcmin$ &$9.6\arcmin\times9.6\arcmin$ \\
Gain:        &10 $e^-$/ADU                    &1.0 $e^-$/ADU             & 1.0 $e^-$/ADU        & 1.0 $e^-$/ADU               &2.687 $e^-$/ADU   \\
Read Out Noise:         &5.3 $e^-$ rms                   &2.0 $e^-$ rms    & 9.0 $e^-$ rms      & 8.5 $e^-$ rms           &8.14 $e^-$ rms       \\
Binning used:&$2\times2$                      &$1\times1$                & $1\times1$            & $3\times3$                   &$1\times1$  \\
Typical seeing : & 1\arcsec to 2.8\arcsec  & 1.5\arcsec to 3.5\arcsec & 2\arcsec to 4\arcsec  & 1.5\arcsec to 3.5\arcsec & 1\arcsec to 2\arcsec   \\\hline
\end{tabular} \\
\noindent
$^{\rm a}$ With a binning factor of $1\times1$
\end{table*}

\begin{table*}
\caption{Observation log of optical photometric observations of 1ES 1959$+$650}
\textwidth=6.0in
\textheight=9.0in
 \centering
\vspace*{0.2in}
\noindent
\begin{tabular}{lcl} \hline
 Date of             & Telescope   & Data Points \\
observation          &             & Filters            \\
(yyyy mm dd)         &             & (B,V,R,I)            \\\hline

2009 07 23           & B         & 1,1,1,1  \\
2009 07 26           & D         &56,2,56,2  \\
2009 07 27           & D         &58,1,58,1  \\
2009 07 28           & D         & 57,1,57,1  \\
2009 07 29           & D         &56,2,55,2  \\
2009 07 30           & D         &76,1,37,1  \\
2009 08 03           & E         &42,2,43,2  \\
2009 08 05           & E         &102,2,102,2 \\
2009 08 17           & D         &57,1,56,1  \\
2009 08 20           & D         &50,2,49,2  \\
2009 08 21           & D         &62,2,61,2 \\
2009 08 27           & E         &90,2,90,2  \\
2009 09 22           & C         &90,2,90,2 \\
2009 10 03           & E         &65,2,67,2  \\
2009 10 09           & A         &1,1,81,1  \\
2009 10 10           & A         &1,1,1,1 \\
2009 11 13           & D         &2,2,2,2 \\
2009 11 14           & D         &2,2,2,2 \\
2009 11 16           & C         &2,2,2,2  \\
2009 11 17           & C         &2,2,2,2  \\
2009 11 19           & C         &2,2,2,2  \\
2009 11 20           & C         &2,2,2,2  \\
2009 11 21           & C         &2,2,2,2  \\
2009 11 25           & B         &2,2,2,2  \\
2010 05 09           & C         &3,2,96,4  \\
2010 05 13           & C         &2,2,2,2  \\
2010 06 08           & C         &2,2,2,2  \\
2010 06 10           & C         &2,2,2,2  \\
2010 06 12           & C         &2,2,2,2  \\
2010 06 19           & E         &61,2,61,2  \\
2010 07 15           & D         &20,2.19,2  \\
2010 07 16           & D         &32,2,32,2  \\
2010 07 17           & E         &78,2,78,2  \\
2010 07 17           & C         &2,2,80,2  \\
2010 07 19           & E         &62,2,62,2  \\
2010 07 21           & E         &72,2,72,2  \\
2010 08 05           & D         &21,2,21,2  \\
2010 08 06           & C         &2,2,2,2  \\
2010 08 07           & C         &2,2,2,2  \\
2010 08 08           & C         &2,2,2,2  \\
2010 08 09           & C         &2,2,2,2  \\
2010 08 10           & D         &15,1,15,1  \\
2010 11 04           & C         &2,2,2,2  \\
2010 11 05           & C         &2,2,2,2  \\ \hline

\end{tabular}
\end{table*}

\begin{table*}
\caption{Observation log of optical photometric observations of 1ES 2344+514}
\textwidth=6.0in
\textheight=9.0in

\vspace*{0.2in}
\noindent
\begin{tabular}{lccl} \\\hline
 Date of             & Telescope    & Data Points \\
observations         &              & Filters     \\
(yyyy mm dd)         &              & (B,V,R,I)      \\\hline
2009 07 23         &   B       &1,1,1,1  \\
2009 08 04         &   E       &52,2,52,2  \\
2009 08 18         &   D       &45,8,41,8 \\
2009 08 19         &   D       &42,6,40,6 \\ 
2009 08 21         &   C       &1,1,1,1  \\
2009 08 22         &   C       &36,1,36,1  \\
2009 08 25         &   C       &44,2,44,2  \\
2009 08 26         &   E       &46,3,48,3  \\
2009 08 28         &   C       &45,2,46,2 \\
2009 08 29         &   C       &54,2,54,2  \\
2009 09 18         &   E       &42,2,42,2 \\
2009 09 21         &   C       &29,1,32,2  \\
2009 10 03         &   E       &32,2,32,2  \\
2009 10 10         &   A        &1,1,65,1  \\
2009 11 13         &   D       &2,2,2,2   \\
2009 11 14         &   D       &2,2,2,2   \\
2009 11 16         &   C       &2,2,2,2   \\
2009 11 17         &   C       &2,2,2,2   \\
2009 11 19         &   C       &2,2,2,2  \\
2009 11 20         &   C       &2,2,2,2  \\
2009 11 21         &   C       &2,2,2,2  \\ 
2009 11 25         &   B        &2,2,2,2  \\
2010 01 10         &   A        &1,1,49,1 \\
2010 01 11         &   A        &1,1,50,1  \\
2010 01 20         &   A        &1,1,30,1  \\ 
2010 06 09         &   C        &2,2,2,2   \\
2010 06 11         &   C        &2,2,2,2   \\
2010 06 13         &   C        &2,2,2,2   \\
2010 07 18         &   C        &2,2,114,2  \\
2010 07 18         &   E        &62,4,62,4 \\
2010 07 20         &   E        &86,2,86,2  \\
2010 07 22         &   E        &17,2,17,2  \\
2010 08 06         &   C        &2,2,2,2  \\
2010 09 08         &   C        &2,2,2,2  \\
2010 09 09         &   C        &2,2,2,2  \\
2010 10 31         &   C        &2,2,2,2  \\
2010 11 01         &   C        &2,2,2,2  \\
2010 11 04         &   C        &2,2,2,2  \\
2010 11 05         &   C        &2,2,2,2  \\
2010 11 06         &   C        &2,2,2,2  \\\hline

\end{tabular} \\
A  : 1.04 meter Sampuranand Telescope, ARIES, Nainital, India  \\
B  : 2-m Ritchey-Chretien Telescope at National Astronomical Observatory, Rozhen, Bulgaria \\
C  : 50/70-cm Schmidt Telescope at National Astronomical Observatory, Rozhen, Bulgaria   \\
D  : 60-cm Cassegrain Telescope at Astronomical Observatory Belogradchik, Bulgaria \\
E  : 1.3-m Skinakas Observatory, Crete, Greece \\
\end{table*}

\begin{table*}
\caption{ Standard stars in the blazar fields }

\begin{tabular}{lcccccc} \hline
Source      & Standard    & B magnitude$^a$  &  V magnitude  &  R magnitude  &  I magnitude$^a$  & References    \\
Name        & star        &  (error)       & (error)        &  (error)       &  (error)          &  \\\hline
1ES 1959+650  &1          &13.37(0.02)     &12.67(0.04)     &12.29(0.02)     &11.92(0.01)       &1, 3   \\
              &2          &13.45(0.02)     &12.86(0.02)     &12.53(0.02)     &12.22(0.01)       &1, 3   \\
              &3          &14.93(0.02)     &13.18(0.02)     &12.27(0.02)     &11.37(0.01)       &1, 3   \\
              &4          &15.28(0.02)     &14.53(0.03)     &14.08(0.03)     &13.62(0.01)       &1, 3   \\
              &5          &15.60(0.03)     &14.54(0.03)     &14.00(0.02)     &13.36(0.02)       &1, 3   \\
              &6          &15.97(0.02)     &15.20(0.03)     &14.78(0.03)     &14.37(0.01)       &1, 3   \\
              &7          &16.01(0.02)     &15.24(0.03)     &14.79(0.03)     &14.37(0.01)       &1, 3   \\
1ES 2344+514  &C1         &-               &12.61(0.04)    &12.25(0.04)   &11.90(0.04)    &2  \\
              &C2         &-                 &14.62(0.06)    &14.20(0.05)   &13.84(0.04)    &2   \\
              &C3         &-                 &15.89(0.08)    &15.40(0.08)   &14.89(0.08)    &2    \\ \hline

\end{tabular} \\
1. Villata et al. 1998;
2. Fiorucci, M., Tosti, G., Rizzi N., et al. 1998. \\
$^a$3. Doroshenko et al. 2007.
\end{table*}

\clearpage
\begin{table*}
\vspace{-0.2in}
\caption{ Results of intra-day variability observations of 1ES 1959+650  }
\vspace{-0.1in}
\begin{tabular}{lccccccc} \hline

Date       &Band       &N      &C-test           &F-test           &$\chi^{2}$test       &ANOVA  &Variable\\
           &           &       &$C_{1},C_{2}$    &$F_{1},F_{2},F_{c}(0.99),F_{c}(0.999)$  &$\chi^{2}_{1},
\chi^{2}_{2},\chi^{2}_{0.99}, \chi^{2}_{0.999}$    &$F_{1},F_{2},F_{c}(0.99),F_{c}(0.999)$  &  \\ \hline
26.07.09  &B   &55   &0.76, 1.02 &0.57, 1.04, 1.90, 2.36    &29.17, 18.20, 81.07, 91.87   &1.33, 3.30, 2.75, 3.78   &NV    \\
          &R   &56   &0.48, 0.96 &0.23, 0.92, 1.89, 2.34    &94.25, 89.98, 82.29, 93.16   &0.92, 3.91, 2.75, 3.78   &NV     \\
         &(B-R)&55   &0.79, 0.78 &0.63, 0.96, 1.90, 2.36    &28.60, 20.94, 81.07, 91.78   &1.18, 1.30, 2.75, 3.78   &NV      \\
27.07.09  &B   &58   &0.86, 0.85 &0.73, 0.72, 1.87, 2.30    &22.31, 24.06, 84.73, 95.75   &0.79, 2.31, 2.66, 3.62   &NV      \\ 
          &R   &58   &0.88, 1.09 &0.77, 1.19, 1.87, 2.30    &30.45, 19.58, 84.73, 95.75   &0.69, 3.30, 2.66, 3.62
&NV      \\
         &(B-R)&58   &0.89, 0.82 &0.80, 0.66, 1.87, 2.30    &18.95, 23.79, 84.73, 95.75   &0.85, 1.57, 2.66, 3.62 
&NV     \\
28.07.09  &B   &51   &0.91, 0.83 &0.83, 0.68, 1.95, 2.44    &17.43, 21.84, 76.15, 86.66   &0.99, 1.22, 2.89, 4.02   &NV     \\ 
          &R   &57   &0.98, 0.94 &0.96, 0.88, 1.88, 2.32    &19.08, 21.29, 83.51, 94.46   &2.10, 3.37, 2.75, 3.78
&NV      \\
         &(B-R)&51   &0.91, 0.82 &0.83, 0.67, 1.95, 2.44    &19.15, 24.50, 76.15, 86.66   &1.02, 1.11, 2.89, 4.02   &NV      \\
29.07.09  &B   &54   &0.74, 0.88 &0.55, 0.78, 1.91, 2.38    &25.96, 19.05, 79.84, 90.57   &1.47, 0.51, 2.76, 3.80   &NV      \\
          &R   &51   &0.75, 0.89 &0.56, 0.80, 1.95, 2.44    &23.67, 17.62, 76.15, 86.66   &1.98, 2.85, 2.89, 4.02   &NV     \\ 
         &(B-R)&53   &0.69, 0.79 &0.47, 0.62, 1.92, 2.40    &21.59, 17.33, 78.62, 89.27   &0.82, 0.25, 2.89, 4.02
&NV      \\
30.07.09  &B   &33   &0.88, 0.79 &0.77, 0.62, 2.32, 3.09    &15.41, 21.59, 53.49, 62.49   &1.07, 2.58, 3.90, 5.98   &NV      \\
          &R   &36   &1.02, 0.94 &1.03, 0.88, 2.23, 2.93    &12.02, 14.14, 57.34, 66.62   &2.91, 0.57, 3.53, 5.24
&NV      \\
         &(B-R)&34   &0.65, 0.78 &0.43, 0.60, 2.29, 3.04    &29.43, 22.30, 54.78, 63.87   &0.70, 0.58, 3.56, 5.31   &NV      \\ 
03.08.09  &B   &42   &0.74, 0.65 &0.55, 0.42, 2.09, 2.69    &30.37, 39.63, 64.95, 74.74   &4.63, 2.59, 3.26, 4.72   &NV       \\
          &R   &42   &0.81, 0.74 &0.65, 0.55, 2.09, 2.69    &18.80, 22.45, 64.95, 74.74   &0.66, 1.57, 3.26, 4.72   &NV       \\
         &(B-R)&42   &0.73, 0.63 &0.53, 0.40, 2.09, 2.69    &37.65, 49.91, 64.95, 74.74   &4.15, 2.13, 3.26, 4.72   &NV       \\
05.08.09 &B   &102   &0.84, 0.83 &0.70, 0.68, 1.59, 1.86    &51.19, 56.30, 136.97, 150,67 &1.50, 1.47, 2.14, 2.72   &NV       \\
          &R   &102  &0.87, 0.80 &0.75, 0.63, 1.59, 1.86    &47.91, 57.05, 136.97, 150.67 &1.34, 1.09, 2.14. 2.72   &NV       \\
         &(B-R)&102  &0.80, 0.80 &0.64, 0.64, 1.59, 1.86    &57.75, 62.16, 136.97, 150.67 &1.21, 1.62, 2.
14. 2.72   &NV       \\
17.08.09 &B   &56    &0.85, 0.94 &0.71, 0.88, 1.89, 2.34    &67.24, 60.91, 82.29, 93.17   &0.37, 1.03, 2.75, 3.78   &NV       \\
         &R   &56    &0.95, 1.00 &0.91, 0.99, 1.89, 2.34    &41.80, 36.99, 82.29, 93.17   &0.43, 1.14, 2.
75, 3.78   &NV       \\
         &(B-R)&54   &0.82, 0.81 &0.67, 0.65, 1.91, 2.38    &52.40, 58.65, 79.84, 90.57   &0.24, 1.49, 2.76, 3.80  &NV       \\
20.08.09 &B   &44    &0.98, 1.08 &0.97, 1.16, 2.06, 2.63    &34.34, 37.20, 67.46, 77.42   &0.52, 0.52, 3.07, 4.36  &NV       \\
         &R   &49    &0.98, 0.82 &0.96, 0.68, 1.98, 2.49    &30.25, 38.70, 73.68, 84.03   &2.93, 2.73, 2.90, 4.05  &NV       \\
         &(B-R)&44   &1.52, 1.20 &1.33, 1.44, 2.06, 2.63    &36.38, 41.95, 67.46, 77.42   &0.54, 0.67, 3.07, 4.36  &NV       \\
21.08.09 &B   &58    &0.67, 0.75 &0.45, 0.56, 1.87, 2.30    &39.05, 36.87, 84.73, 95.75   &1.09, 0.63, 2.75, 3.78  &NV       \\
         &R   &58    &0.92, 0.87 &0.85, 0.75, 1.87, 2.30    &35.54, 39.42, 84.73, 95.75   &1.47, 1.32, 2.75, 3.78  &NV       \\
         &(B-R)&57   &0.62, 0.82 &0.39, 0.68, 1.88, 2.32    &44.06, 28.95, 83.51, 94.46   &0.92, 0.97,  2.75, 3.78  &NV       \\ 
27.08.09 &B   &90    &1.23, 1.21 &1.52, 1.47, 1.64, 1.94    &62.78, 66.28, 122.94, 135.98 &2.79, 2.69, 2.23, 2.86   &NV       \\
         &R   &90    &1.08, 1.17 &1.63, 1.36, 1.64, 1.94    &81.52, 69.82, 122.94, 135.98 &2.96, 2.49, 2.23, 2.8
6   &NV       \\
         &(B-R)&90   &1.15, 1.16 &1.32, 1.34, 1.64, 1.94    &74.46, 74.04, 122.94, 135.98 &1.41, 1.64, 2.23, 2.8
6   &NV       \\
22.09.09 &B   &48    &0.86, 0.81 &0.75, 0.65, 1.99, 2.51    &31.11, 37.60, 72.44, 82.72   &3.81, 0.81, 3.05, 4.33   &NV       \\
         &R   &49    &0.94, 0.75 &0.88, 0.56, 1.98, 2.49    &25.84, 43.25, 73.68, 84.03   &1.81, 1.04, 3.05, 4.3
3   &NV       \\
         &(B-R)&48   &0.98, 0.83 &0.96, 0.69, 1.99, 2.51    &46.49, 65.99, 72.44, 82.72   &0.77, 1.28, 3.05, 4.33   &NV       \\
03.10.09 &B   &60    &0.88, 0.98 &0.78, 0.96, 1.85, 2.27    &35.08, 32.29, 87.17, 98.32   &1.20, 1.03, 2.64, 3.58   &NV       \\
         &R   &60    &1.01, 1.00 &1.03, 1.00, 1.85, 2.27    &42.00, 44.85, 87.17, 98.32   &1.18, 2.16, 2.64, 3.58   &NV       \\
         &(B-R)&60   &0.99, 0.94 &0.97, 0.89, 1.85, 2.27    &35.12, 41.80, 87.17, 98.32   &0.88, 0.66, 2.64, 3.5
8   &NV       \\
09.10.09 &R   &80    &1.03, 0.97 &1.06, 0.93, 1.70, 2.02    &52.18, 56.23, 111.14, 123.59  &2.17, 2.90, 2.33, 3.04      &NV       \\
09.05.10 &R   &96    &1.42, 1.32 &2.03, 1.74, 1.62, 1.90    &41.08, 44.48, 129.97, 143.34  &1.09, 0.66, 2.18, 2.79      &NV       \\
19.06.10 &B   &61    &1.08, 1.08 &1.18, 1.16, 1.84, 2.25    &34.68, 34.60, 88.38, 99.61   &0.66, 0.85, 2.64, 3.58    &NV       \\
         &R   &61    &1.04, 1.07 &1.07, 1.14, 1.84, 2.25    &34.33, 30.73, 88.38, 99.61   &0.86, 1.74, 2.64, 3.5
8    &NV       \\
        &(B-R)&61    &0.97, 0.93 &0.95, 0.87, 1.84, 2.25    &35.47, 35.16, 88.38, 98.61   &0.53, 1.35, 2.64, 3.5
8    &NV       \\
15.07.10 &B   &20    &0.76, 0.99 &0.58, 0.99, 3.03, 4.47    &23.79, 9.21, 36.19, 43.82    &1.37, 0.44, 5.42, 9.34    &NV       \\
         &R   &18    &1.04, 1.50 &1.09, 0.25, 3.24, 4.92    &3.51, 15.06, 33.41, 40.79    &0.02, 0.17, 6.93, 12.97  &NV       \\
        &(B-R)&19    &0.85, 1.15 &0.72, 1.32, 3.13, 4.68    &28.38, 10.55, 34.81, 42.31   &1.37, 0.44, 5.42, 9.34    &NV       \\
16.07.10 &B   &28    &1.26, 1.27 &1.58, 1.60, 2.51, 3.44    &21.17, 15.18, 46.96, 55.48   &0.33, 1.71, 3.99, 6.19    &NV       \\
         &R   &30    &1.20, 1.19 &1.43, 1.43, 2.42, 3.29    &16.59, 16.66, 49.59, 58.30   &1.45, 3.50, 3.90, 5.98    &NV       \\
        &(B-R)&25    &1.18, 1.18 &1.39, 1.39, 2.66, 3.74    &10.39, 07.40, 42.98, 51.18   &0.24, 0.91, 4.43, 7.10    &NV       \\
17.07.10 &R   &80    &1.06, 0.90 &1.12, 0.80, 1.70, 2.02    &57.98, 83.01, 111.14, 123.59  &1.19, 0.82, 2.33, 3.04      &NV       \\ 
17.07.10 &B   &78    &1.38, 1.04 &1.91, 1.08, 1.71, 2.04    &44.26, 52.82, 108.77, 121.10 &2.16, 1.63, 2.39, 3.15    &NV       \\
         &R   &78    &1.24, 1.17 &1.55, 1.36, 1.71, 2.04    &52.97, 43.59, 108.77, 121.10 &1.49, 2.40, 2.39, 3.15    &NV       \\
        &(B-R)&78    &1.36, 1.14 &1.84, 1.31, 1.71, 2.04    &74.62, 76.24, 108.77, 121.10 &3.28, 2.73, 2.39, 3.15    &NV       \\
19.07.10 &B   &62    &1.18, 0.90 &1.40, 0.81, 1.83, 2.24    &28.47, 35.09, 89.59, 100.89  &4.06, 1.93, 2.64, 3.58    &NV       \\
         &R   &62    &1.02, 1.02 &1.03, 1.05, 1.83, 2.24    &34.71, 24.71, 89.59, 100.89  &0.82, 1.92, 2.64, 3.58    &NV       \\
        &(B-R)&62    &1.14, 0.95 &1.31, 0.91, 1.83, 2.24    &43.18, 43.04, 89.59, 100.89  &2.23, 1.51, 2.64, 3.58    &NV       \\
21.07.10 &B   &72    &1.16, 1.05 &1.35, 1.09, 1.75, 2.11    &43.77, 37.64, 101.62, 113.58 &1.49, 0.77, 2.47, 3.27    &NV       \\
         &R   &72    &1.29, 1.08 &1.65, 1.16, 1.75, 2.11    &49.95, 51.39, 101.62, 113.58 &1.80, 1.44, 2.47, 3.27    &NV       \\
        &(B-R)&72    &1.20, 1.05 &1.45, 1.10, 1.75, 2.11    &57.21, 50.85, 101.62, 113.58 &1.58, 0.93, 2.47, 3.27    &NV       \\
05.08.10 &B   &19    &2.31, 2.05 &5.34, 4.19, 3.13, 4.68    &130.68, 85.15, 34.81, 42.31  &5.00, 12.38, 5.42, 9.34   &NV       \\
         &R   &20    &1.91, 1.96 &3.64, 3.84, 3.03, 4.47    &19.83, 14.16, 36.19, 43.82   &2.90, 4.54, 5.29, 9.01    &NV       \\
        &(B-R)&20    &2.15, 1.83 &4.64, 3.36, 3.03, 4.47    &96.74, 69.81, 36.19, 43.82   &4.95, 11.4, 5.29, 9.01    &NV       \\
10.08.10 &B   &12    &1.38, 0.99 &1.91, 0.98, 4.46, 7.76    &14.21, 16.80, 24.72, 31.26   &0.001, 0.02, 11.26, 25.41   &NV       \\
         &R   &14    &1.34, 0.94 &1.80, 0.89, 3.91, 6.41    &5.70, 10.91, 27.69, 34.53    &0.20, 0.68, 7.21, 13.81     &NV       \\
        &(B-R)&14    &1.43, 1.17 &2.05, 1.36, 3.91, 6.41    &21.69, 27.11, 27.69, 34.53    &0.29, 0.18, 7.21, 13.81    &NV       \\ \hline
\end{tabular}

$ $V : Variable, NV : Non-Variable    \\
\end{table*}

\clearpage
\begin{table*}
\caption{ Results of intra-day variability observations of 1ES 2344+514  }

\begin{tabular}{lccccccc} \hline

Date       &Band       &N      &C-test           &F-test           &$\chi^{2}$test       &ANOVA  &Variable\\
           &           &       &$C_{1},C_{2}$    &$F_{1},F_{2},F_{c}(0.99),F_{c}(0.999)$  &$\chi^{2}_{1},
\chi^{2}_{2},\chi^{2}_{0.99}, \chi^{2}_{0.999}$    &$F_{1},F_{2},F_{c}(0.99),F_{c}(0.999)$  &  \\ \hline

04.08.09   &B    &52  &1.03, 1.61   &1.05, 2.60, 1.94, 2.42   &33.86, 32.56, 77.39, 87.97   &3.88, 2.89, 2.89, 4.02 &NV  \\
           &R    &52  &1.06, 1.27   &1.11, 1.60, 1.94, 2.42   &58.70, 93.56, 77.39, 87.97   &3.17, 0.95, 2.89, 4.02 &NV  \\
           &(B-R)&52  &0.79, 1.43   &0.63, 2.06, 1.94, 2.42   &65.52, 49.98, 77.39, 87.97   &2.54, 2.24, 2.89, 4.02 &NV  \\ 
18.08.09   &B    &40  &1.02, 1.19   &1.04, 1.43, 2.14, 2.76   &29.53, 42.65, 62.43, 72.05   &1.38, 0.54, 3.26, 4.72 &NV  \\
           &R    &39  &1.02, 1.13   &1.03, 1.27, 2.16, 2.80   &28.04, 53.42, 61.16, 70.70   &1.12, 0.99, 3.28, 4.77 &NV  \\
           &(B-R)&40  &0.95, 1.14   &0.91, 1.30, 2.14, 2.76   &25.29, 34.61, 62.43, 72.05   &1.11, 0.29, 3.26, 4.72 &NV  \\ 
19.08.09   &B    &40  &1.26, 1.54   &1.59, 2.37, 2.14, 2.76   &42.25, 57.83, 62.43, 72.05   &0.79, 0.66, 3.26, 4.72 &NV  \\
           &R    &35  &0.88, 1.13   &0.77, 1.27, 2.16, 2.80   &21.39, 23.44, 61.16, 70.70   &4.47, 1.86, 3.52, 5.24 &NV  \\
           &(B-R)&38  &1.49, 1.60   &2.21, 2.55, 2.18, 2.84   &42.29, 68.94, 59.89, 69.35   &0.63, 0.52, 3.30, 4.82 &NV  \\ 
22.08.09   &B    &36  &0.89, 1.22   &0.79, 1.26, 2.23, 2.93   &~3.30, ~4.02, 57.34, 66.62     &1.35, 1.03, 3.53, 5.24 &NV  \\
           &R    &36  &1.00, 1.14   &0.99, 1.30, 2.23, 2.93   &~5.71, ~9.47, 57.34, 66.62     &8.38, 1.22, 3.53, 5.24 &NV  \\
           &(B-R)&36  &0.91, 1.31   &0.83, 1.72, 2.23, 2.93   &~4.13, ~4.02, 57.34, 66.62     &1.68, 1.14, 3.53, 5.24 &NV  \\

25.08.09   &B    &42  &0.79, 1.14   &0.63, 1.29, 2.09, 2.69   &33.59, 40.27, 64.95, 74.74   &1.34, 0.91, 3.26, 4.72 &NV \\
           &R    &41  &0.81, 1.13   &0.65, 1.28, 2.11, 2.73   &31.64, 34.89, 63.69, 73.40   &3.29, 3.01, 3.26, 4.72 &NV \\
           &(B-R)&40  &0.90, 1.17   &0.81, 1.38, 2.14, 2.76   &30.94, 38.88, 62.43, 72.05    &0.93, 1.61, 3.26, 4.72 &NV \\ 
26.08.09   &B    &45  &0.70, 1.28   &0.49, 1.64, 2.04, 2.60   &45.26, 27.02, 68.71, 78.75    &2.57, 2.98, 3.05, 4.33 &NV \\
           &R    &44  &0.79, 1.21   &0.62, 1.47, 2.06, 2.63   &44.04, 39.80, 67.46, 77.42    &3.19, 1.41, 3.07, 4.36 &NV  \\
           &(B-R)&43  &0.74, 1.28   &0.54, 1.63, 2.08, 2.66   &47.39, 32.06, 66.21, 76.08    &2.25, 2.51, 3.09, 4.40 &NV  \\ 
28.08.09   &B    &41  &0.91, 1.08   &0.82, 1.67, 2.11, 2.73   &30.99, 47.19, 63.69, 73.40    &2.89, 0.45, 3.26, 5.72 &NV  \\
           &R    &42  &0.79, 1.31   &0.62, 1.73, 2.09, 2.69   &58.94, 48.06, 64.95, 74.74    &2.08, 1.27, 3.26, 4.72 &NV   \\
           &(B-R)&40  &0.84, 1.19   &0.70, 1.43, 2.14, 2.76   &45.68, 51.62, 63.69, 73.40    &1.10, 0.21, 3.26, 4.72 &NV   \\
29.08.09   &B    &54  &1.19, 1.08   &1.41, 1.67, 1.91, 2.38   &65.84, 184,10, 70.84, 90.57   &23.90, 5.31, 2.76, 3.80 &NV   \\
           &R    &54  &0.96, 0.53   &0.87, 0.28, 1.91, 2.38   &23.77, 168.72, 70.84, 90.57   &45.68, 0.46, 2.78, 3.80 &NV   \\
           &(B-R)&54  &0.82, 1.37   &0.67, 1.86, 1.92, 2.40   &74.12, 63.93, 78.62, 89.27    &4.31, 6.01, 2.78, 3.82 &NV    \\
18.09.09   &B    &41  &0.84, 1.03   &0.71, 1.05, 2.11, 2.73   &18.36, 26.20, 63.69, 73.40    &1.86, 0.44, 3.26, 4.72 &NV   \\
           &R    &41  &0.77, 1.24   &0.60, 1.55, 2.11, 2.73   &40.81, 30.21, 63.69, 73.40    &2.51, 2.65, 3.26, 4.72 &NV   \\
           &(B-R)&41  &0.85, 1.17   &0.72, 1.37, 2.11, 2.73   &28.46, 33.43, 63.69, 73.40    &2.11, 0.59, 3.26, 4.72 &NV  \\
21.09.09   &B    &32  &1.35, 1.74   &1.83, 3.03, 2.35, 3.15   &26.68, 36.38, 52.19, 61.10    &4.93, 2.99, 3.90, 5.98 &NV   \\
           &R    &29  &0.79, 1.12   &0.63, 1.25, 2.46, 3.36   &17.05, 21.27, 48.29, 56.89    &1.07, 0.13, 3.94, 6.08 &NV   \\
           &(B-R)&24  &0.89, 1.05   &0.80, 1.11, 2.72, 3.85   &21.50, 35.26, 41.64, 49.73    &0.93, 0.37, 4.50, 7.27 &NV   \\
03.10.09   &B    &24  &0.81, 1.34   &0.65, 1.80, 2.72, 3.85   &13.61, 10.86, 41.64, 49.73    &0.17, 0.12, 4.50, 7.27 &NV   \\
           &R    &26  &1.04, 1.49   &1.09, 2.22, 2.60, 3.63   &24.47, 31.23, 44.31, 52.62    &3.06, 2.20, 4.43, 7.10 &NV   \\
           &(B-R)&25  &0.85, 1.35   &0.72, 1.83, 2.66, 3.74   &15.18, 13.49, 42.98, 51.18    &0.31, 0.40, 4.43, 7.10 &
NV    \\
10.10.09   &R    &65  &0.95, 0.14   &0.90, 0.02, 1.80, 2.19   &33.84, 62.33, 93.22, 104.72   &4.27, 2.73, 2.55, 3.41
&NV    \\
10.01.10   &R    &49  &1.45, 1.63   &2.06, 2.66, 1.98, 2.49   &31.55, 56.13, 73.68, 84.04    &6.71, 6.99, 2.90, 4.05 &NV    \\
11.01.10   &R    &50  &0.68, 0.82   &0.46, 0.67, 1.96, 2.46   &17.32, 26.85, 74.92, 85.35    &2.41, 1.07, 2.89, 4.02  &NV    \\
20.01.10   &R    &30  &0.88, 1.10   &0.77, 1.22, 2.42, 3.29   &12.68, 19.40, 49.59, 58.30    &3.47, 0.70, 3.90, 5.98   &NV    \\
18.07.10   &R    &115 &0.84, 1.14   &0.72, 1.31, 1.55, 1.79   &63.09, 67.71, 152.04, 166.41  &0.78, 1.11, 2.04, 2.55   &NV    \\
18.07.10   &B    &62  &0.85, 1.20   &0.72, 1.45, 1.83, 2.24   &31.49, 36.59, 89.59, 100.89   &0.45, 1.89, 2.64, 3.58 &NV    \\
           &R    &62  &1.06, 1.39   &1.13, 1.92, 1.83, 2.24   &60.00, 67.95, 89.59, 100.89   &4.09, 3.75, 2.64, 3.58 &
NV    \\
           &(B-R)&62  &0.82, 1.19   &0.67, 1.41, 1.83, 2.24   &42.62, 42.68, 89.59, 100.89   &0.87, 1.92, 2.64, 3.59 &
NV    \\
20.07.10   &B    &86  &0.69, 1.20   &0.47, 1.43, 1.66, 1.97   &56.45, 50.61, 118.24, 131.04  &0.87, 1.05, 2.28, 2.95 &
NV   \\
           &R    &86  &0.93, 1.12   &0.86, 1.25, 1.66, 1.97   &55.47, 99.15, 118.24, 131.04  &1.36, 1.00, 2.28, 2.95 &
NV   \\
           &(B-R)&86  &0.77, 1.23   &0.59, 1.52, 1.66, 1.97   &63.49, 75.00, 118.24, 131.04  &0.69, 0.70, 2.28, 2.95 &
NV   \\
22.07.10   &B    &16  &0.83, 1.12   &0.68, 1.25, 3.52, 5.54   &~3.22, ~4.71, 30.58, 37.70    &1.28, 0.37, 6.93, 12.97 & NV  \\
           &R    &17  &0.71, 0.78   &0.51, 0.61, 3.37, 5.20   &~4.44, ~8.51, 32.00, 39.25    &0.27, 4.04, 6.93, 12.97 &NV   \\
           &(B-R)&16  &0.69, 1.05   &0.48, 1.10, 3.52, 5.54   &~3.40, ~3.72, 30.58, 37.70    &2.12, 1.90, 6.93, 12.97  &NV \\ \hline
\end{tabular}
$ $V : Variable, NV : Non-Variable    \\
\end{table*}

\clearpage
\begin{table*}
\caption{ Results of short-term variability observations  }

\begin{tabular}{lclcccc} \hline

Source Name             & Band       &N     &C-Test      &F              &Variable &A (\%)\\
                        &            &   &$C_{1},C_{2}$  &$F_{1},F_{2},F_{c}(0.99),F_{c}(0.999)$  & &  \\\hline
1ES 1959+650            &B           &75    &4.25, 4.21     &18.05, 17.69, 1.73, 2.07   &V  &102.4\\
                        &V           &75    &5.48, 5.65     &29.99, 31.91, 1.73, 2.07   &V  &90.4   \\
                        &R           &81    &11.87, 11.83   &140.80, 139.89, 1.69, 2.01 &V  &81.5   \\
                        &I           &78    &9.09, 9.14     &82.59, 83.49, 1.71, 2.04   &V  &74.5   \\
                        &(V-R)       &75    &3.26, 2.84     &10.64, 8.09, 1.73, 2.07    &V  &36.0   \\
1ES 2344+514            &B           &39    &1.50, 0.61     &2.24, 0.37, 2.16, 2.80     &NV &-   \\
                        &V           &39    &1.95, 1.54     &3.78, 2.38, 2.16, 2.80     &NV &-   \\
                        &R           &38    &1.94, 1.37     &3.77, 1.88, 2.18, 2.84     &NV &-   \\
                        &I           &39    &1.72, 1.39     &2.97, 1.94, 2.16, 2.80     &NV &-   \\
                        &(V-R)       &38    &1.08, 0.83     &1.17, 0.69, 2.18, 2.84     &NV &-   \\ \hline

\end{tabular}

$ $V : Variable, NV : Non-Variable    

\end{table*}

\clearpage

\begin{figure*}
\epsfig{figure= 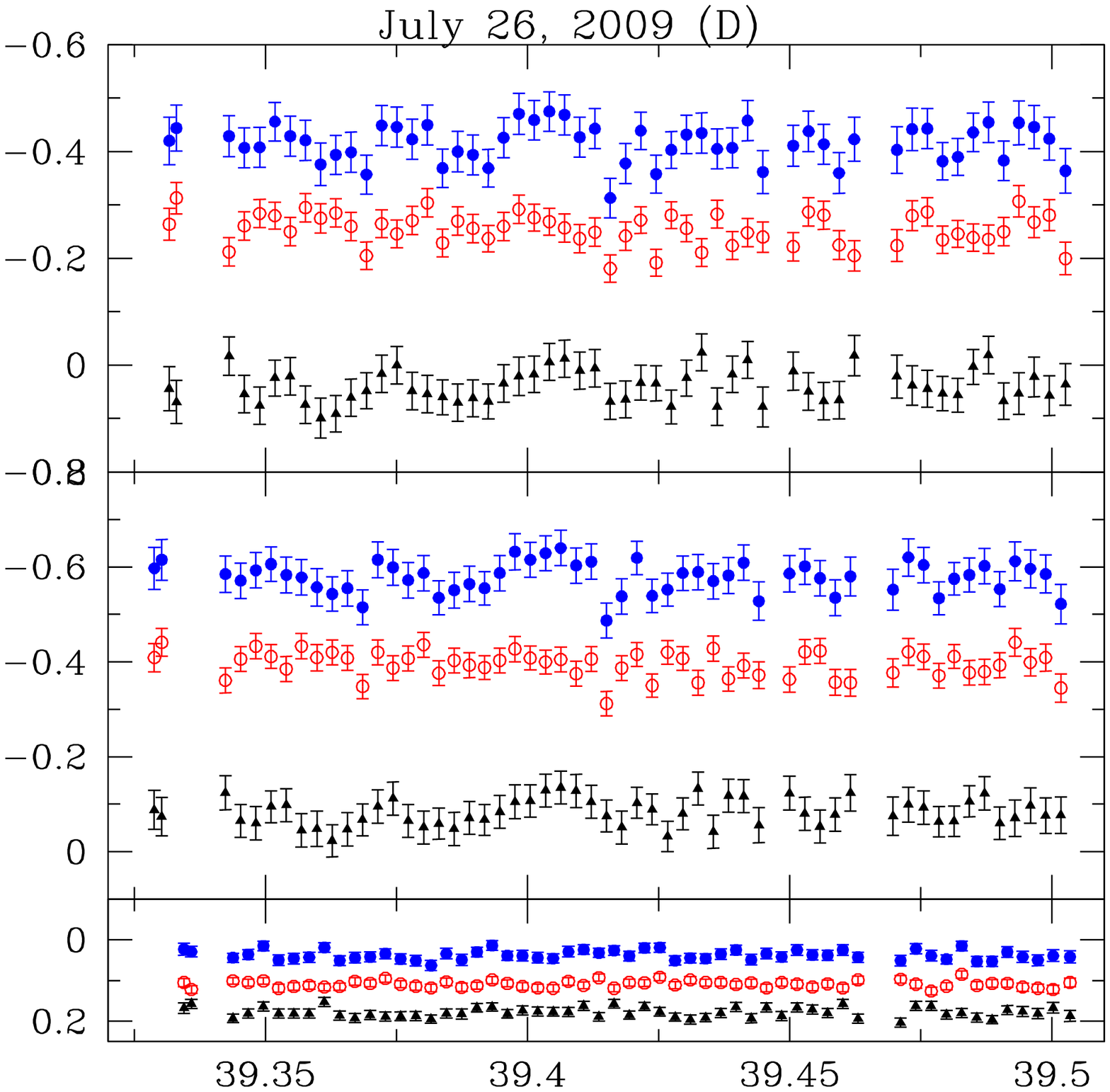,height=2.0in,width=2.2in,angle=0}
\epsfig{figure= 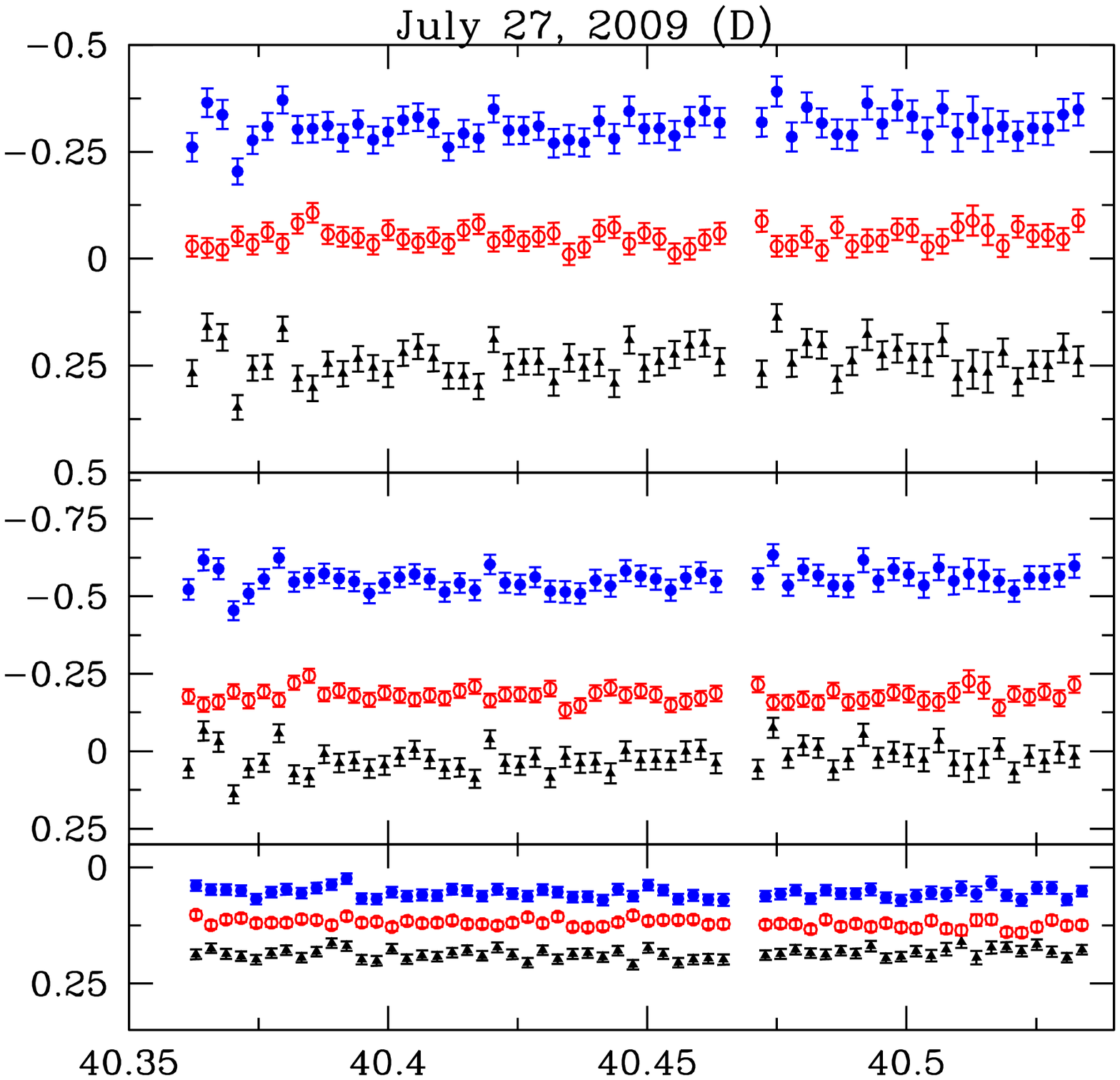,height=2.0in,width=2.2in,angle=0}
\epsfig{figure= 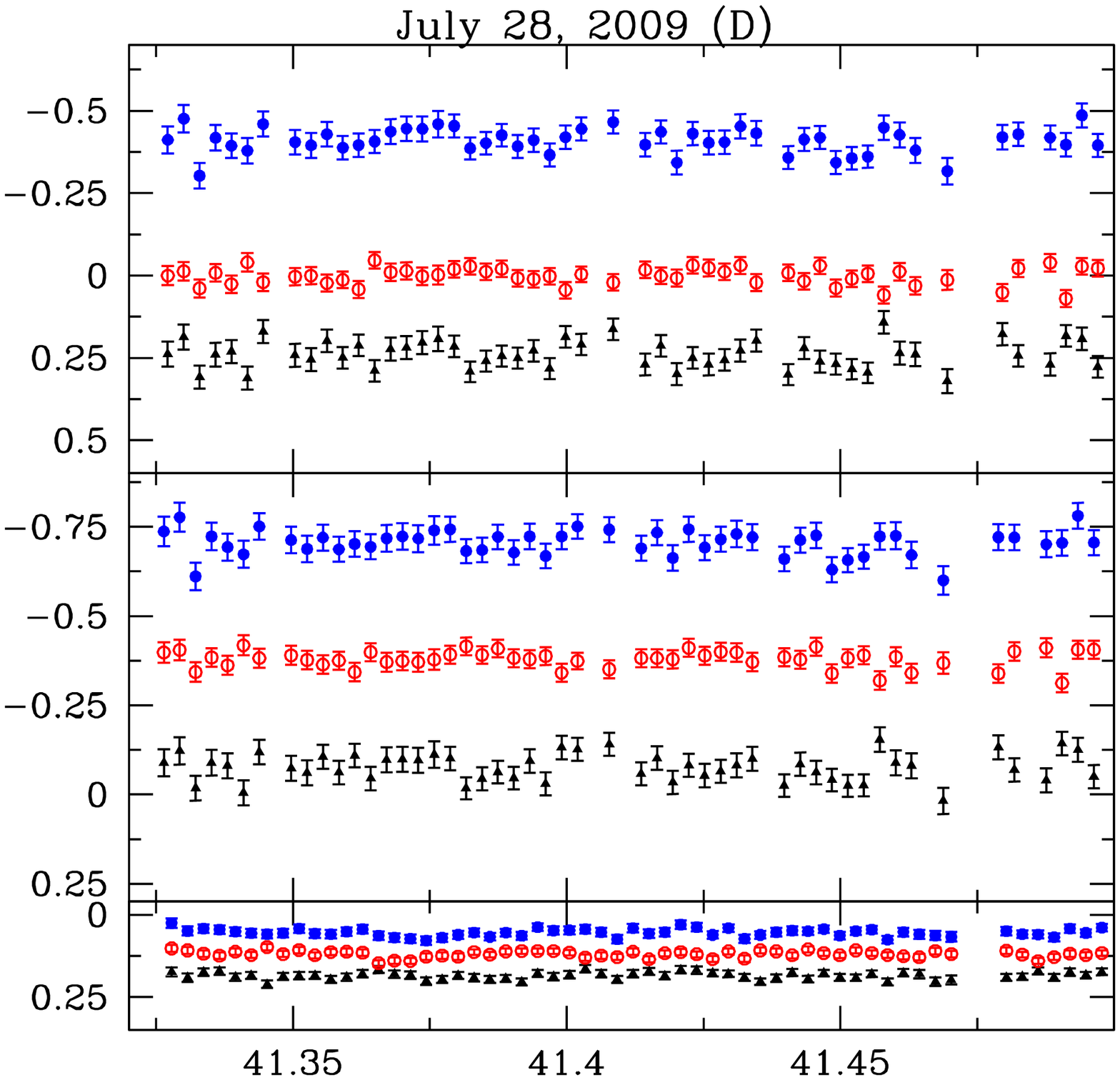,height=2.0in,width=2.2in,angle=0}
\epsfig{figure= 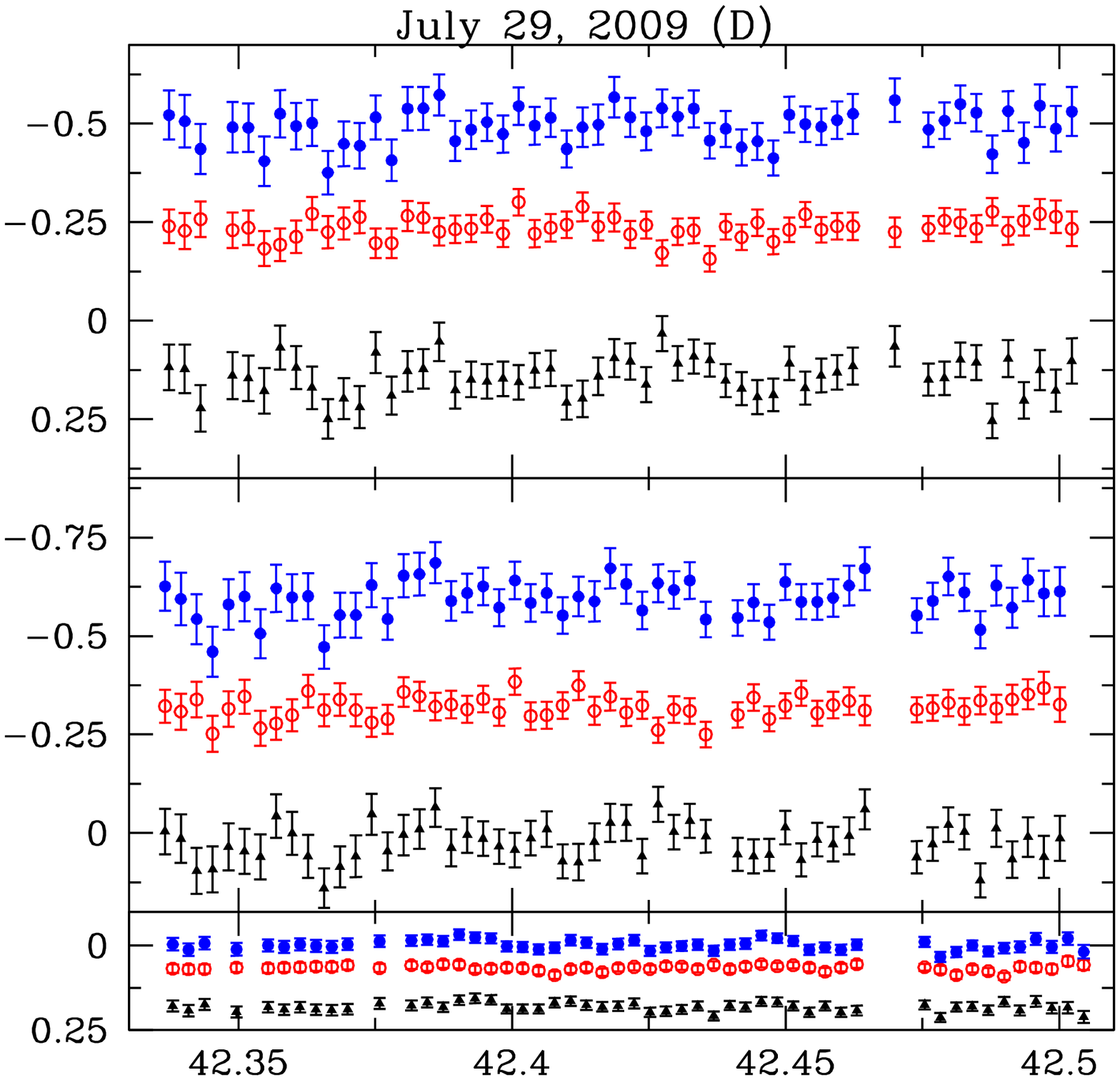,height=2.0in,width=2.2in,angle=0}
\epsfig{figure= 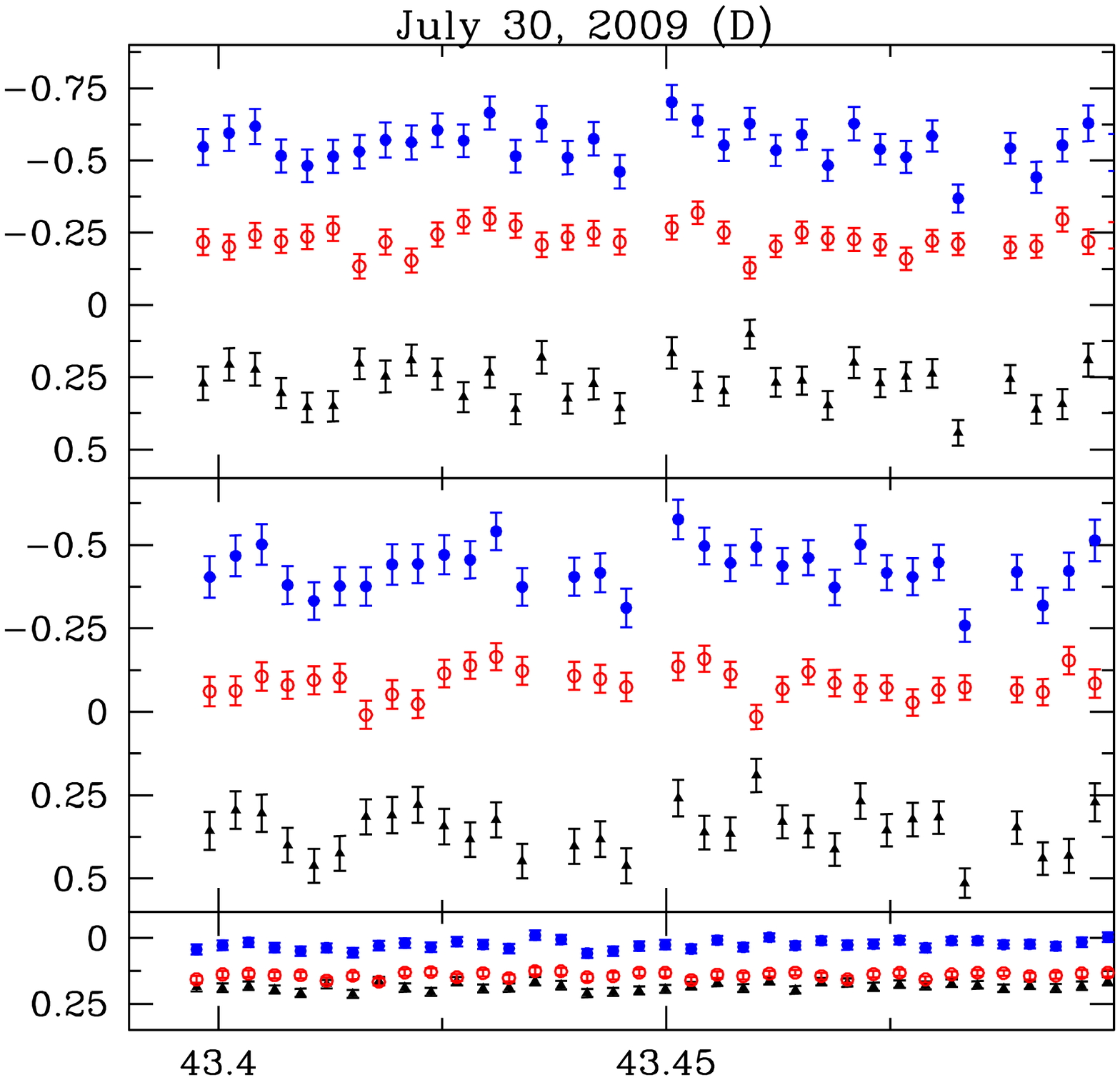,height=2.0in,width=2.2in,angle=0}
\epsfig{figure= 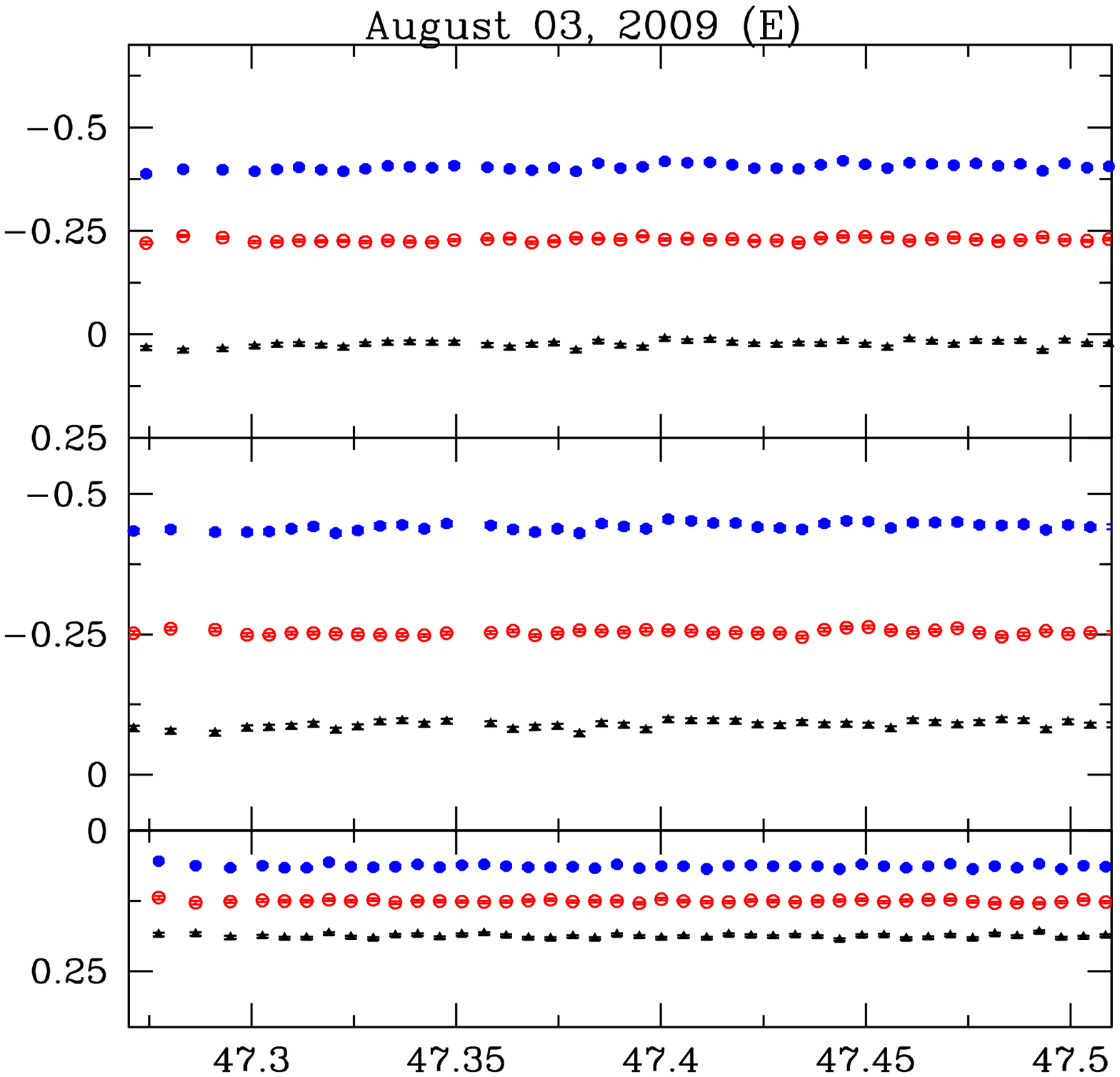,height=2.0in,width=2.2in,angle=0}
\epsfig{figure= 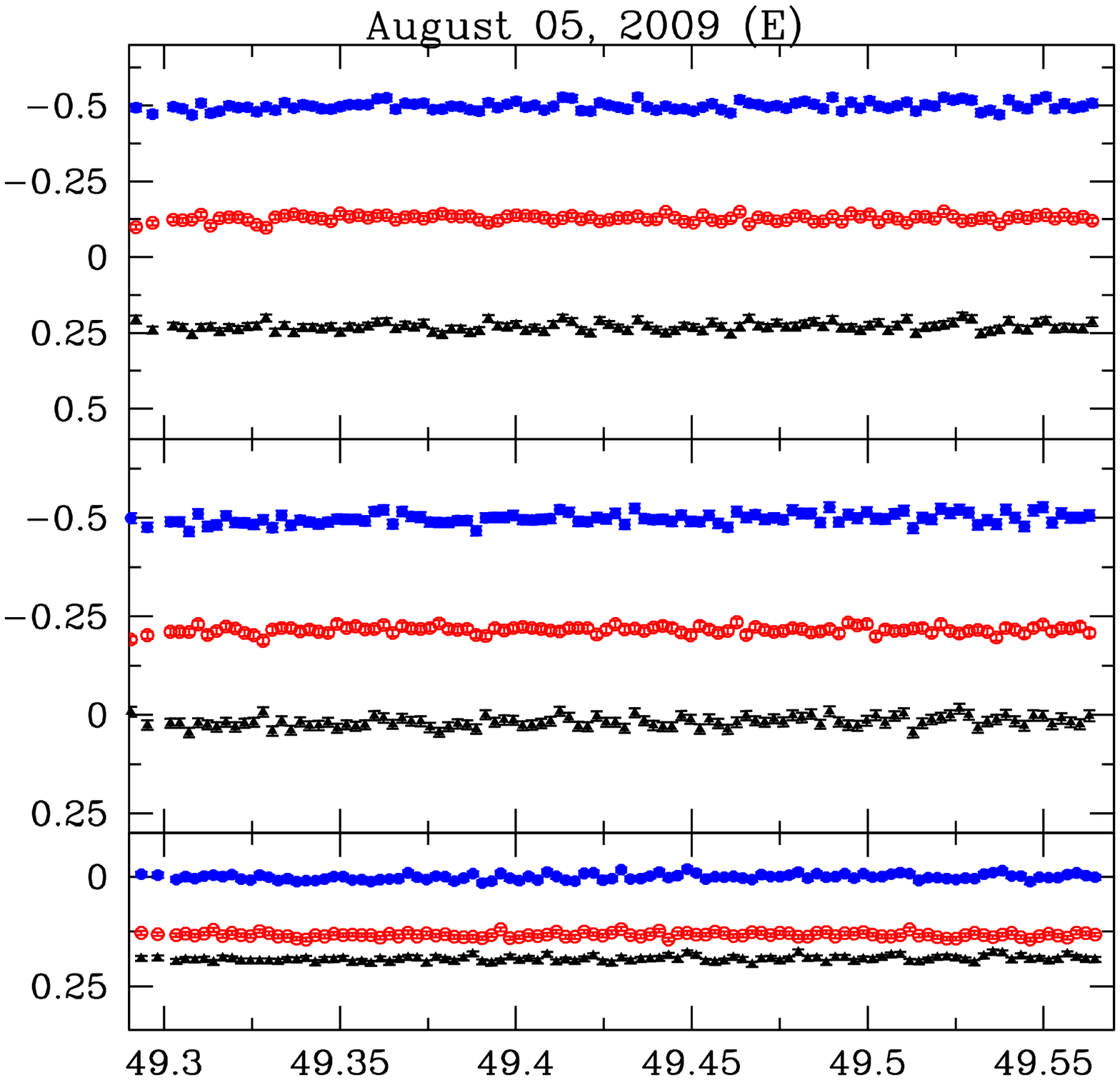,height=2.0in,width=2.2in,angle=0}
\epsfig{figure= 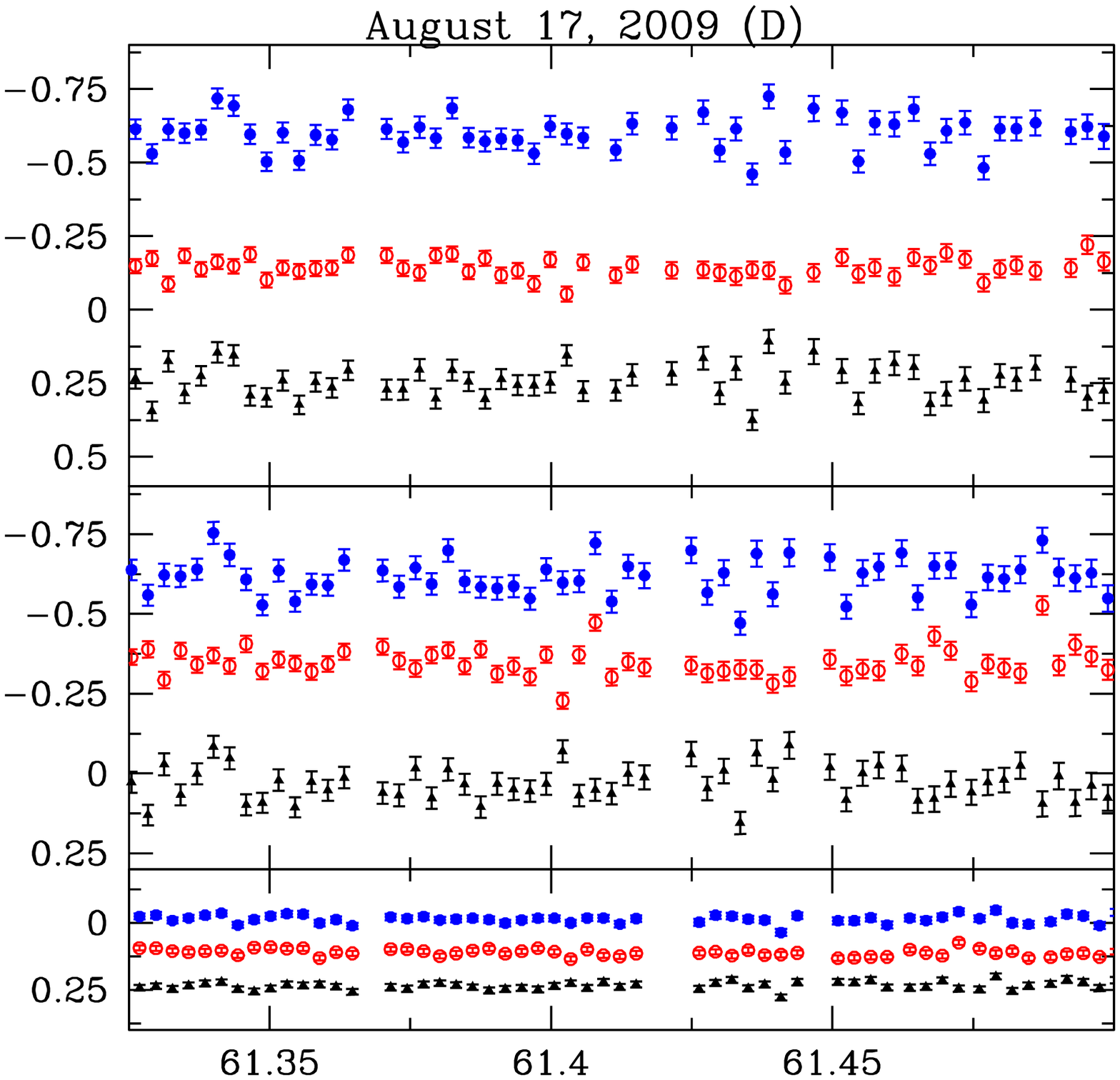,height=2.0in,width=2.2in,angle=0}
\epsfig{figure= 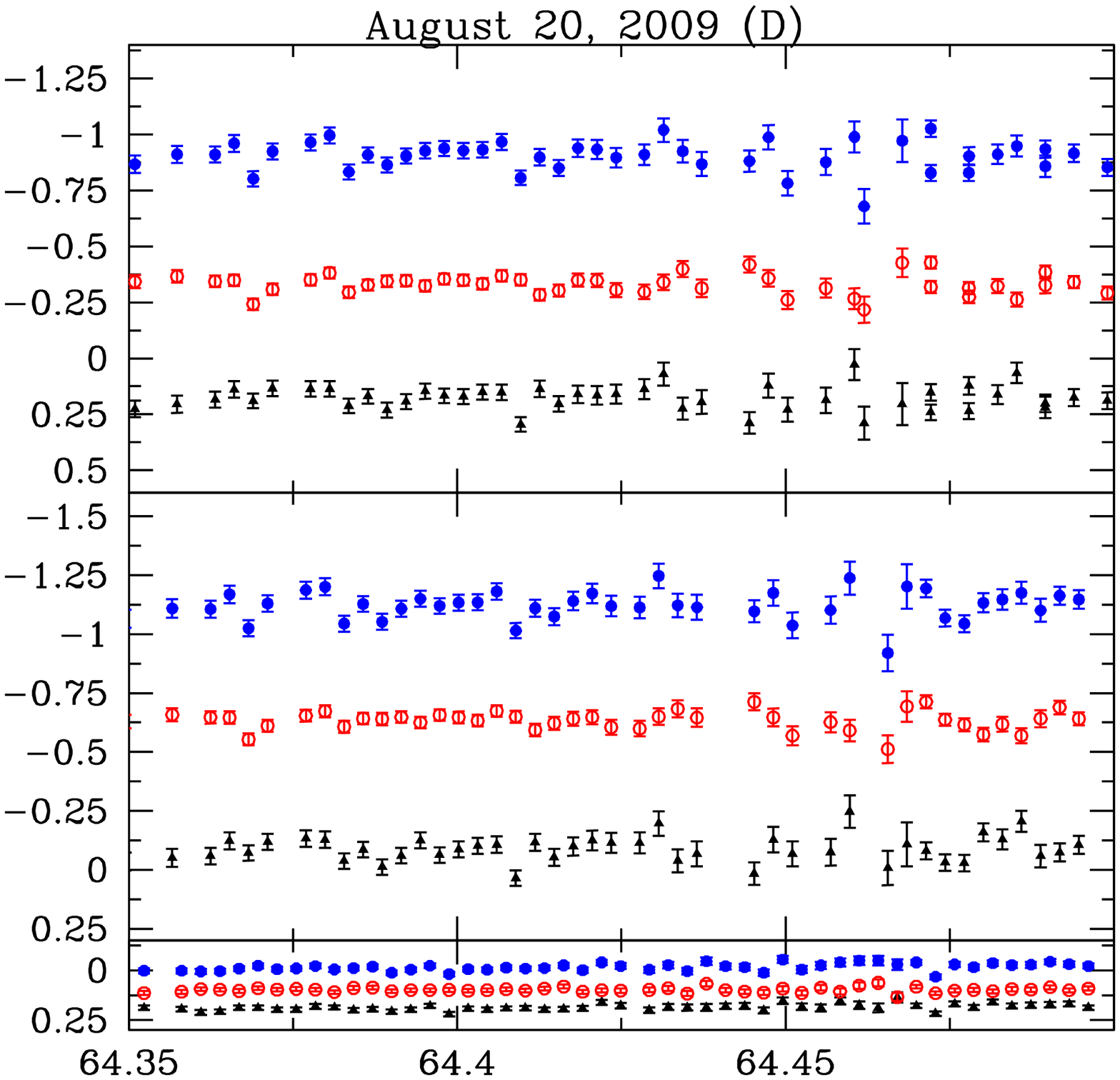,height=2.0in,width=2.2in,angle=0}
\epsfig{figure= 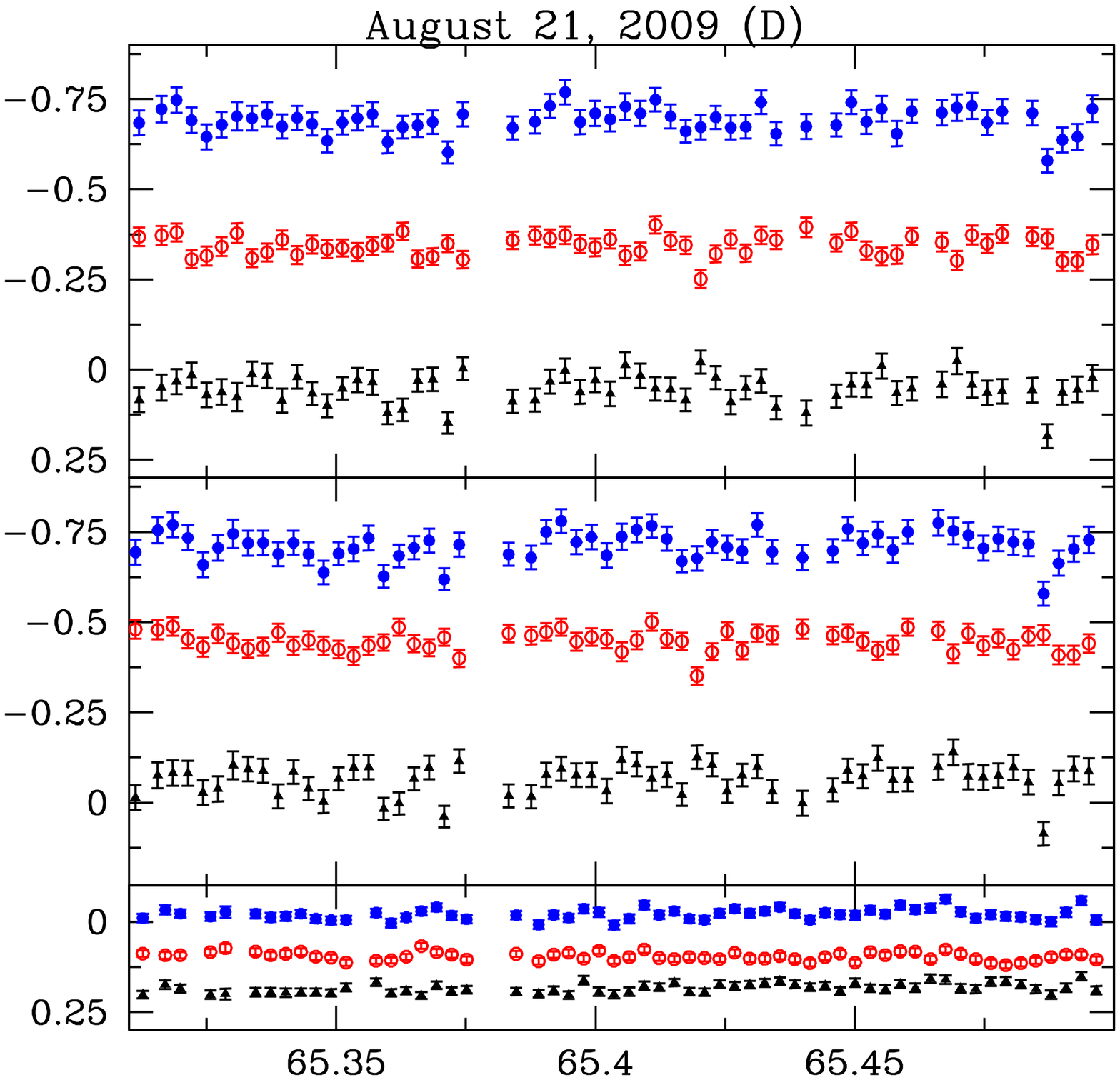,height=2.0in,width=2.2in,angle=0}
\epsfig{figure= 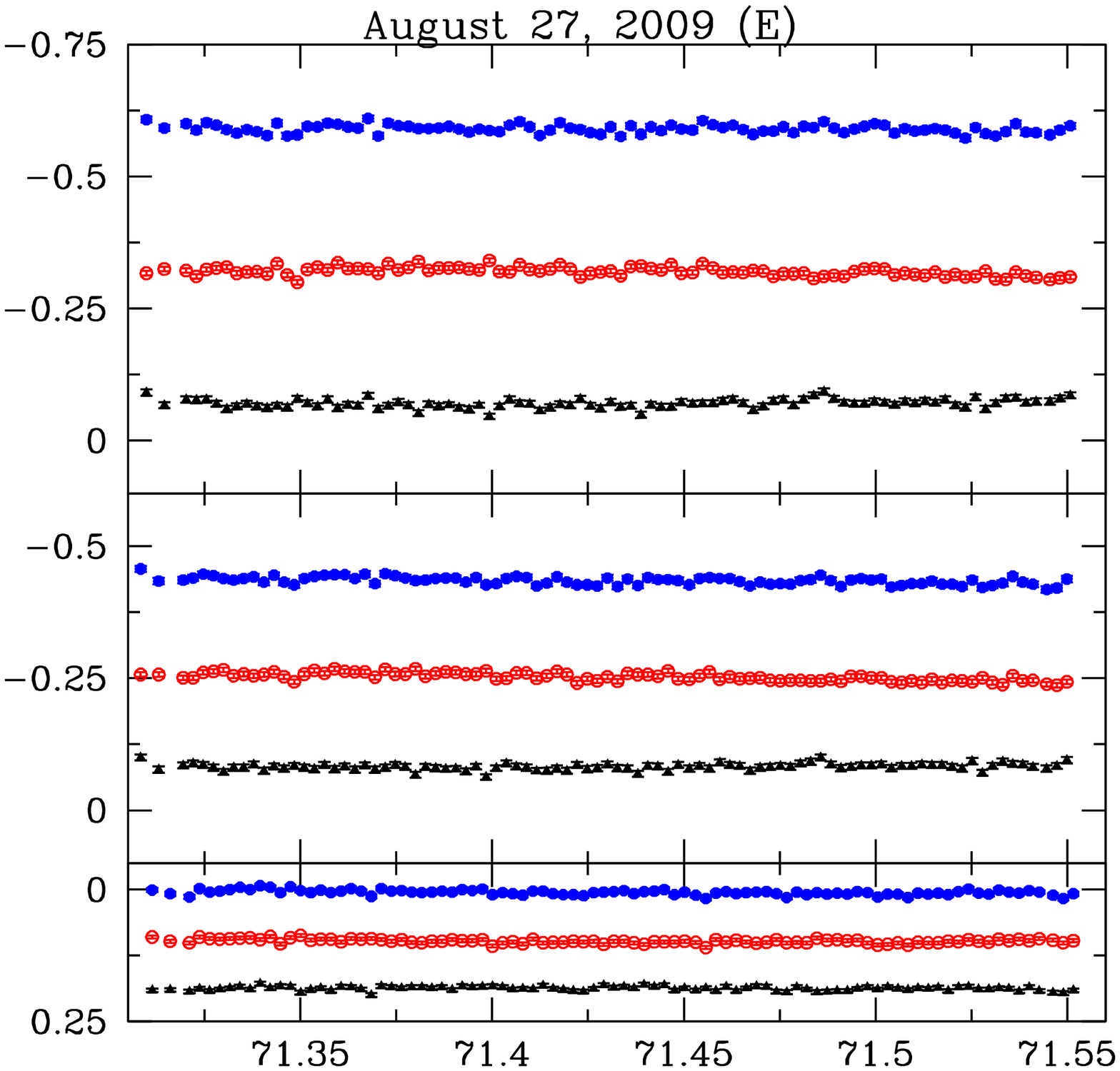,height=2.0in,width=2.2in,angle=0}
\epsfig{figure= 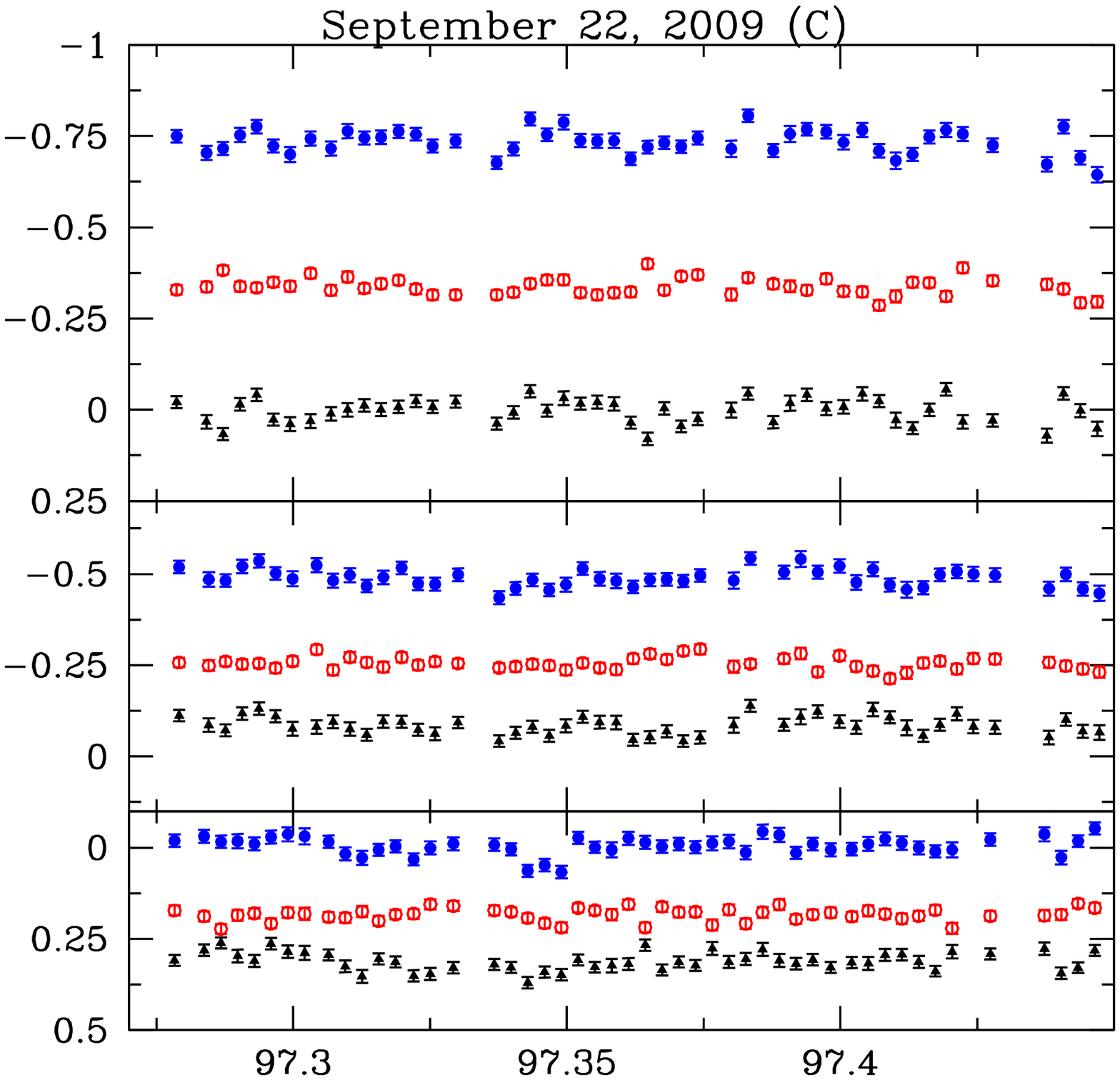,height=2.0in,width=2.2in,angle=0}
\caption{The B (middle panel), R (lower panel) and (B-R)(upper panel) light curves of 1ES 1959$+$650. 
The $x$-axis
is JD (2455000+) and the $y$-axis is the differential instrumental magnitude. Open circles (also in red colour) 
give the DLC of Blazar-Star1, filled circles (in blue colour), the DLC of Blazar-Star2 
while star symbols (in black colour) represents DLC of comparison stars (st1-st2). Dates and telescopes 
are given on top of each plot.  The magnitudes of each band are adjusted with an arbitrary offset (for clarity) 
in each panel of figure.  
} 
\end{figure*}

\clearpage
\begin{figure*}
\epsfig{figure= 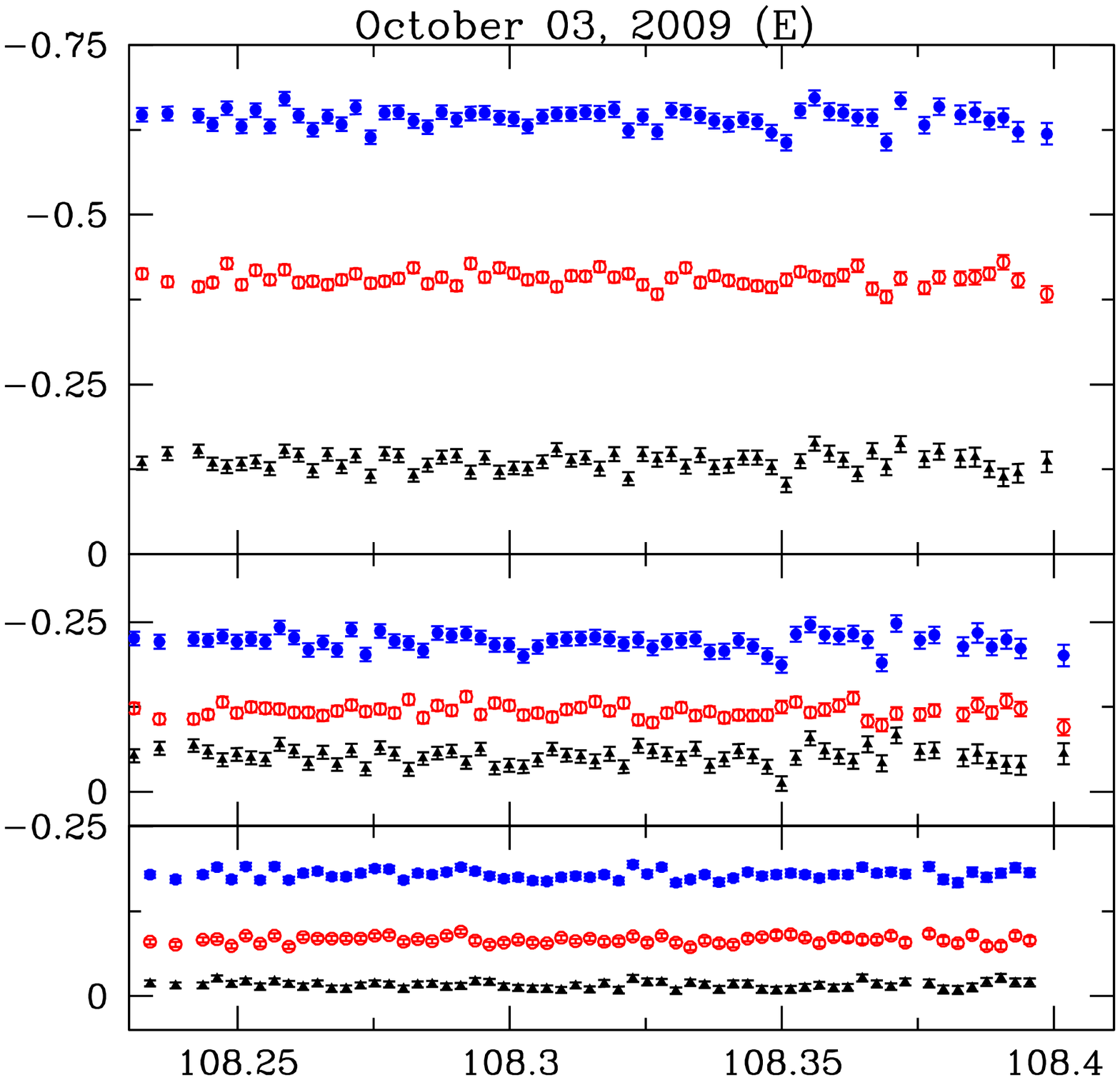 ,height=2.0in,width=2.2in,angle=0}
\epsfig{figure= 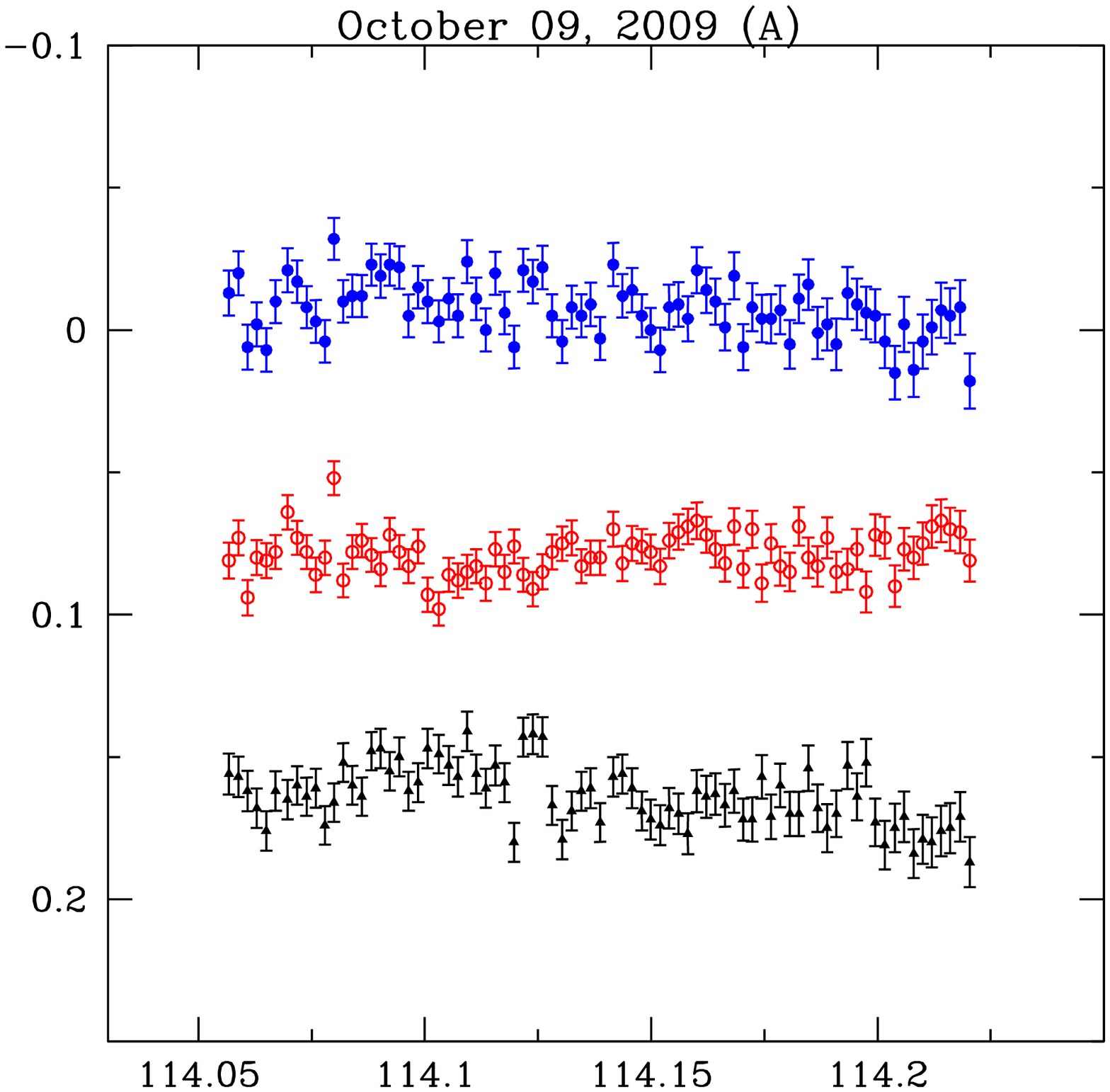,height=2.0in,width=2.2in,angle=0}
\epsfig{figure= 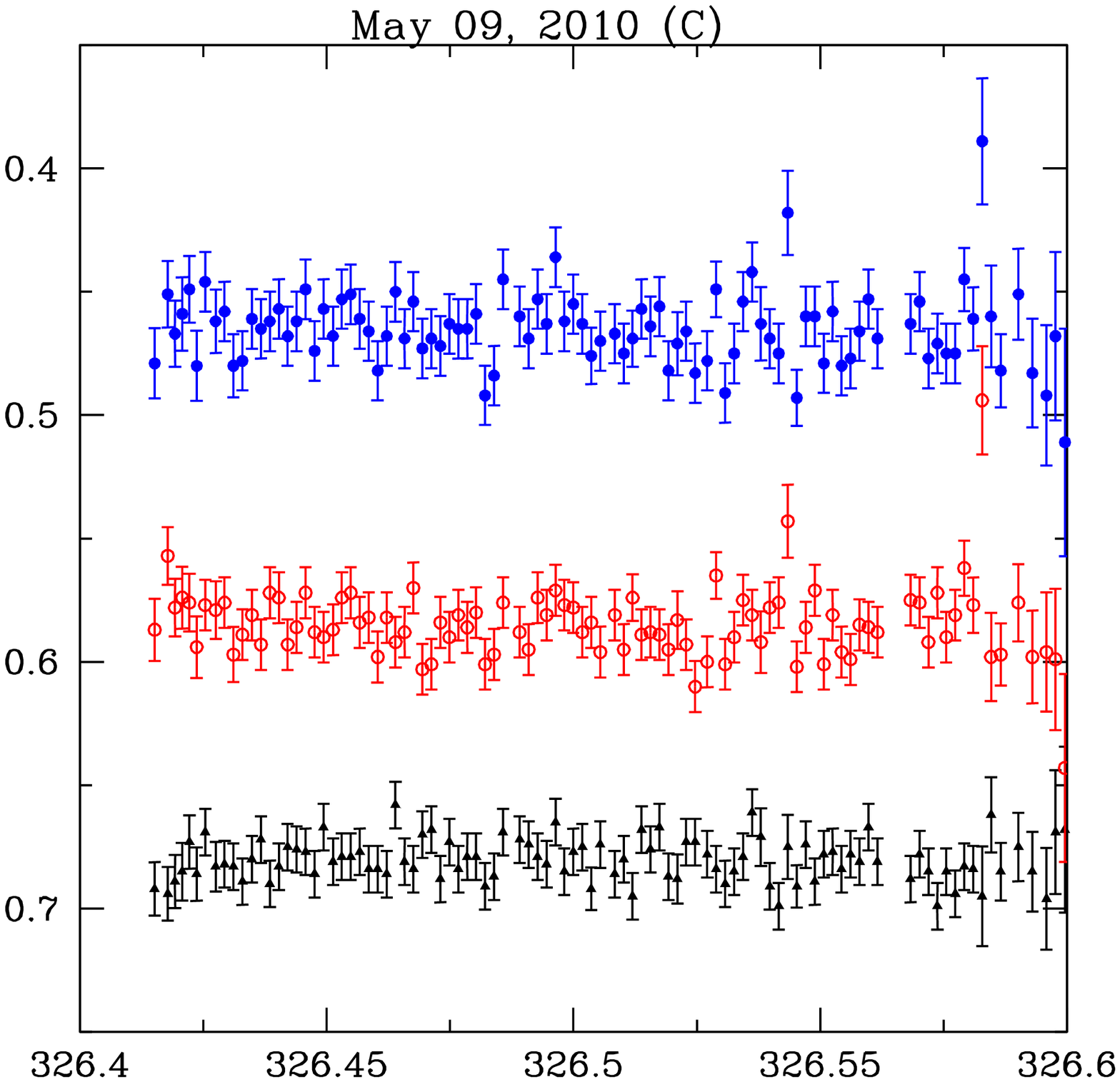,height=2.0in,width=2.2in,angle=0}
\epsfig{figure= 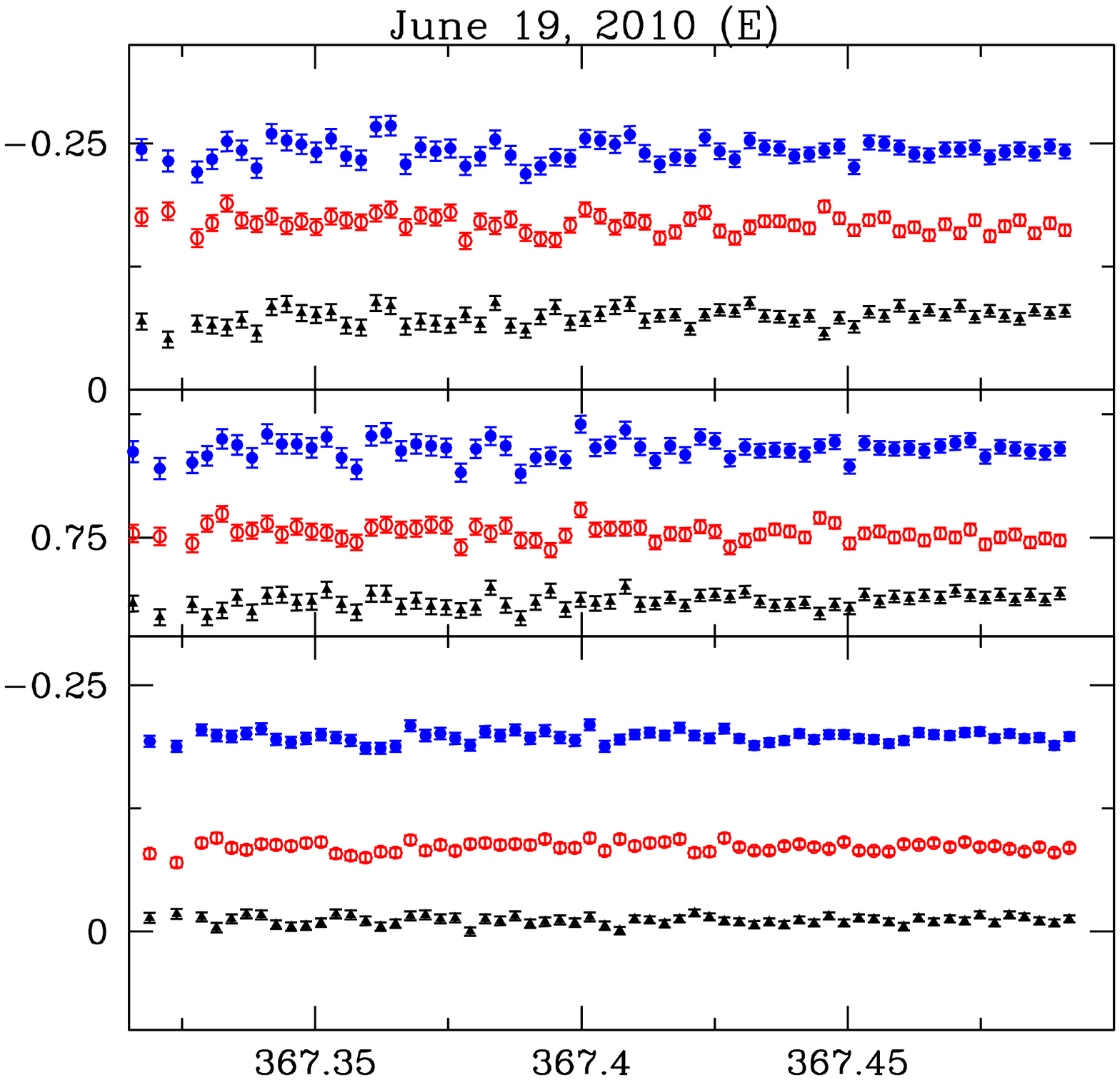,height=2.0in,width=2.2in,angle=0}
\epsfig{figure= 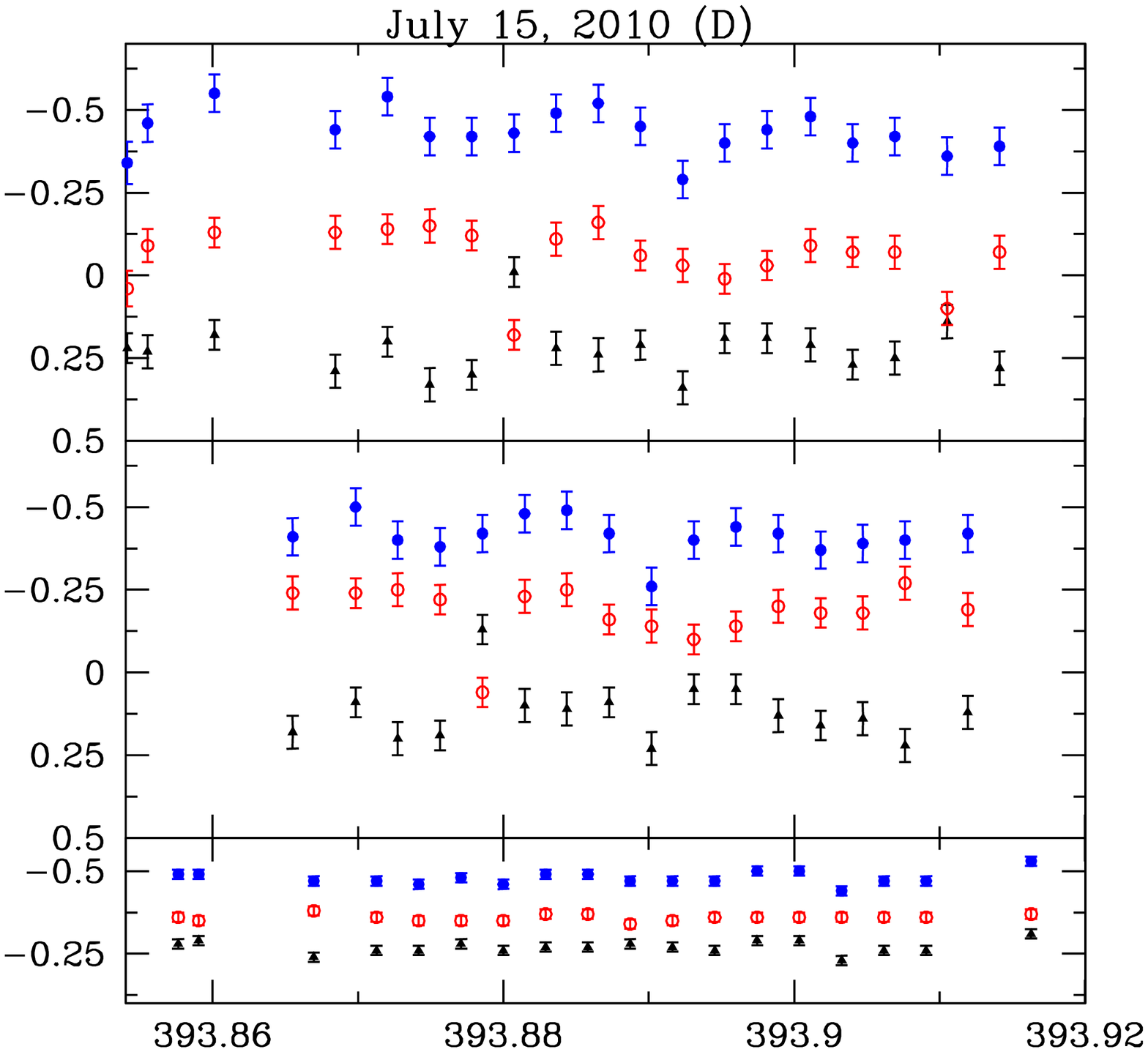,height=2.0in,width=2.2in,angle=0}
\epsfig{figure= 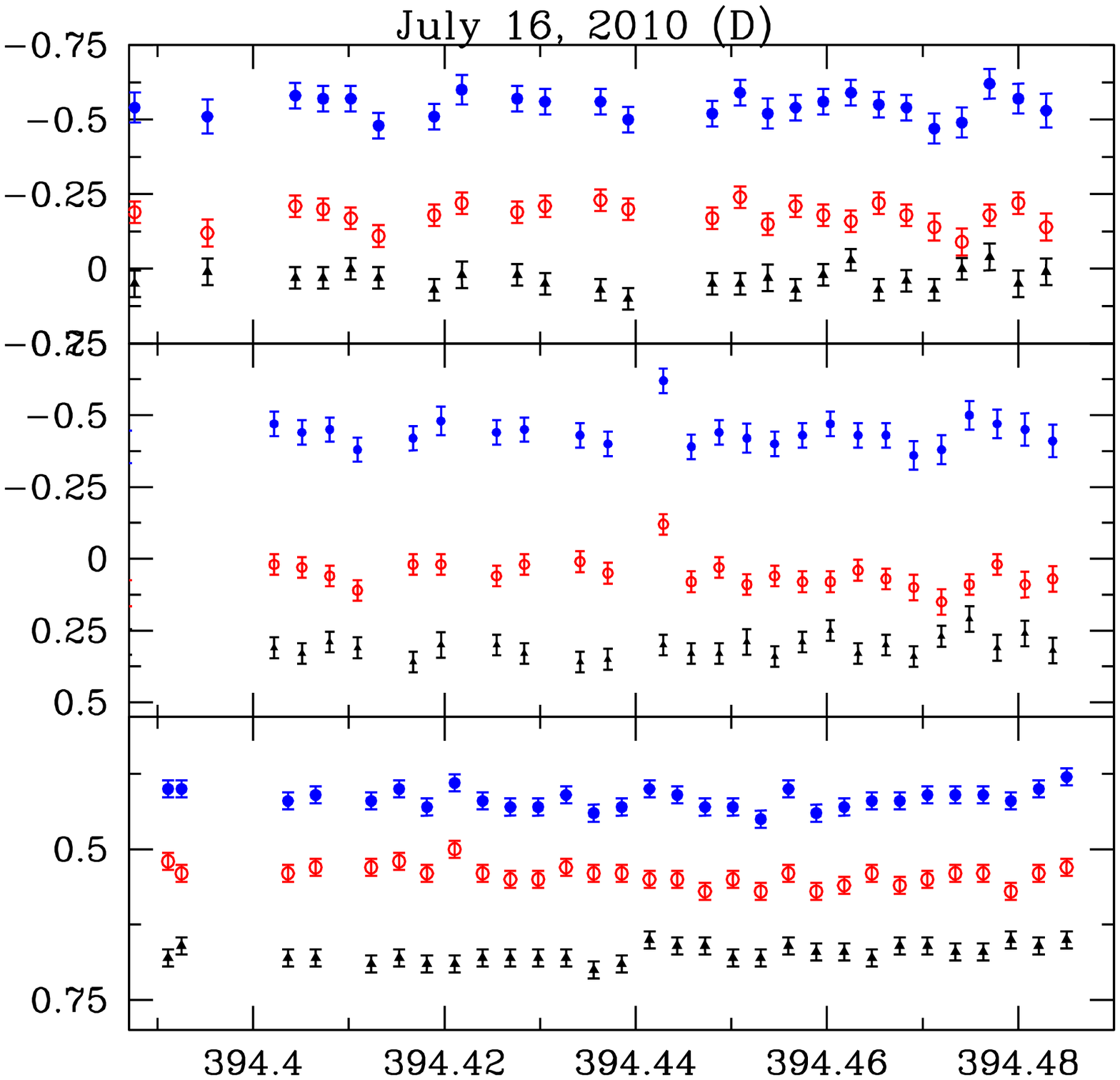,height=2.0in,width=2.2in,angle=0}
\epsfig{figure= 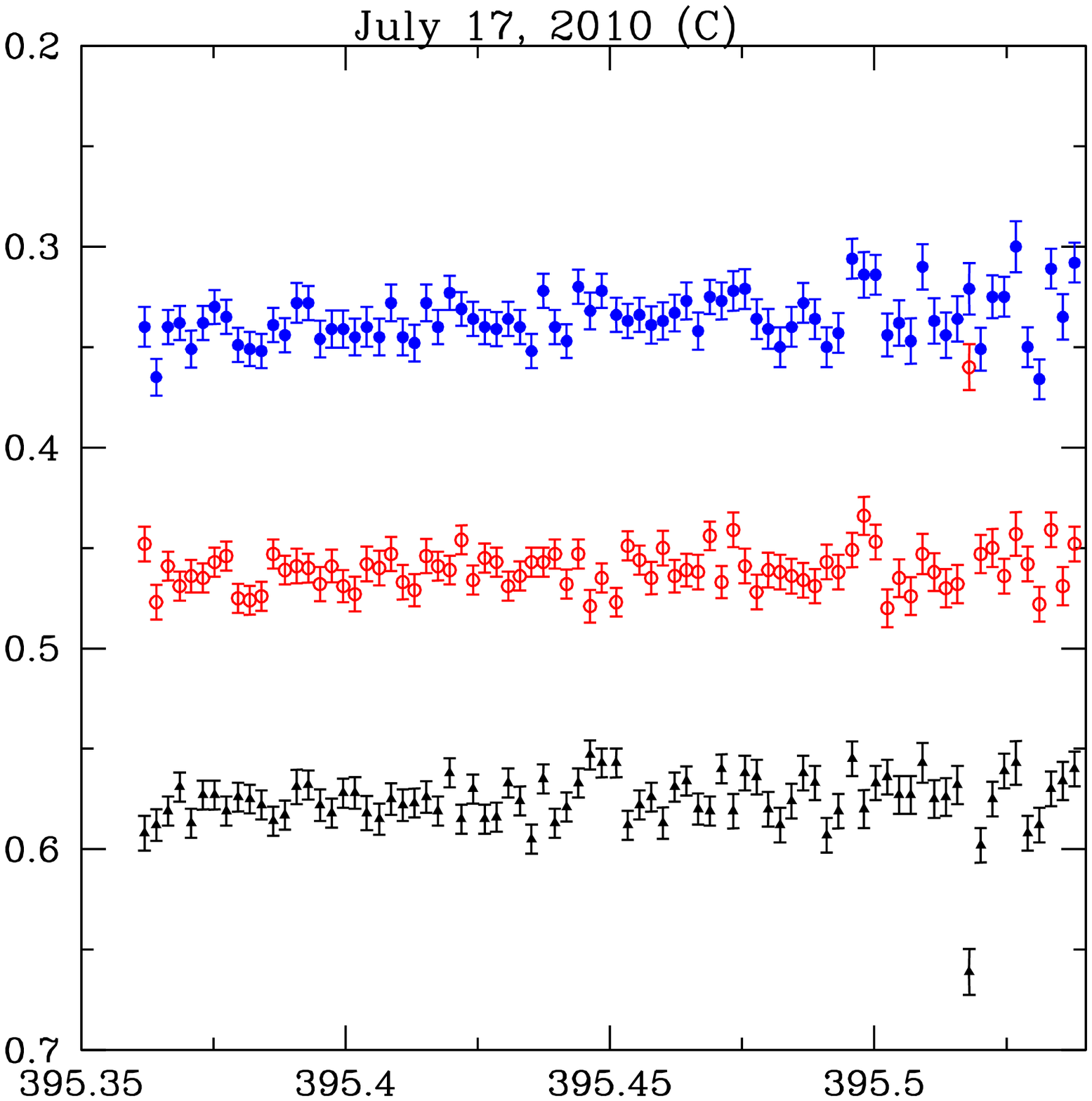,height=2.0in,width=2.2in,angle=0}
\epsfig{figure= 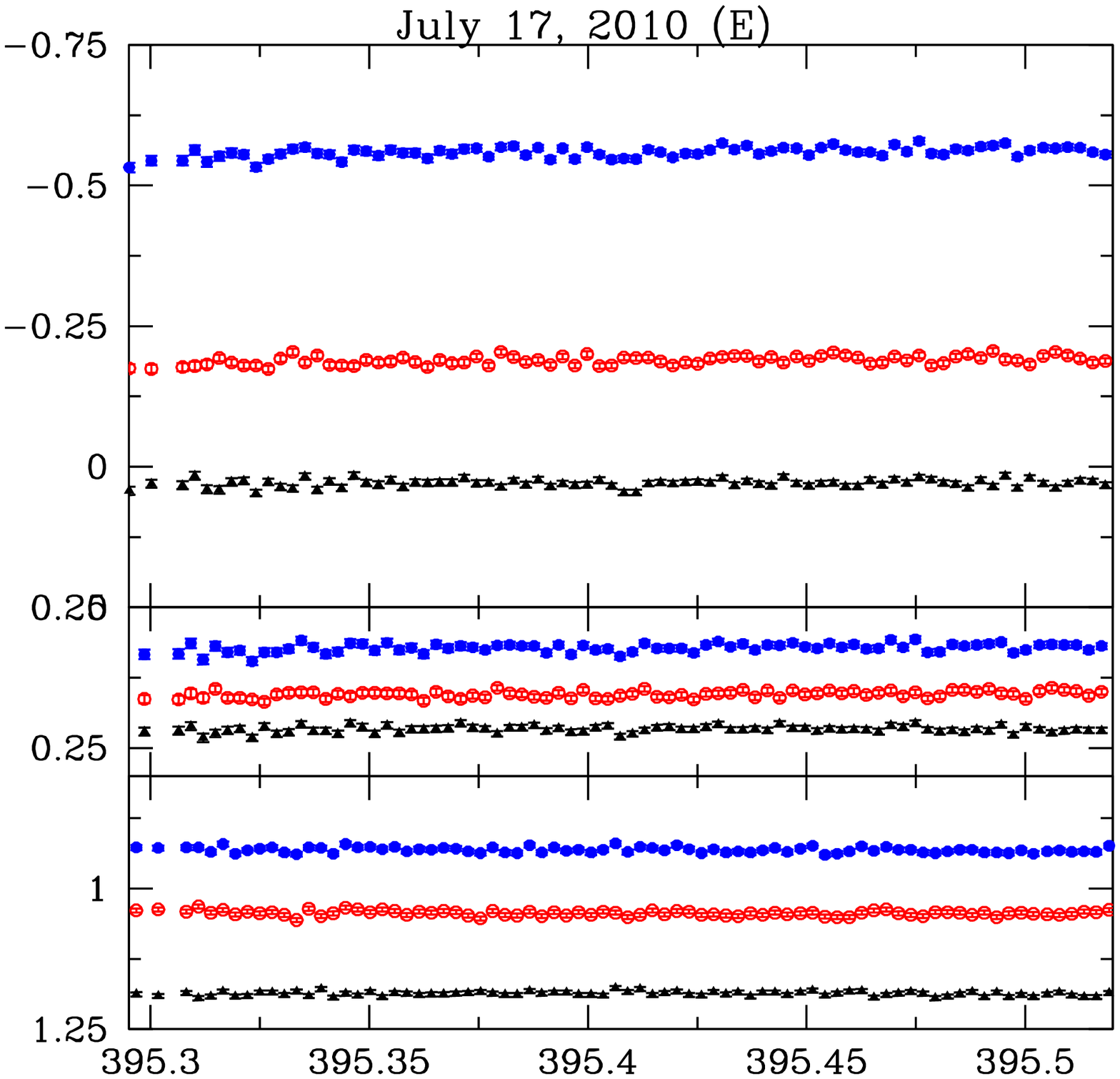,height=2.0in,width=2.2in,angle=0}
\epsfig{figure= 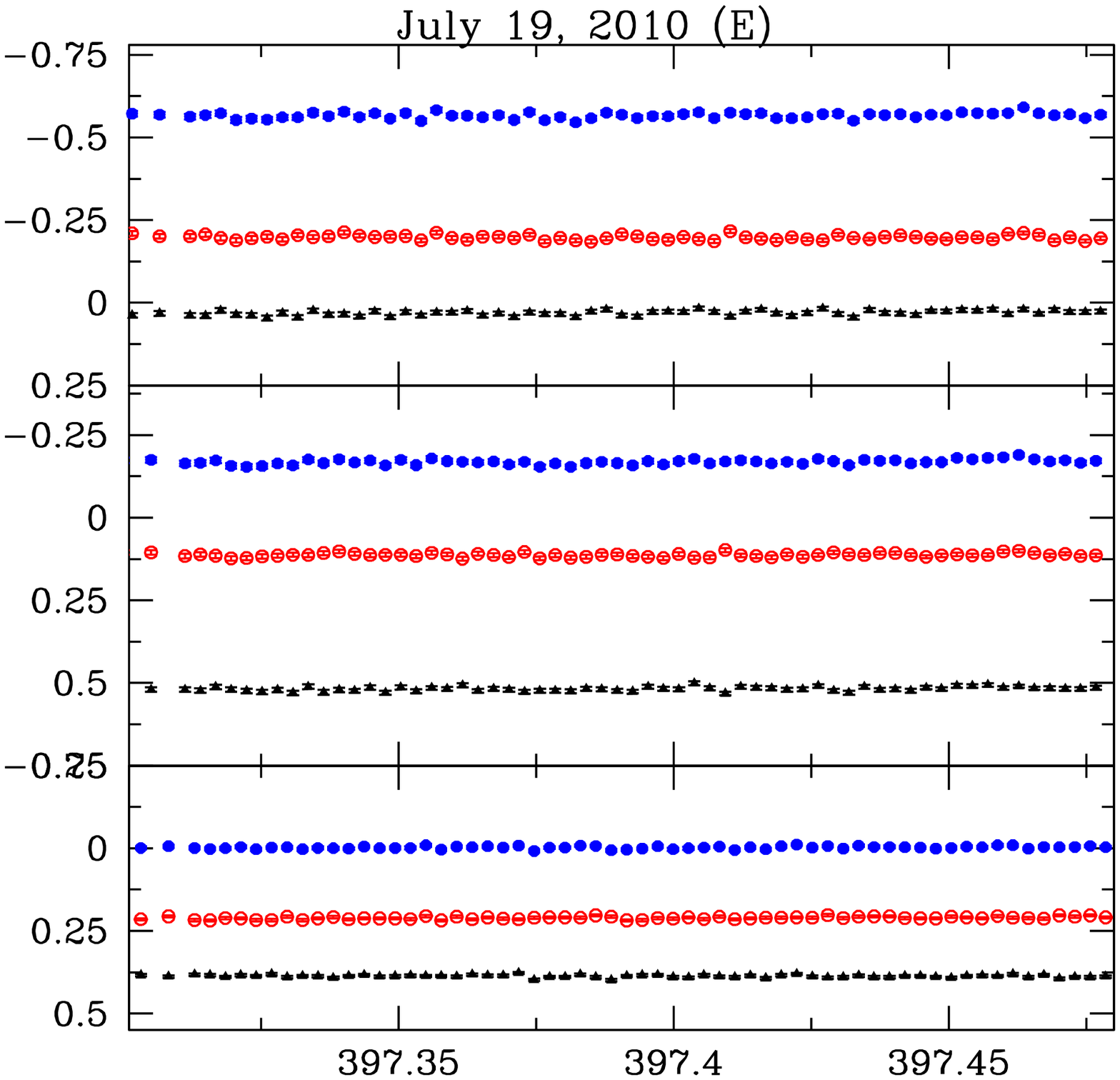,height=2.0in,width=2.2in,angle=0}
\epsfig{figure= 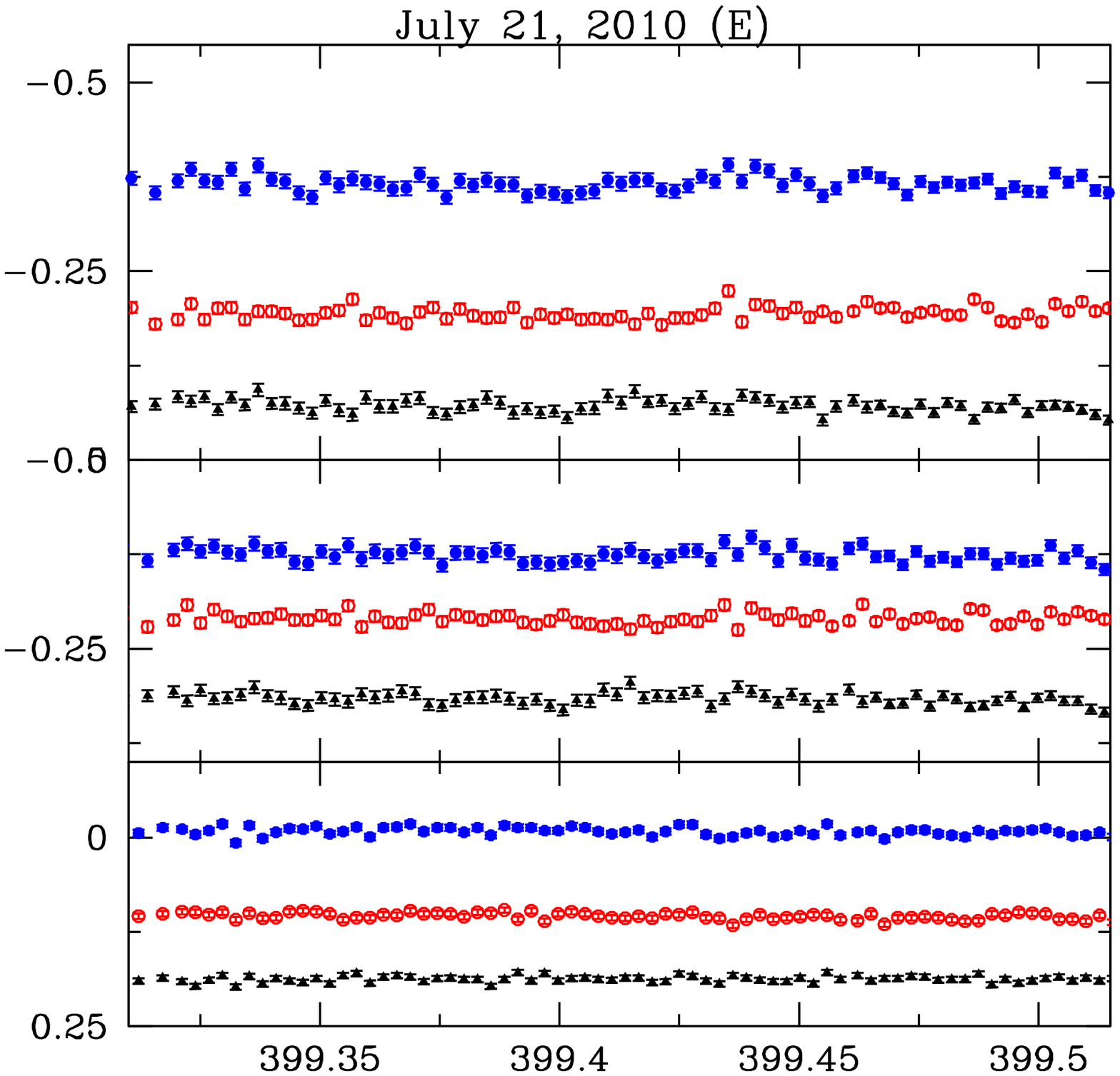,height=2.0in,width=2.2in,angle=0}
\epsfig{figure= 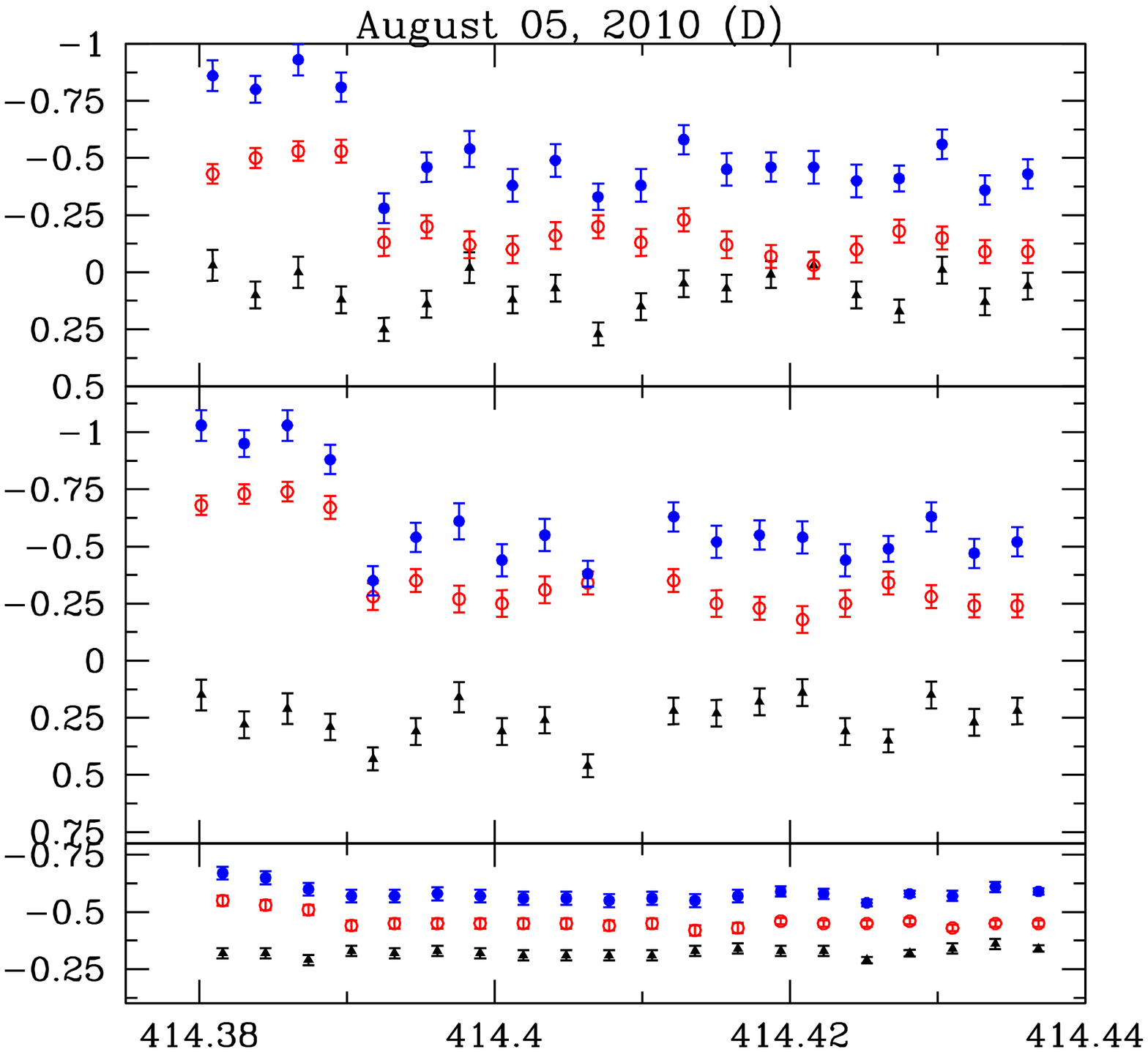,height=2.0in,width=2.2in,angle=0}
\epsfig{figure= 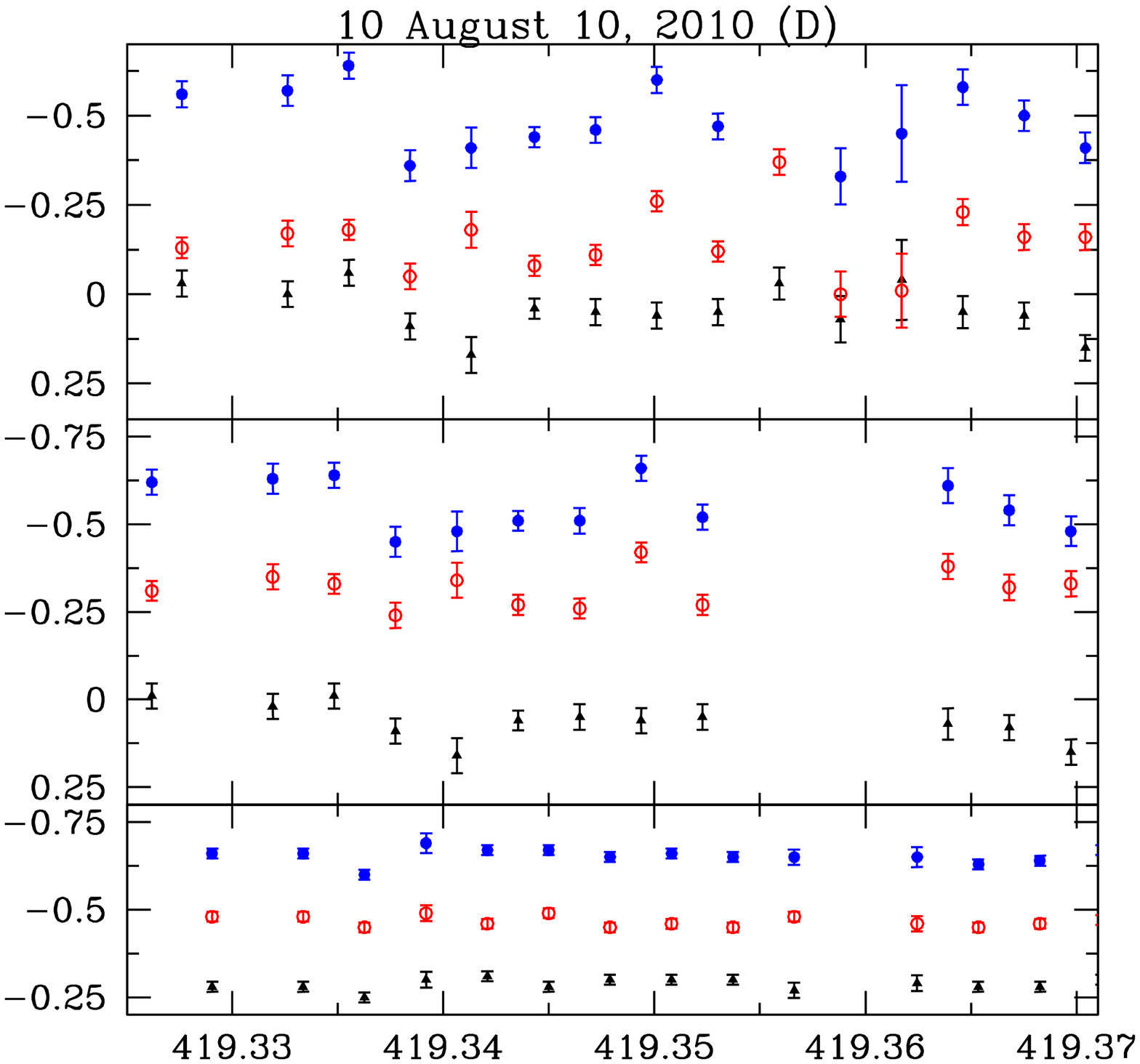,height=2.0in,width=2.2in,angle=0}
\caption{As in Fig.\ 1 for additional nights for 1ES 1959+650.}
\end{figure*}

\clearpage
\begin{figure*}
\epsfig{figure= 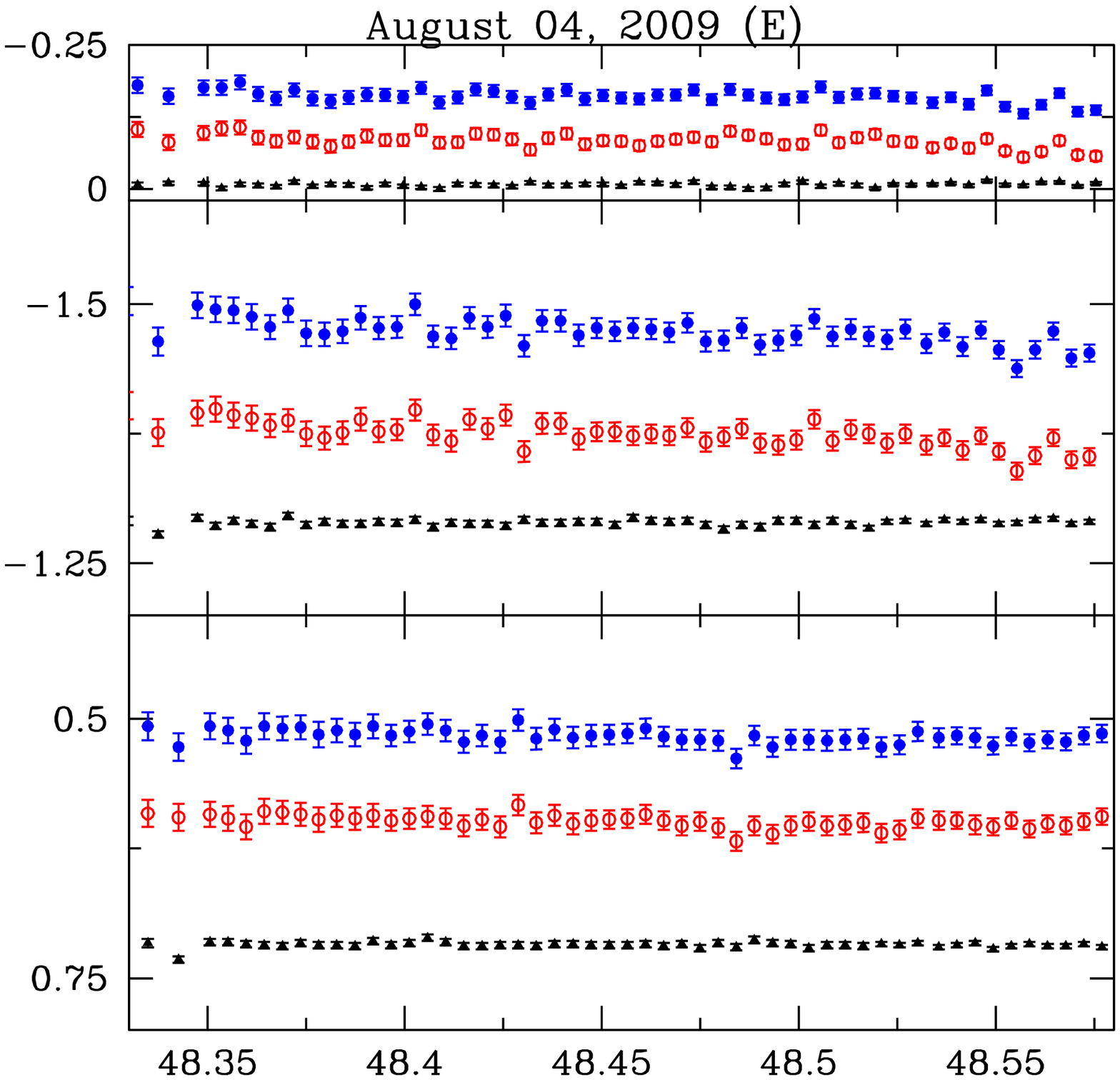,height=2.0in,width=2.2in,angle=0}
\epsfig{figure= 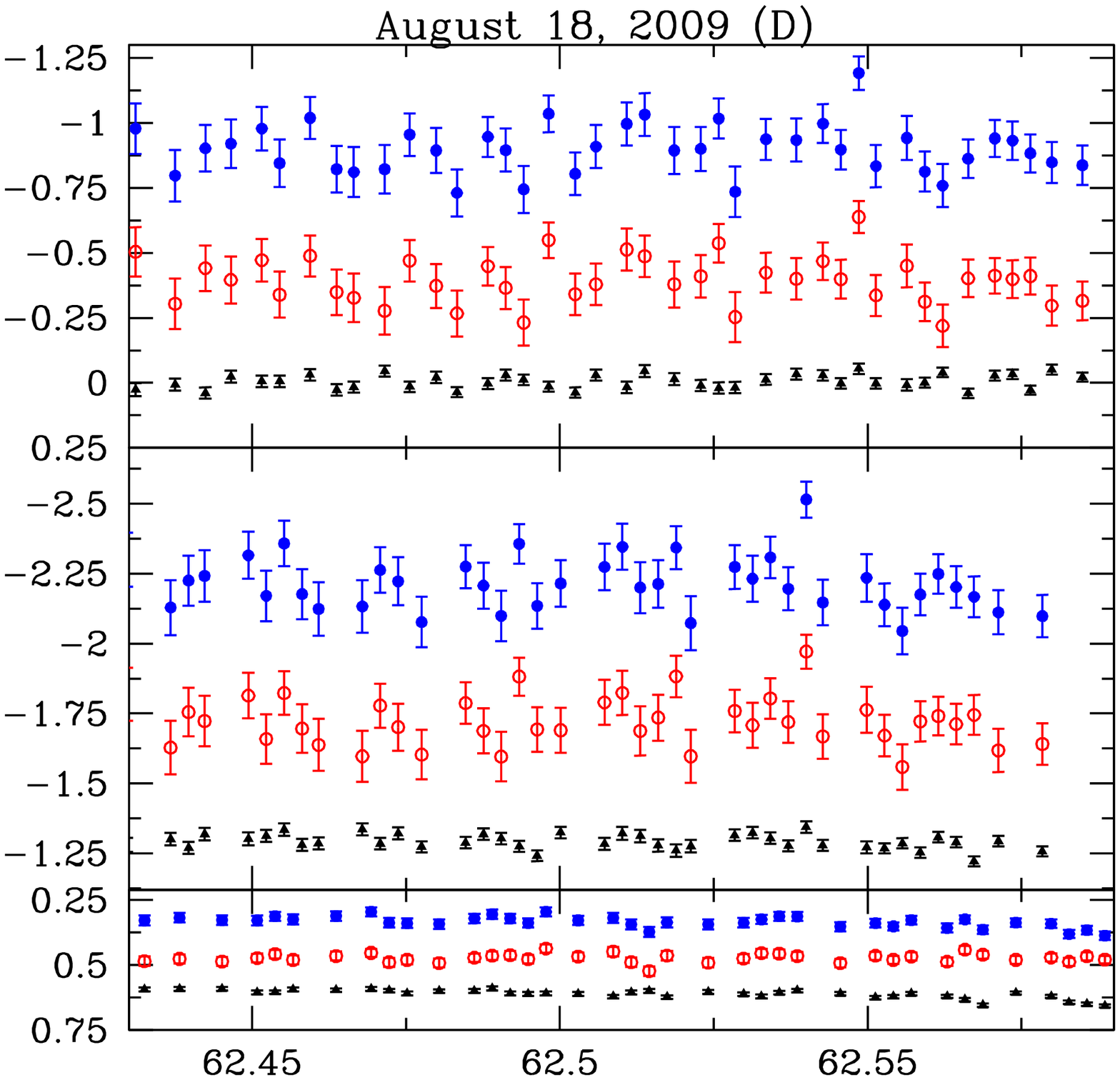,height=2.0in,width=2.2in,angle=0}
\epsfig{figure= 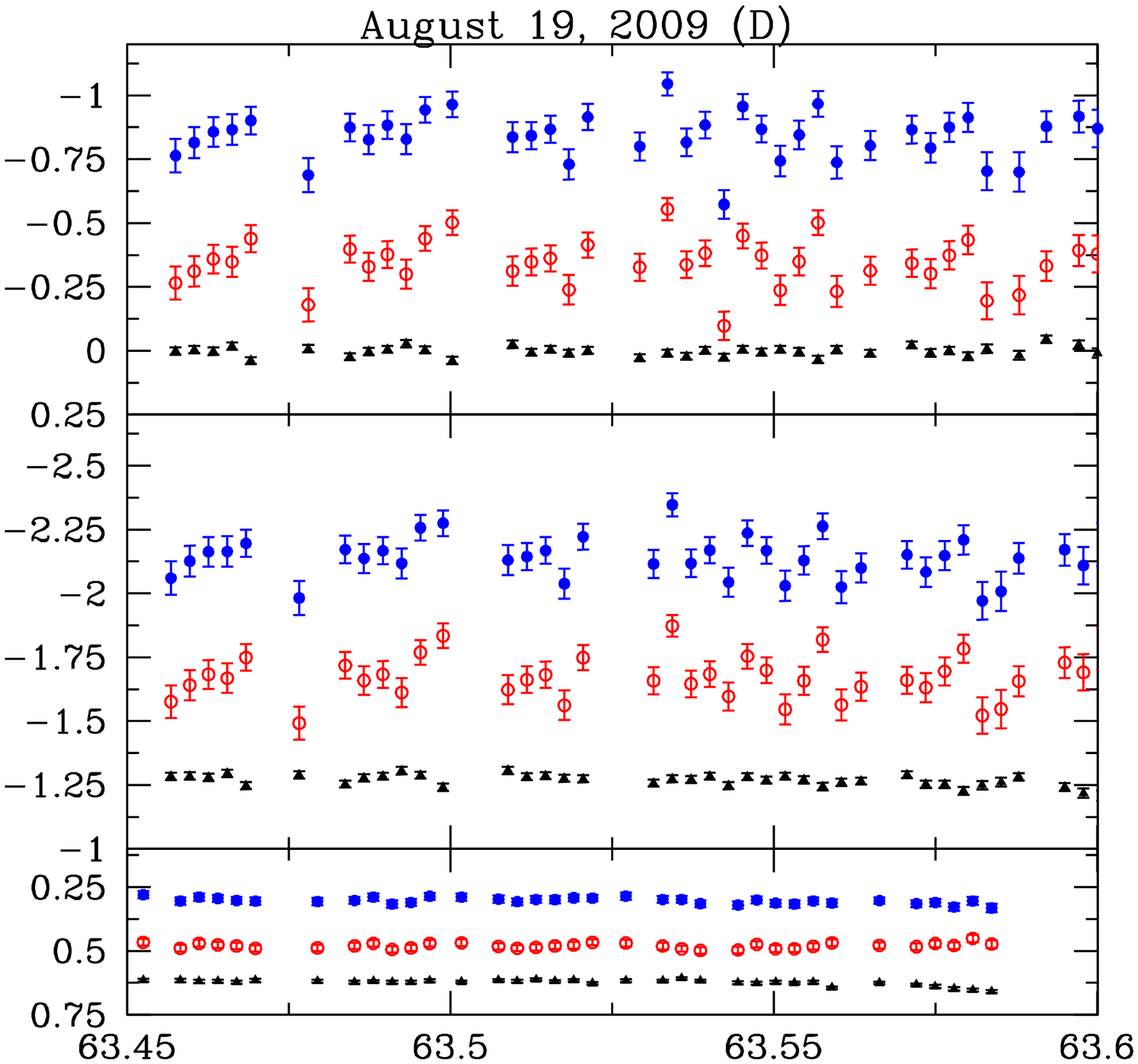,height=2.0in,width=2.2in,angle=0}
\epsfig{figure= 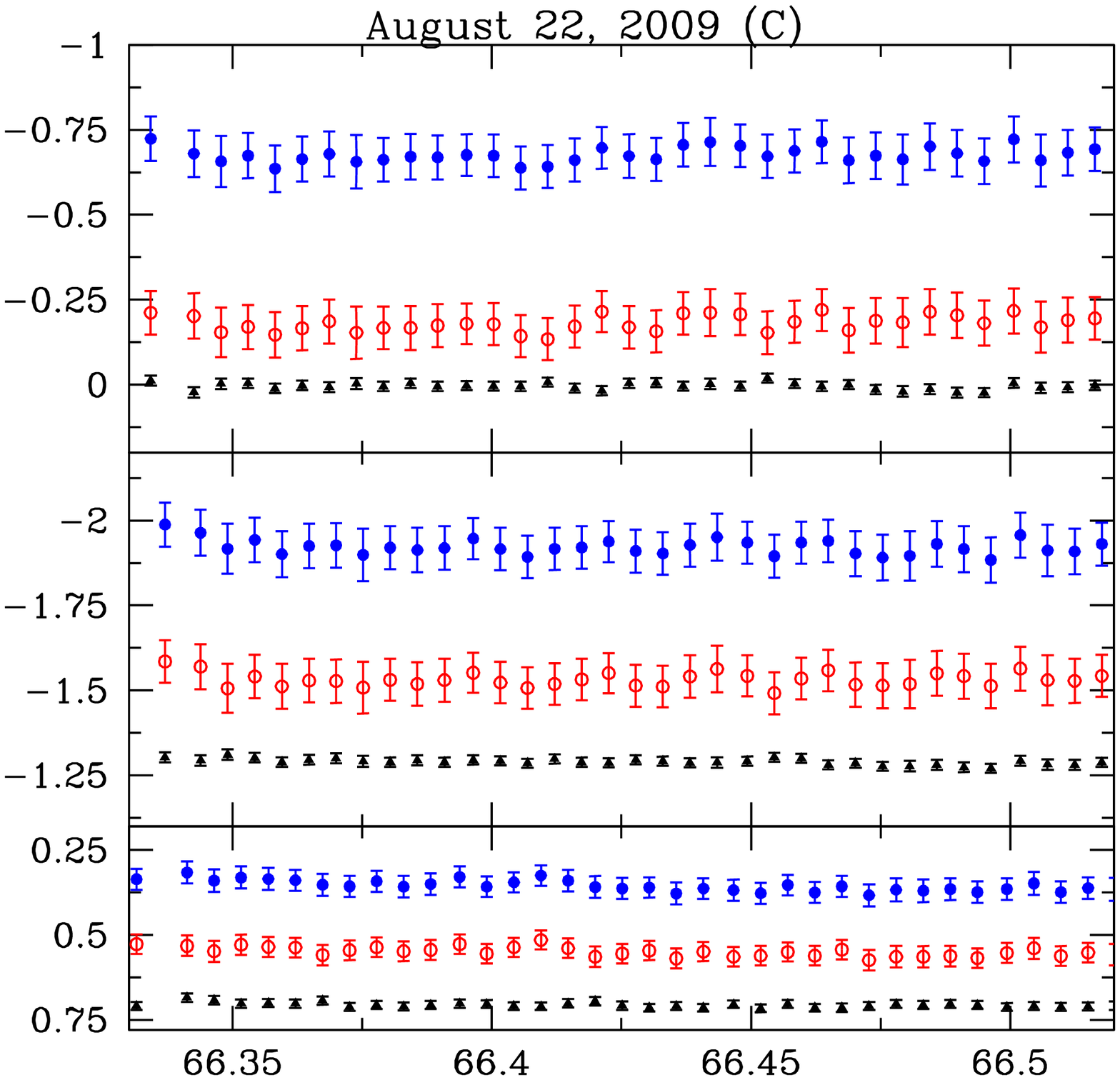,height=2.0in,width=2.2in,angle=0}
\epsfig{figure= 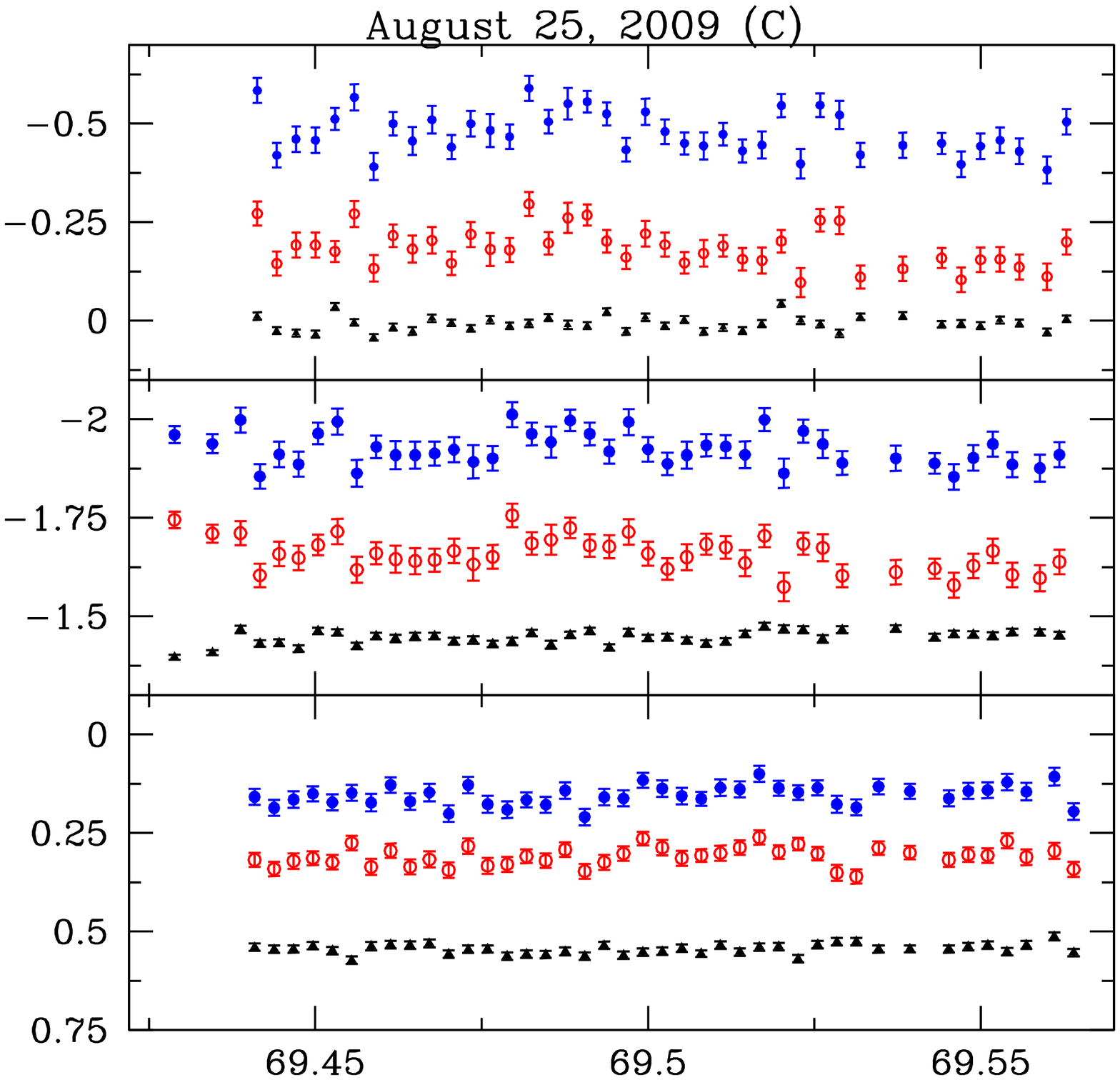,height=2.0in,width=2.2in,angle=0}
\epsfig{figure= 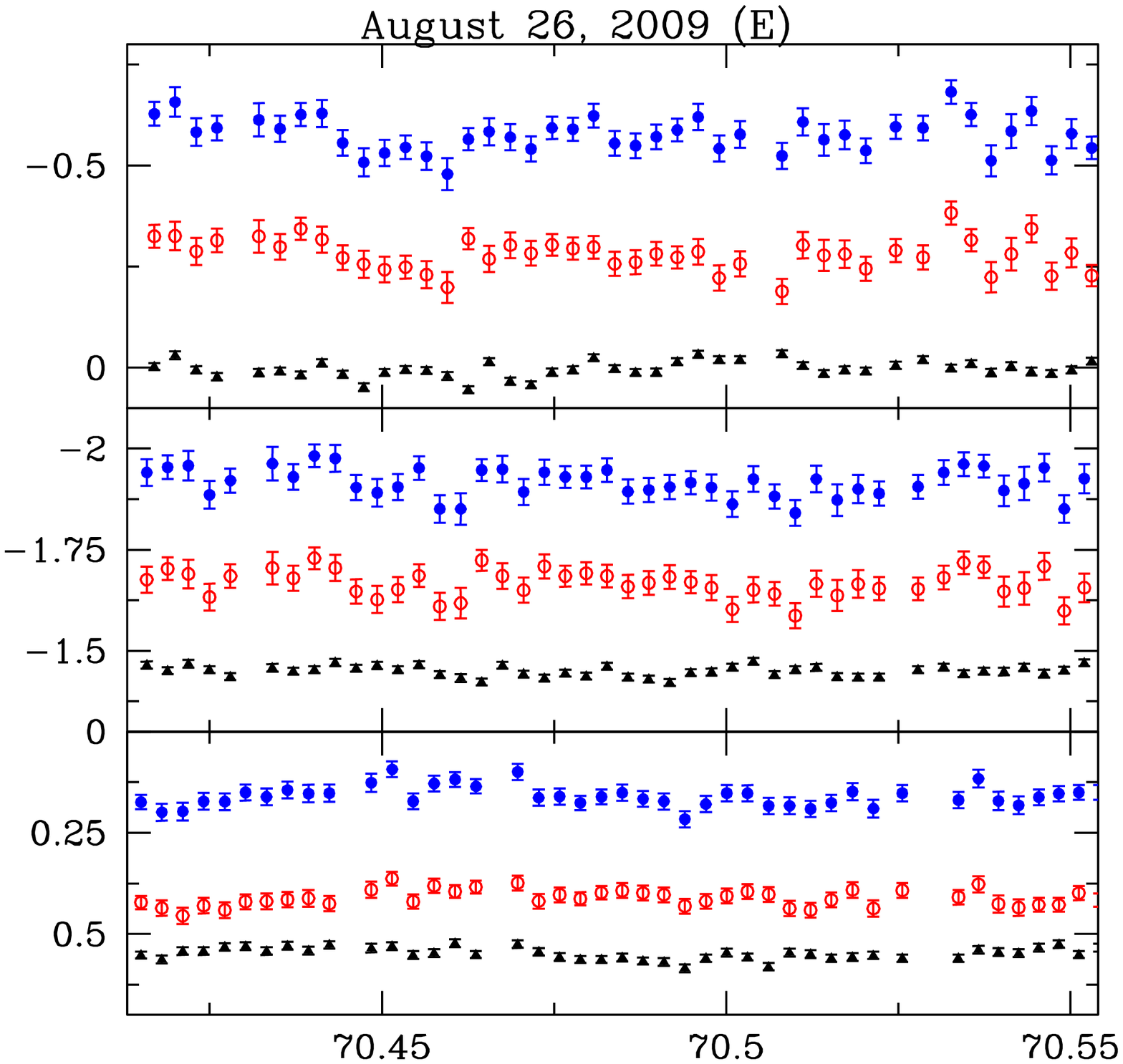,height=2.0in,width=2.2in,angle=0}
\epsfig{figure= 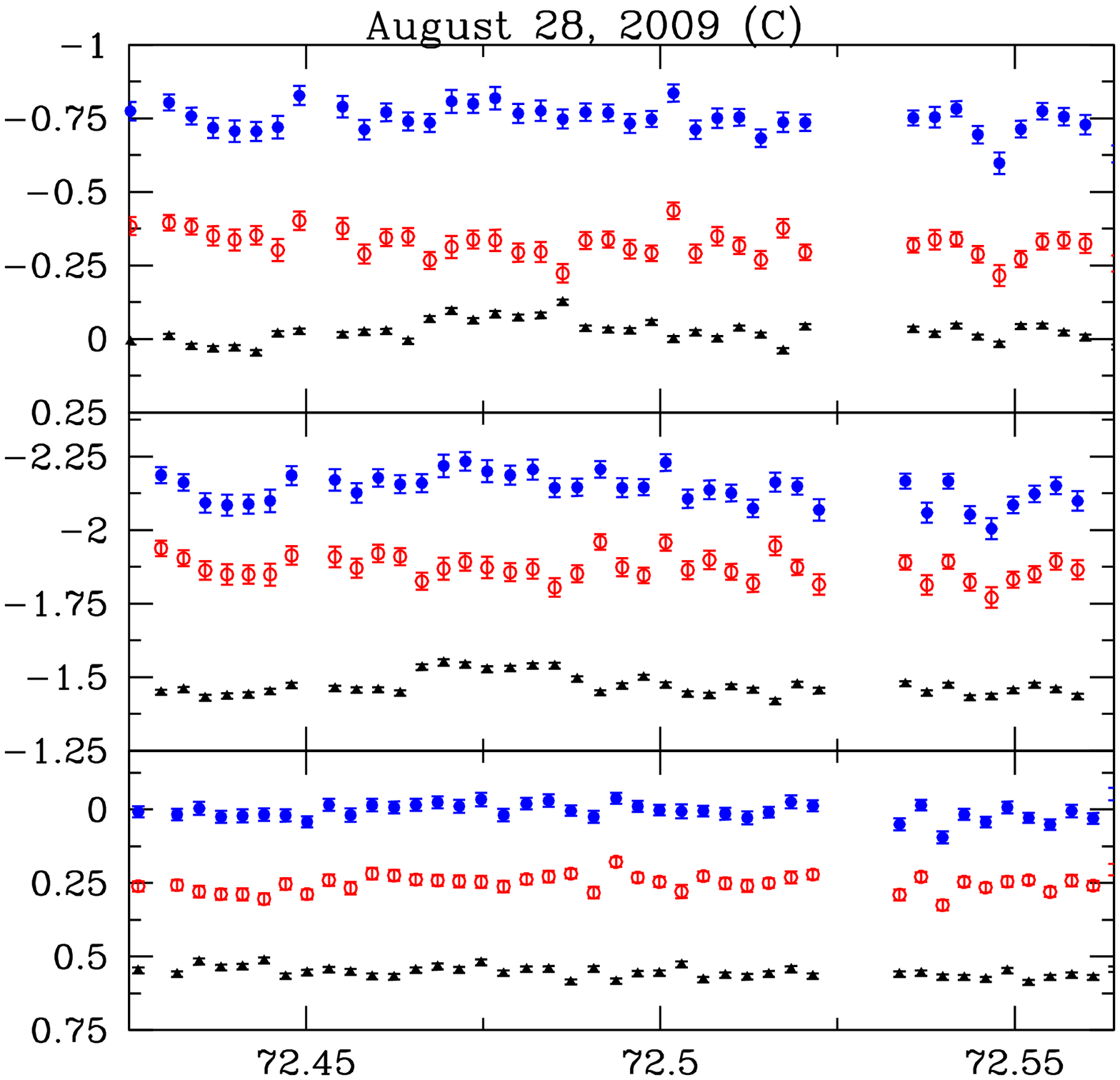,height=2.0in,width=2.2in,angle=0}
\epsfig{figure= 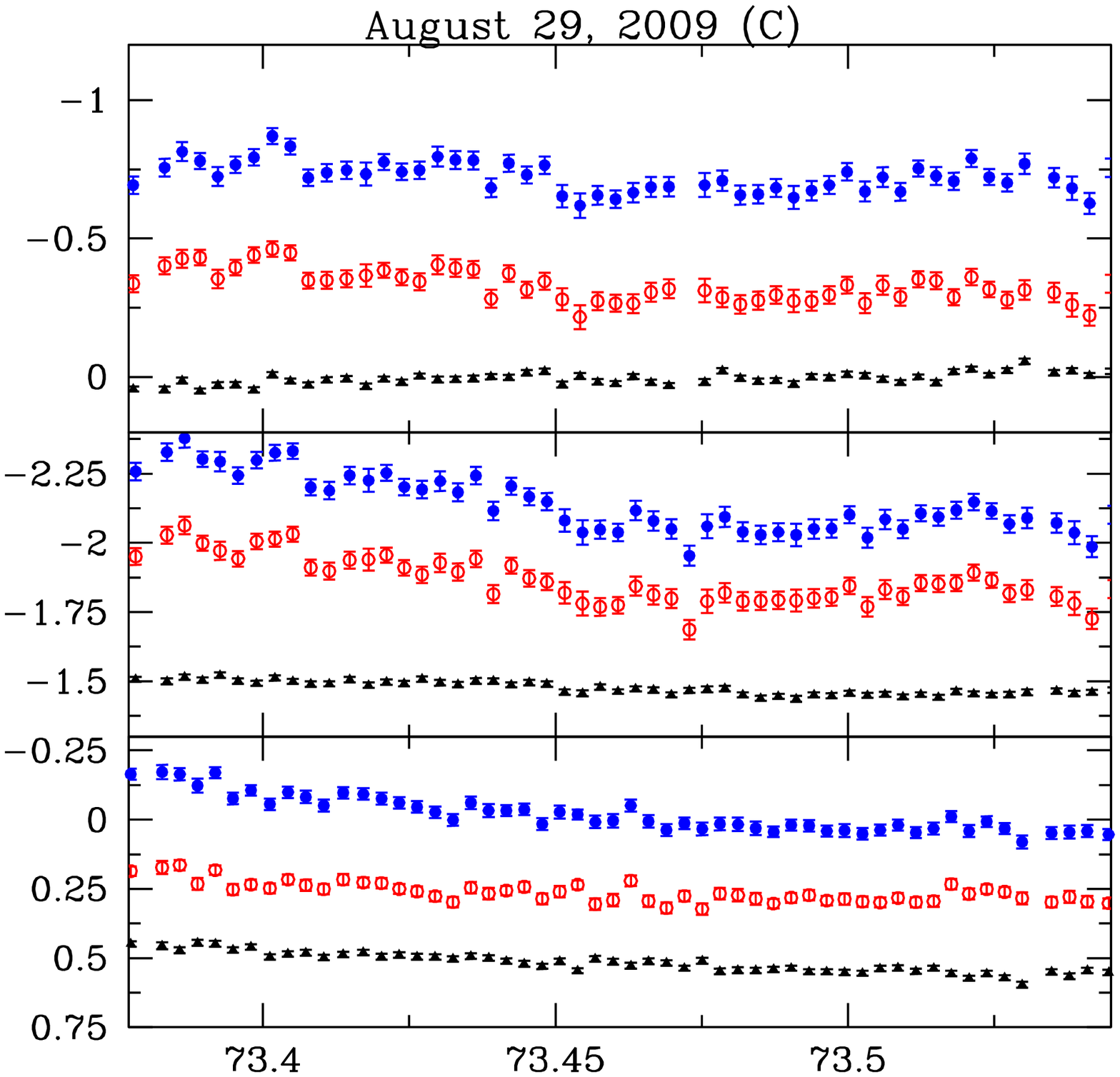,height=2.0in,width=2.2in,angle=0}
\epsfig{figure= 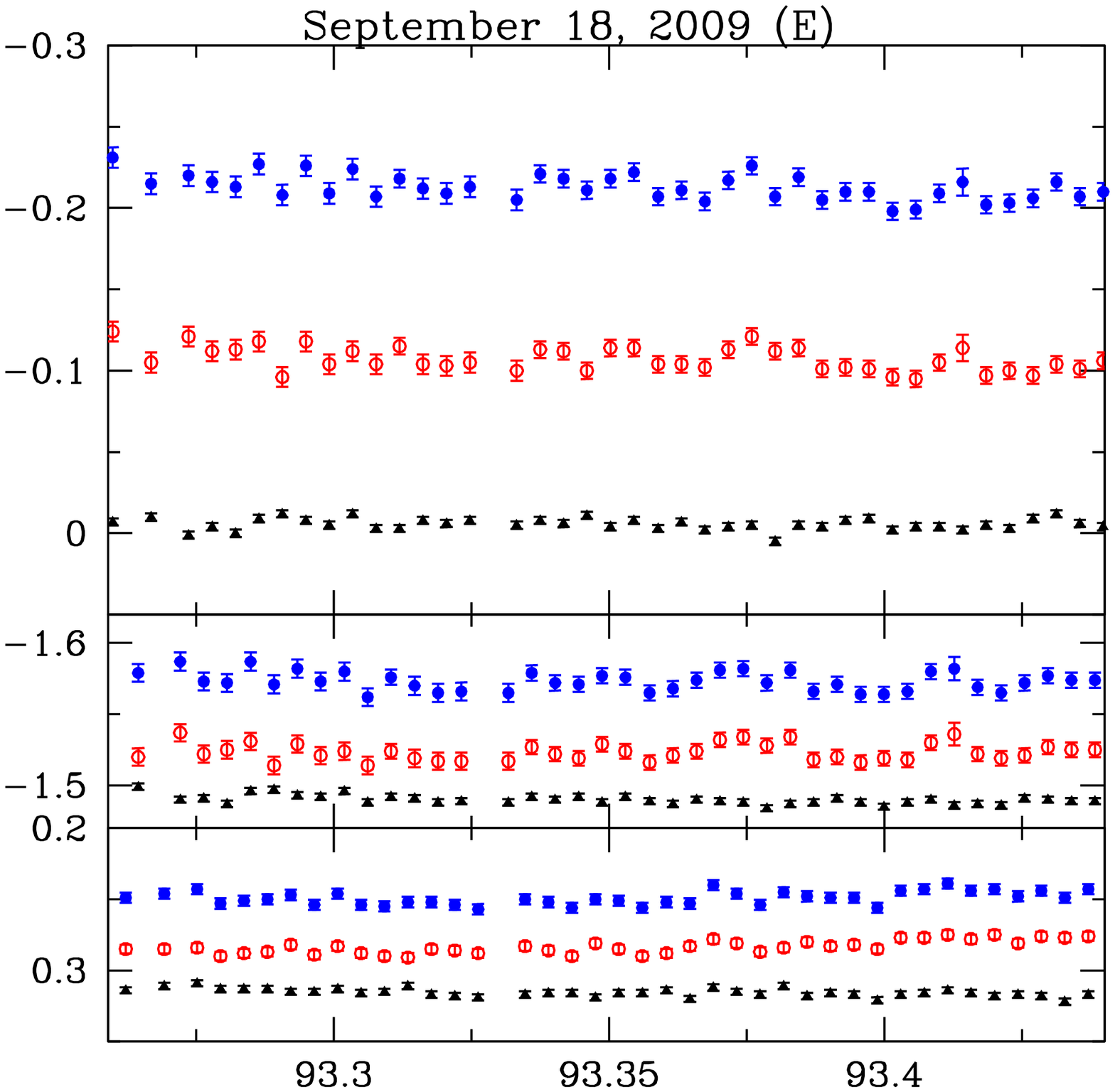,height=2.0in,width=2.2in,angle=0}
\epsfig{figure= 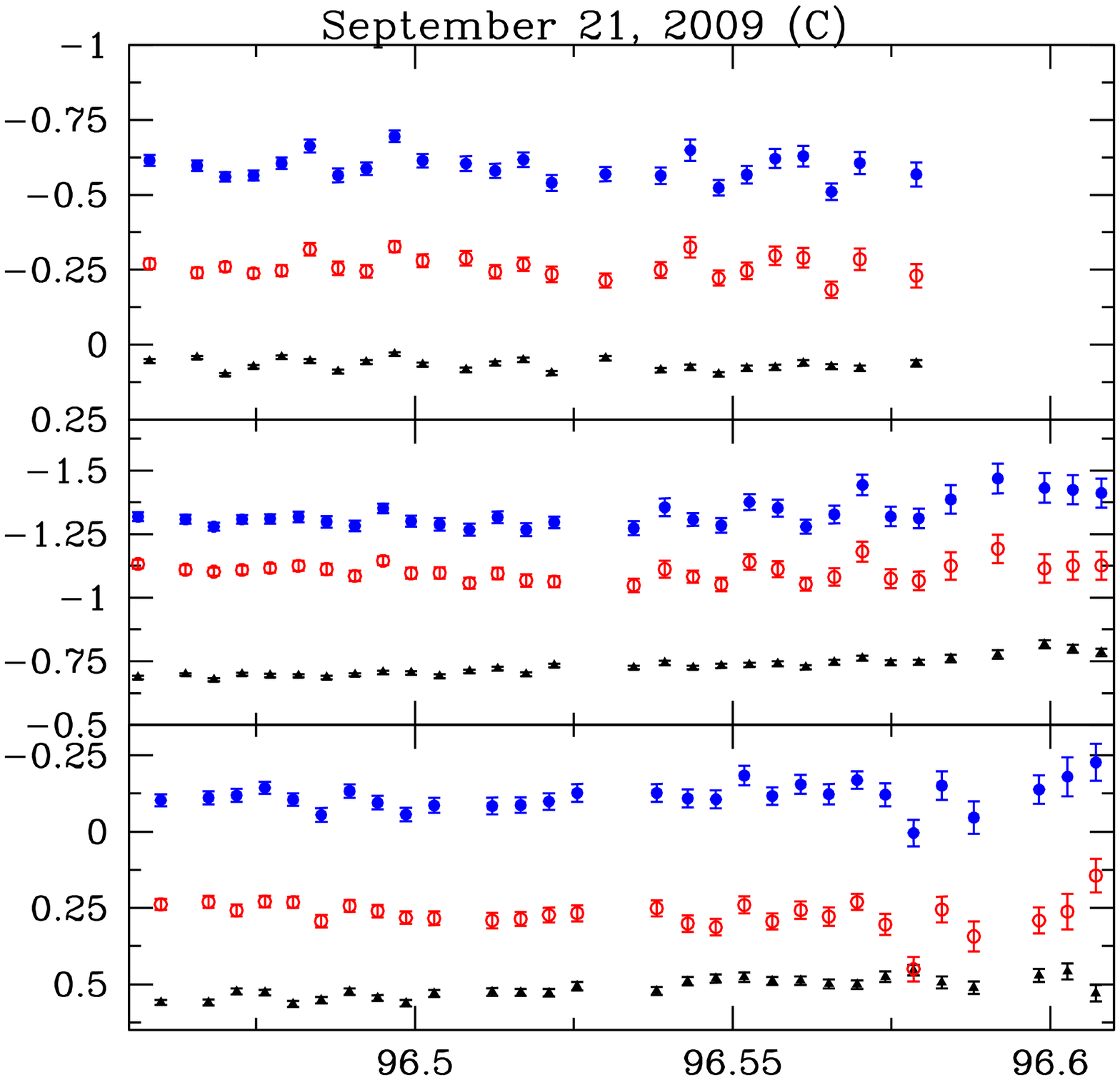,height=2.0in,width=2.2in,angle=0}
\epsfig{figure= 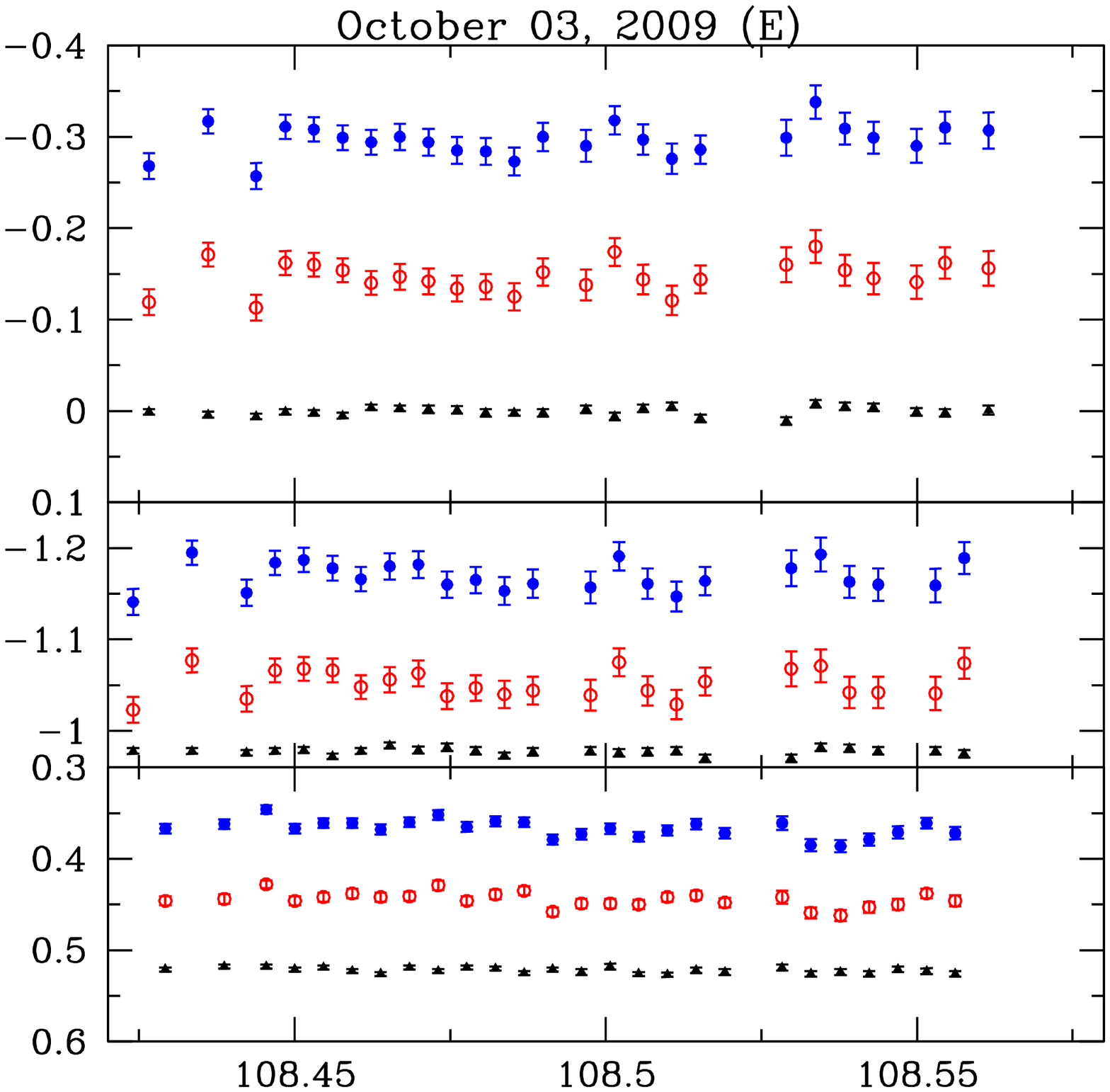,height=2.0in,width=2.2in,angle=0}
\epsfig{figure= 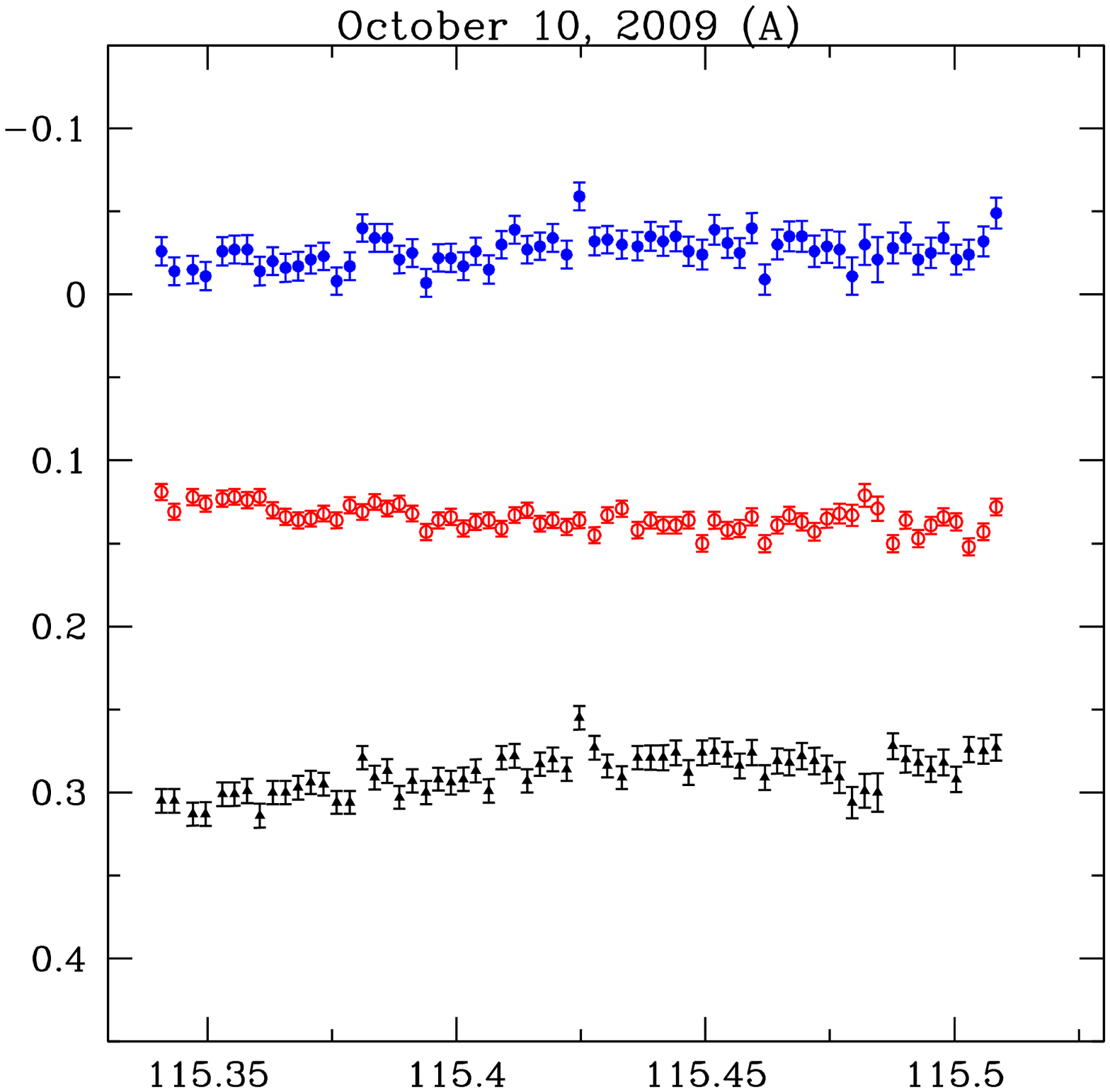,height=2.0in,width=2.2in,angle=0}

\caption{As in Fig.\ 1 for 1ES 2344$+$514.}
\end{figure*}

\clearpage
\begin{figure*}
\epsfig{figure= 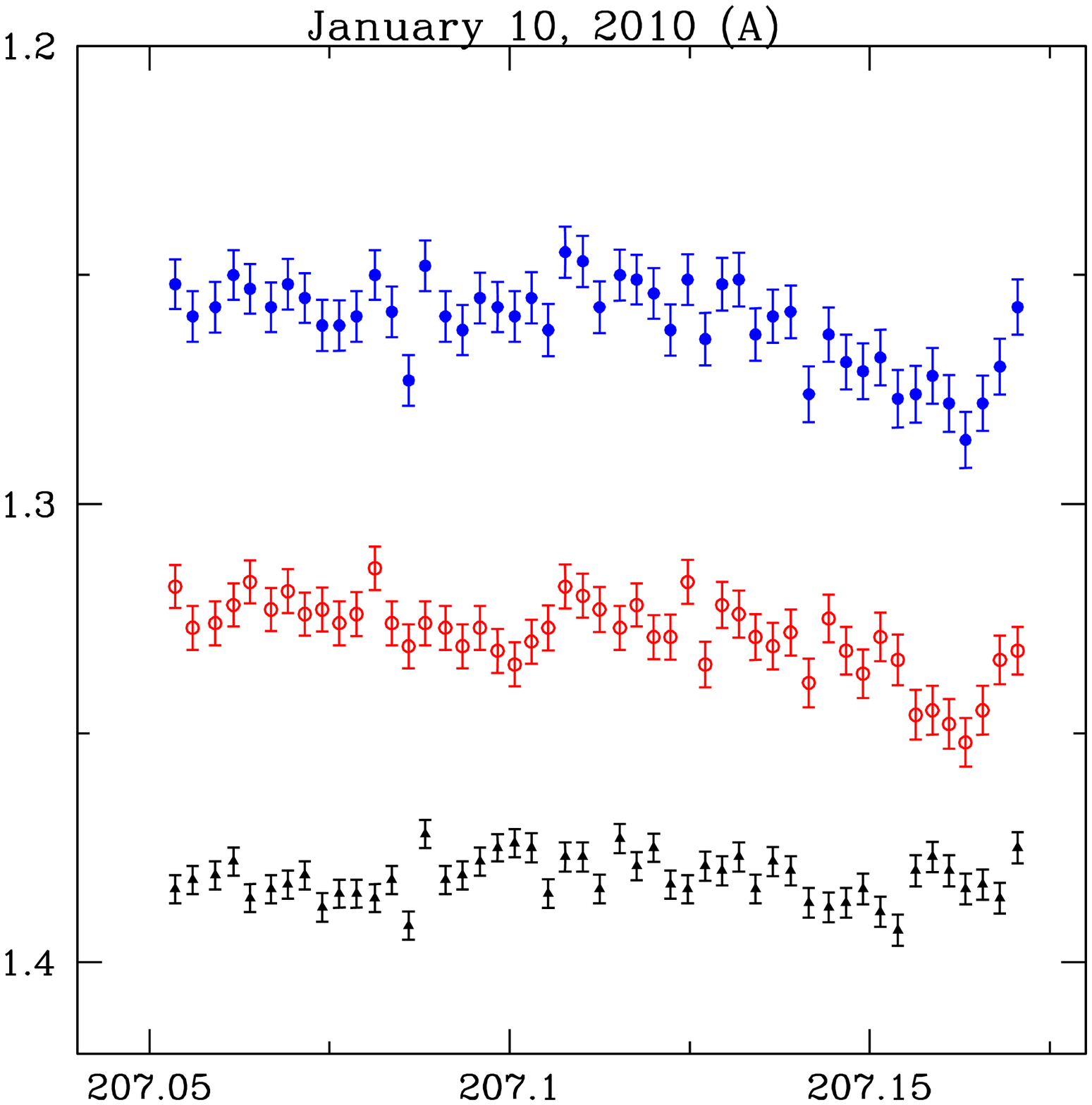,height=2.0in,width=2.2in,angle=0}
\epsfig{figure= 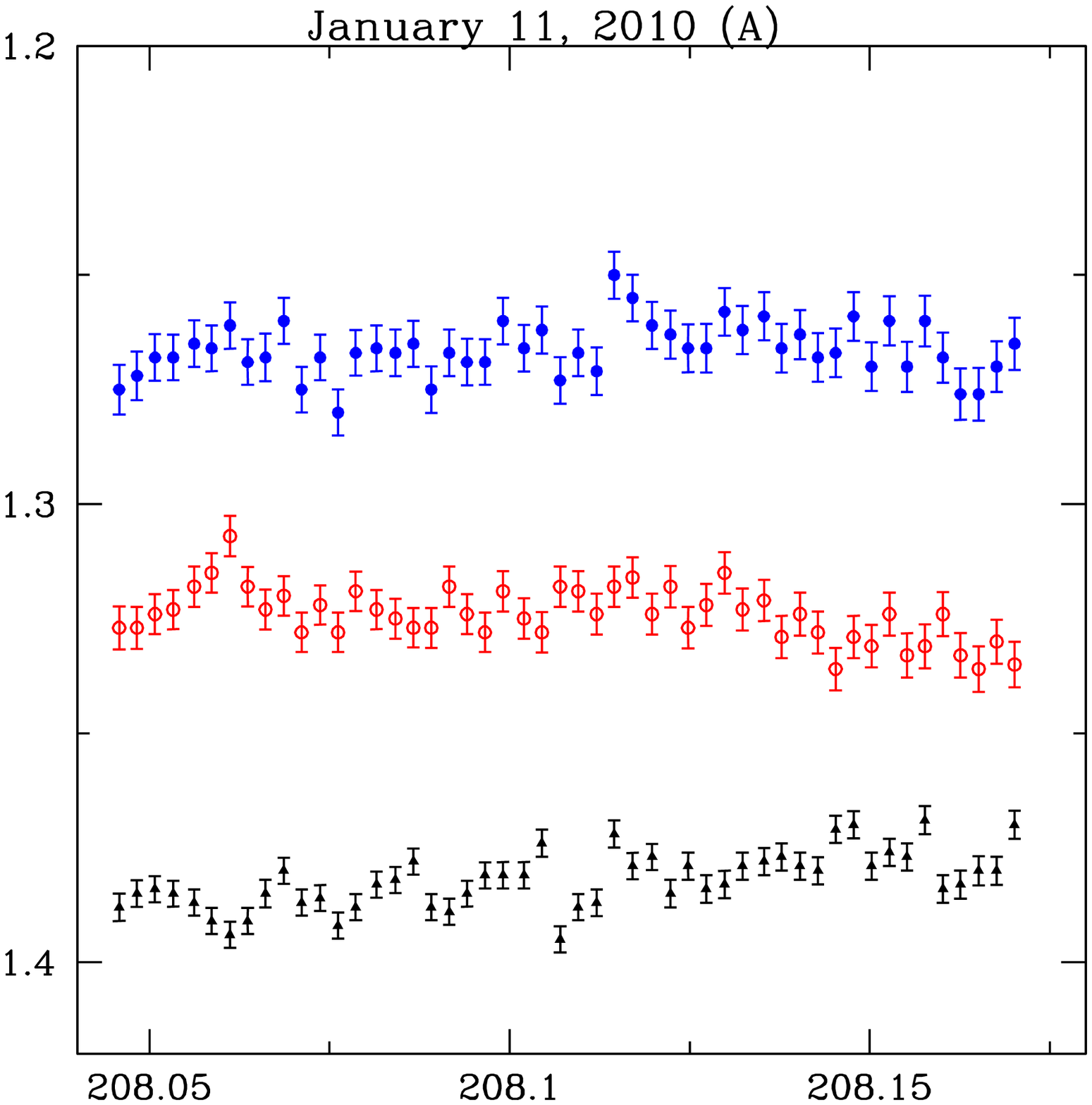,height=2.0in,width=2.2in,angle=0}
\epsfig{figure= 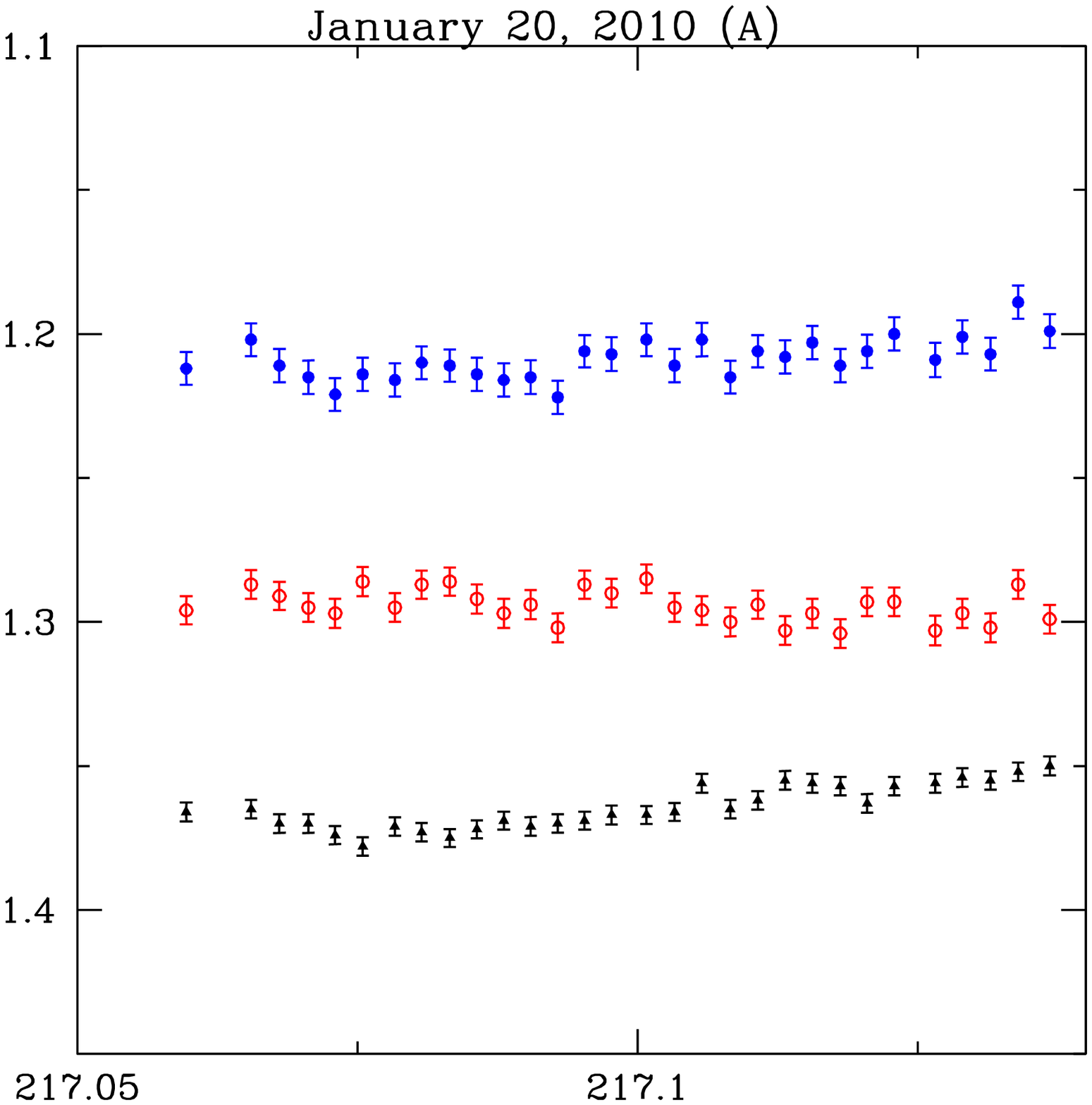,height=2.0in,width=2.2in,angle=0}
\epsfig{figure= 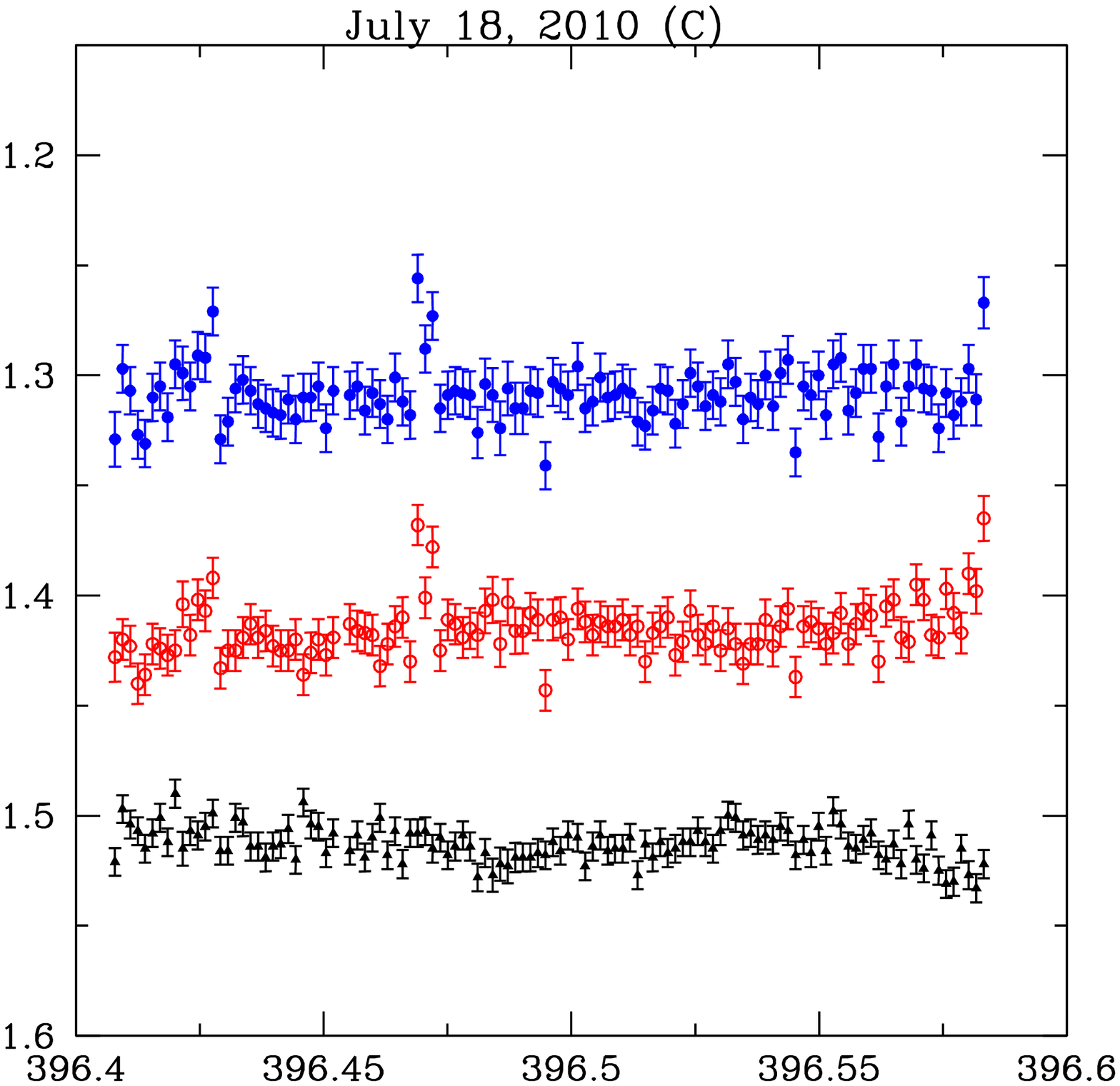,height=2.0in,width=2.2in,angle=0}
\epsfig{figure= 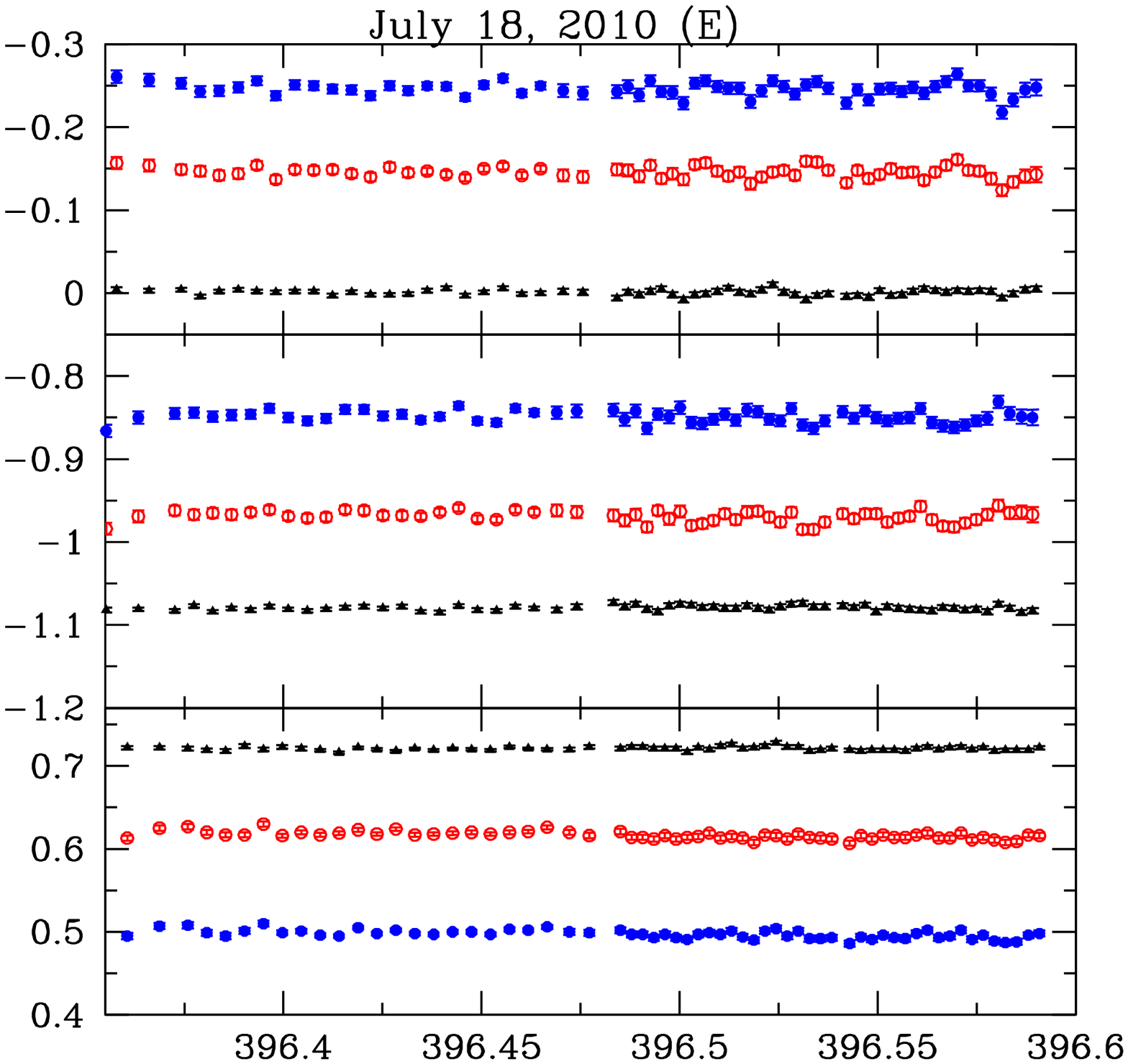,height=2.0in,width=2.2in,angle=0}
\epsfig{figure= 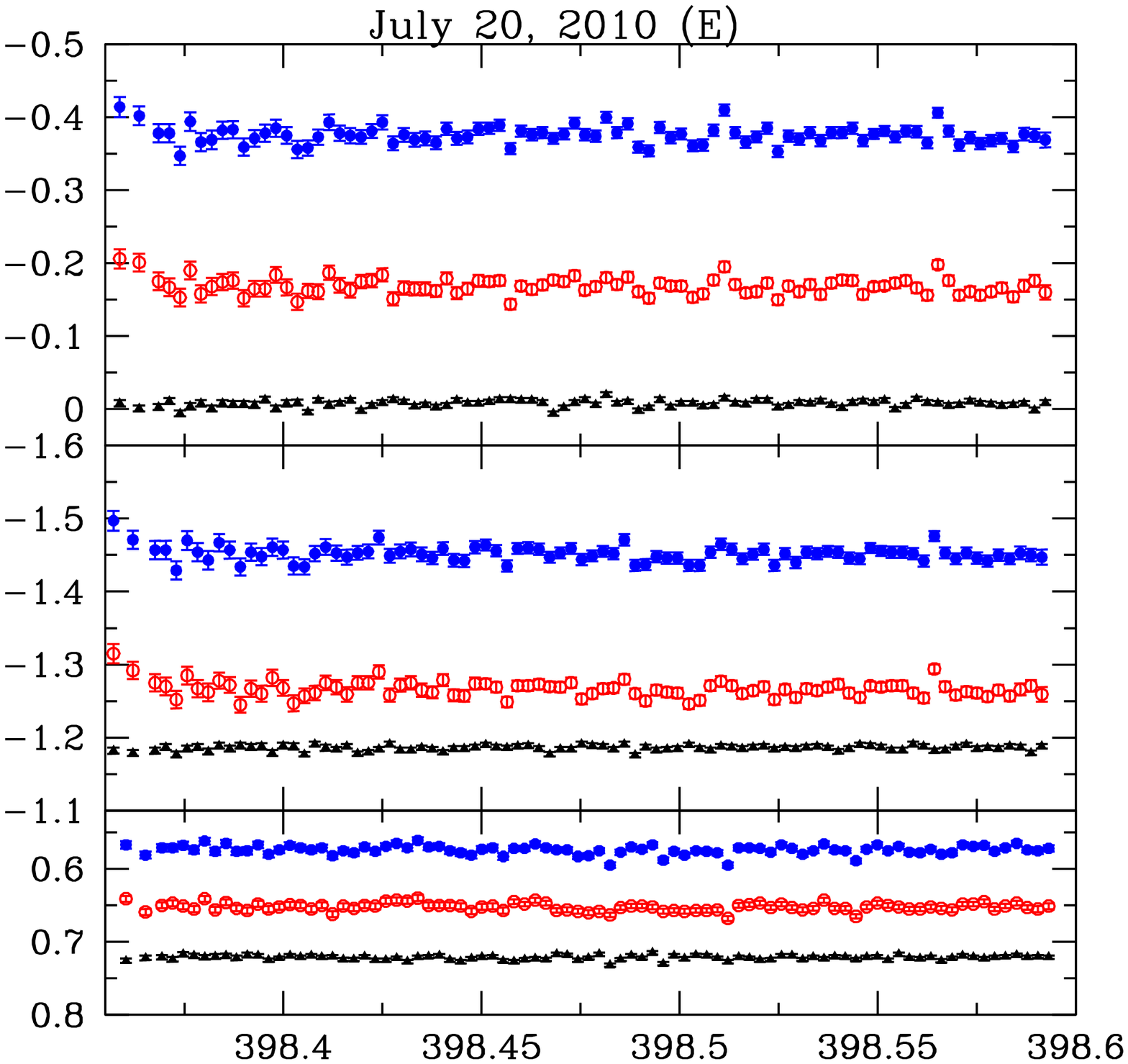,height=2.0in,width=2.2in,angle=0}
\epsfig{figure= 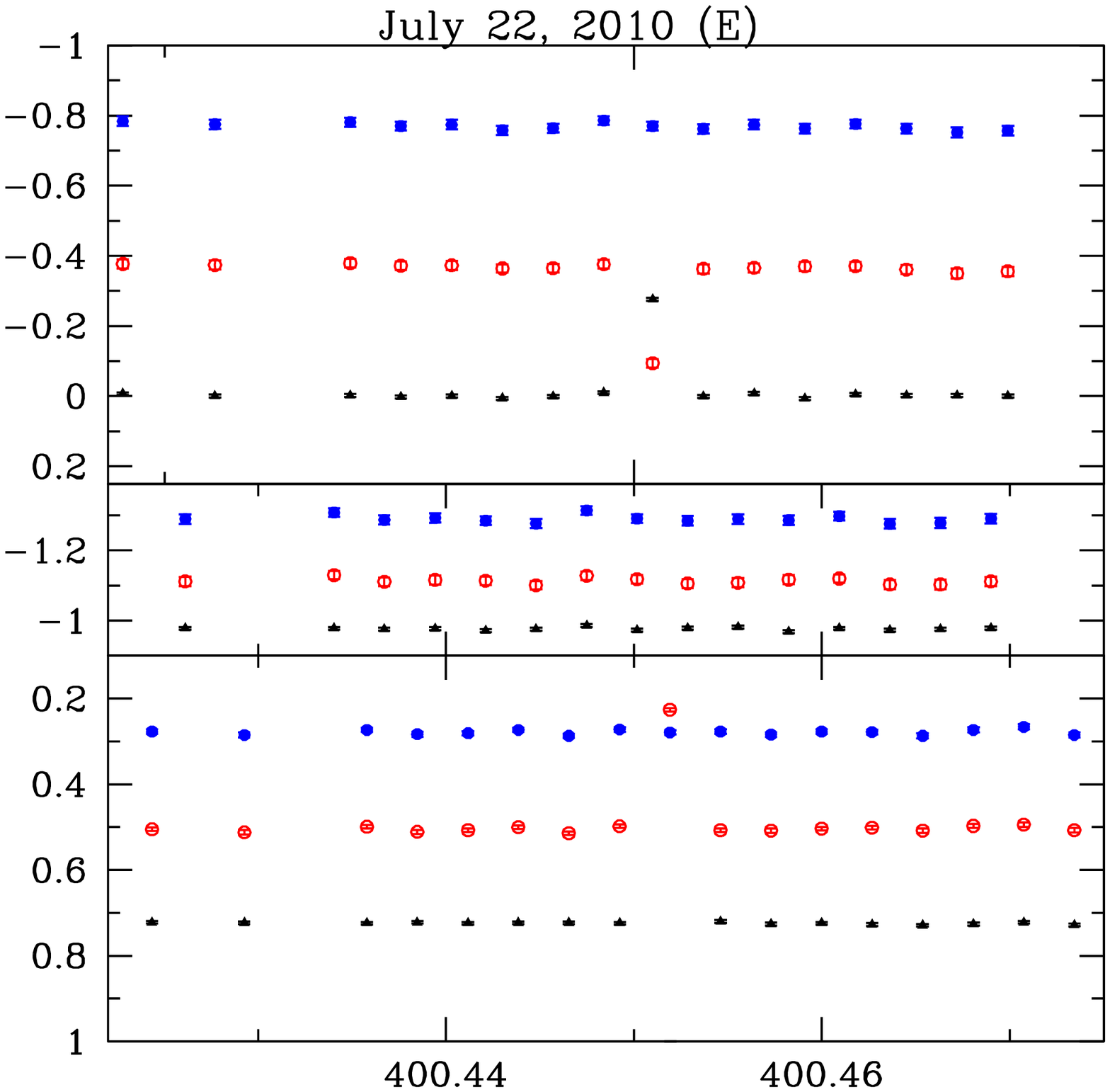,height=2.0in,width=2.2in,angle=0}
\caption{As in Fig.\ 1 for 1 ES 2344$+$514. }
\end{figure*}

\clearpage
\begin{figure*}
\epsfig{figure= 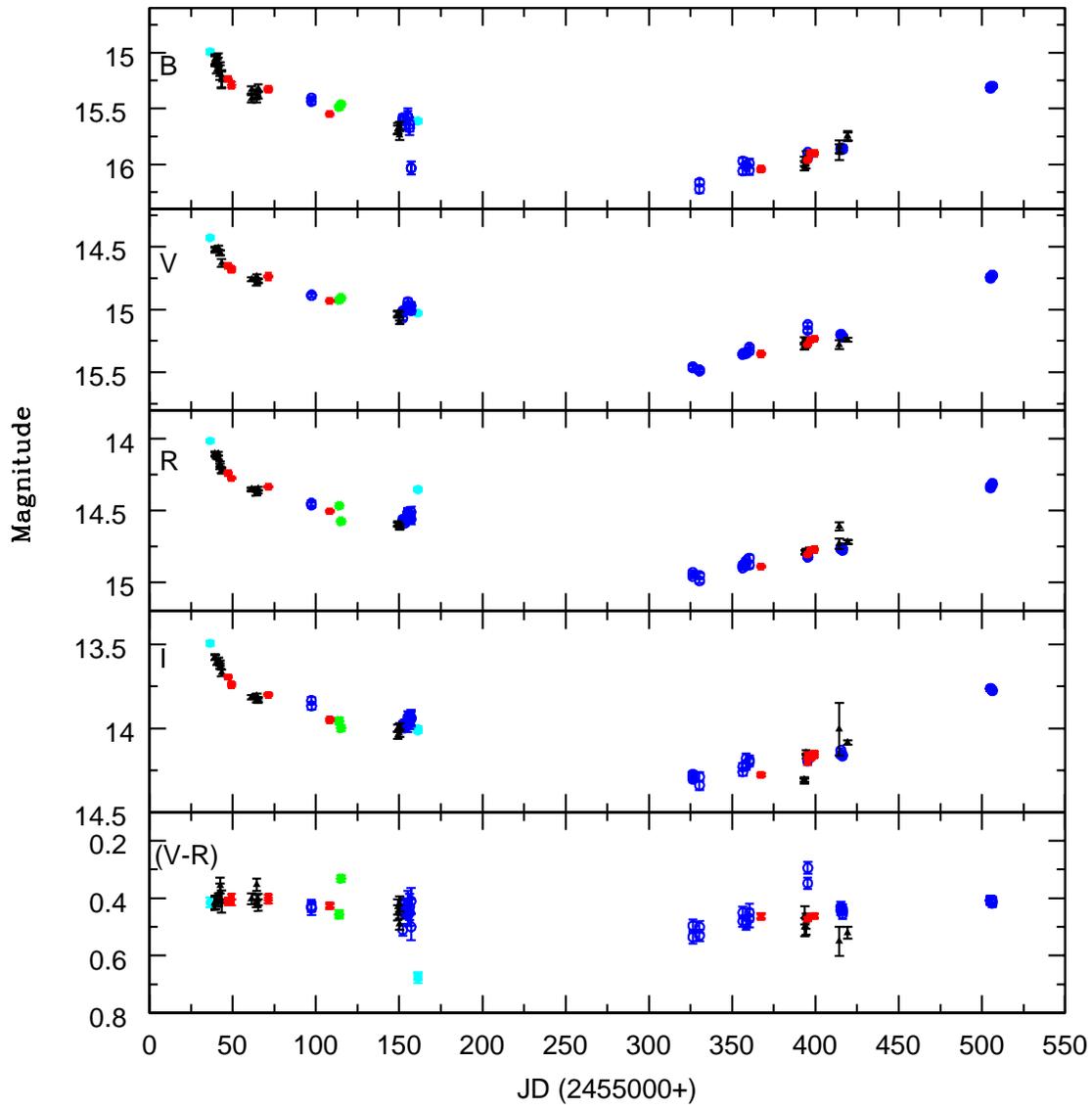,height=16.cm,width=16.cm,angle=0}
\caption{Short$-$term variability light curve of 1ES 1959$+$650. Starred, solid circle, 
open circle, triangle and square symbols represent data from the telescopes A, B, C, D and E respectively.  
 }
\end{figure*}

\clearpage
\begin{figure*}
\epsfig{figure= 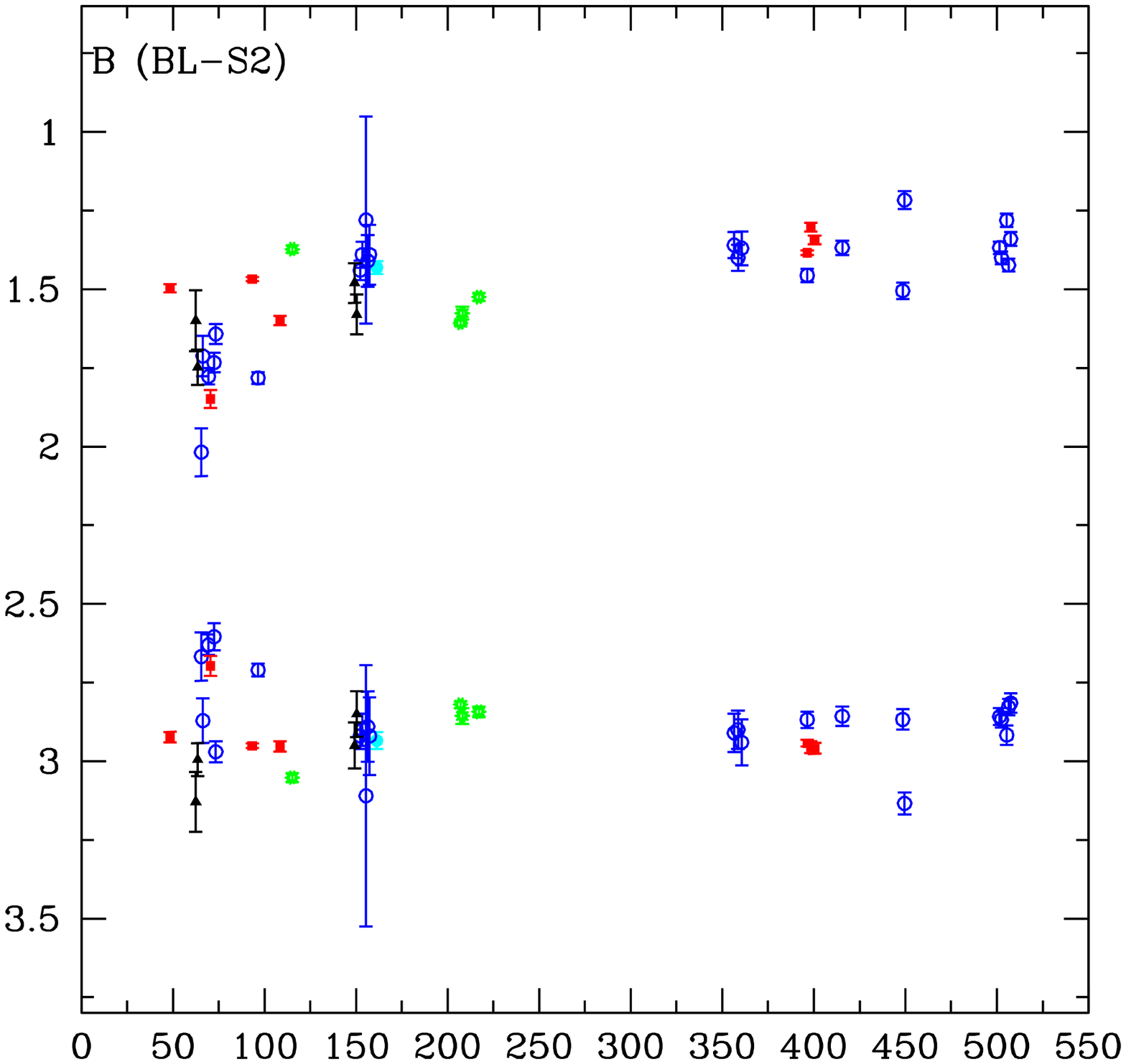,height=2.0in,width=3.2in,angle=0}
\epsfig{figure= 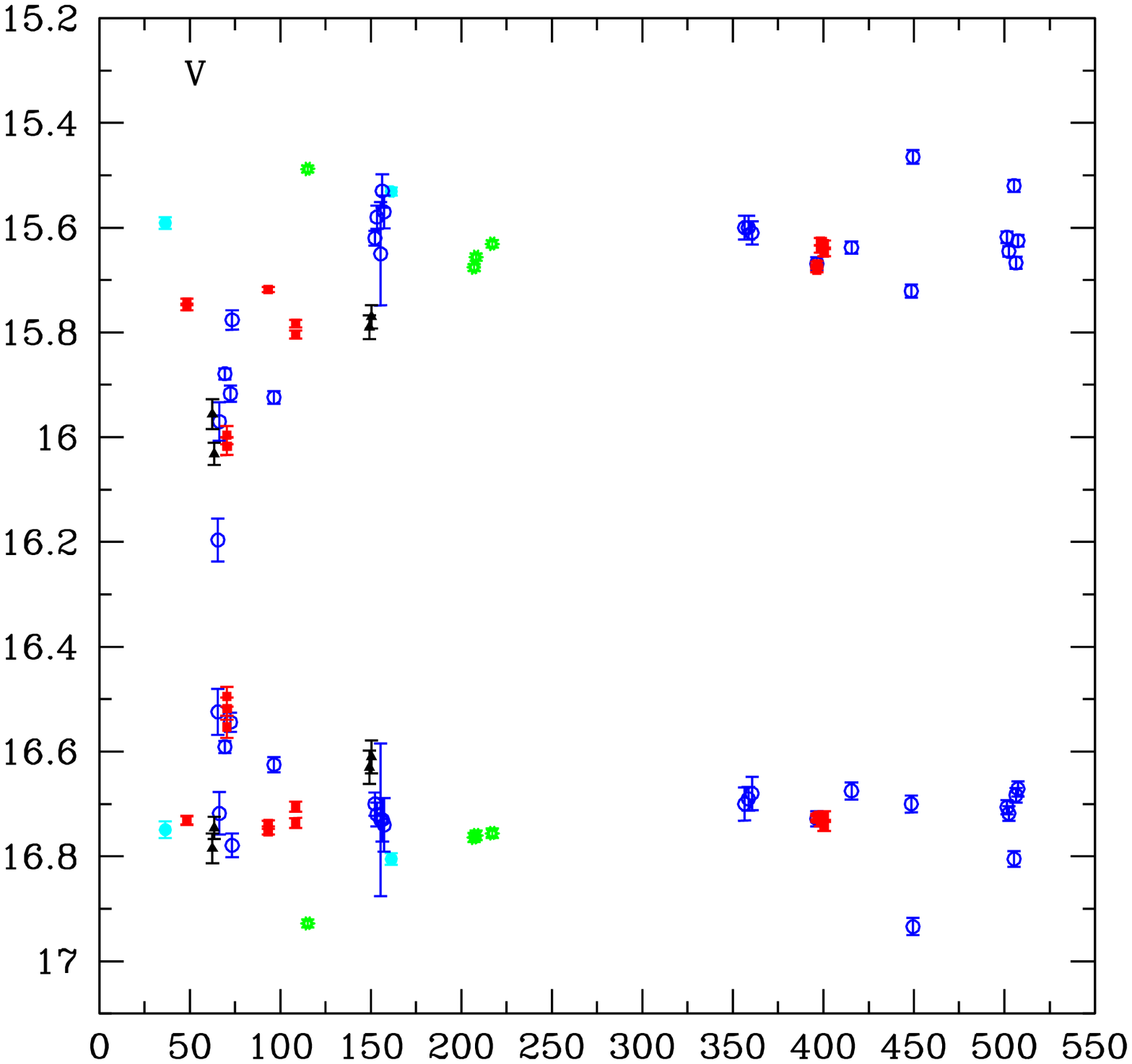,height=2.0in,width=3.2in,angle=0}
\epsfig{figure= 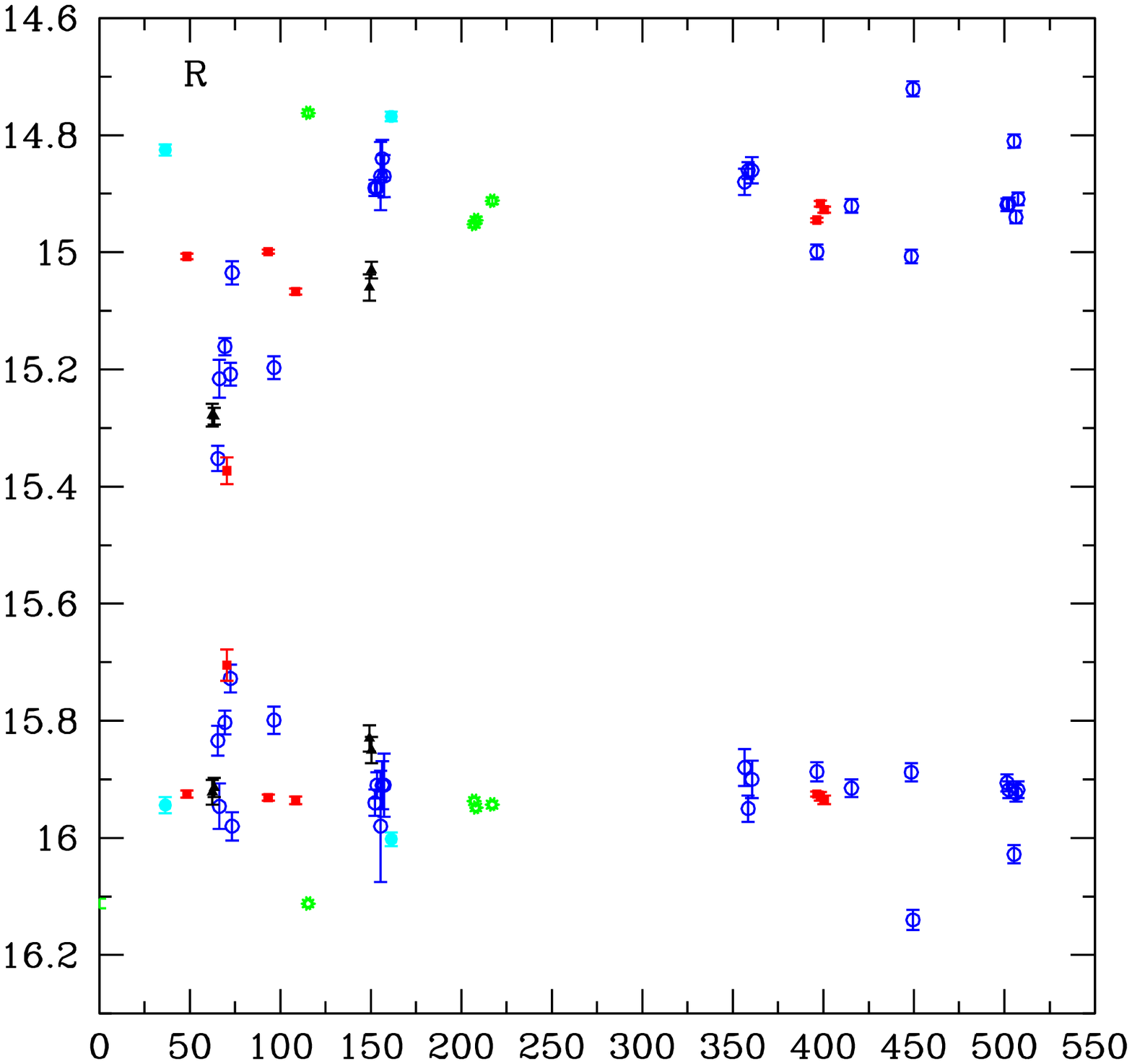,height=2.0in,width=3.2in,angle=0}
\epsfig{figure= 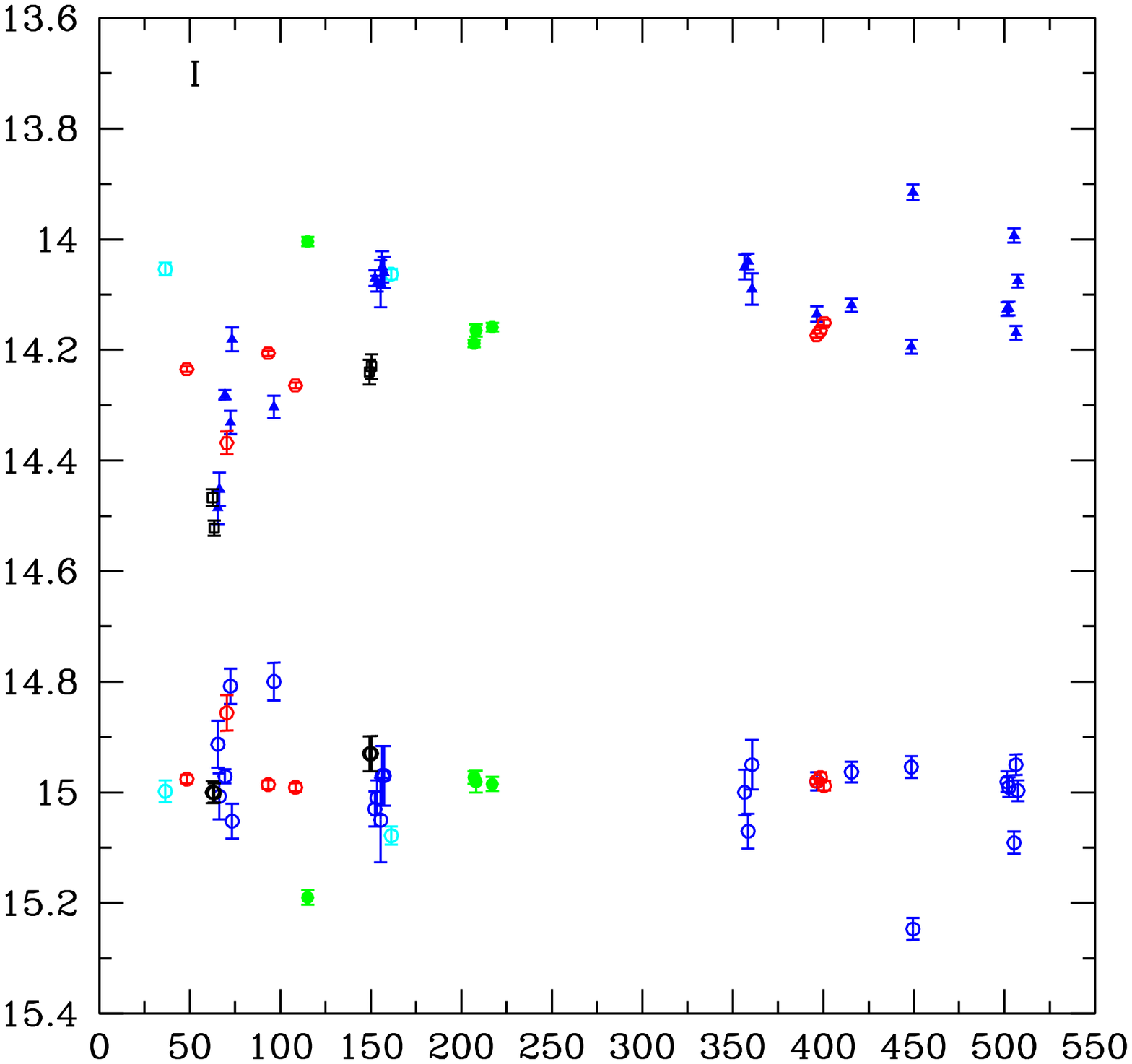,height=2.0in,width=3.2in,angle=0}
\epsfig{figure= 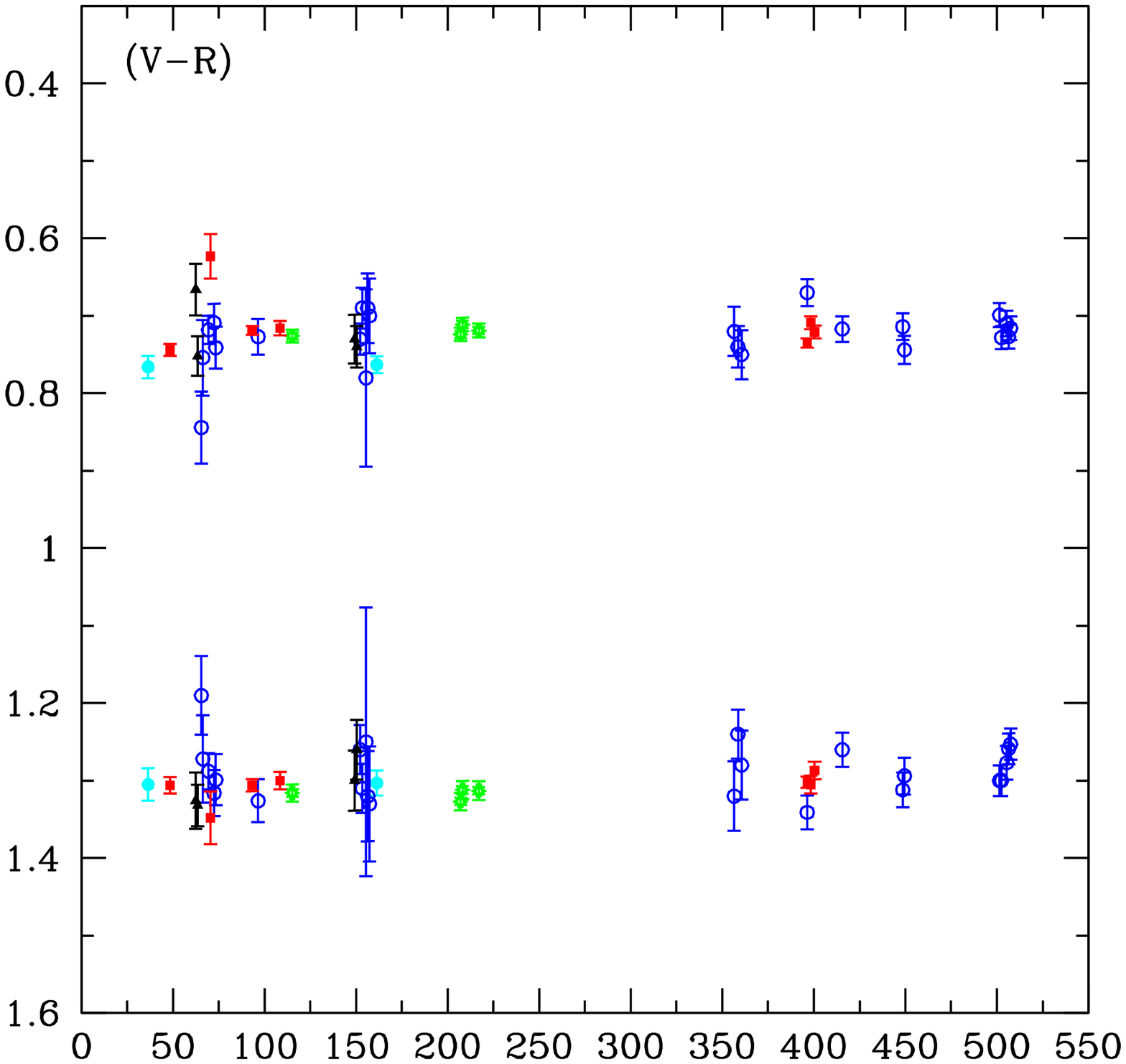,height=2.0in,width=3.2in,angle=0}
\caption{Short$-$term variability light curve of 1ES 2344$+$514. Starred, solid circle, 
open circle, triangle and square symbols represent data from the telescopes A, B, C, D and E, respectively. 
Y-axis is magnitude and X-axis is JD (2455000+) in each plot. In the B-band data set, the
 upper LC is (blazar$-$Star1) and the lower one is  the
differential magnitude  of (Star1$-$Star2).  In the remaining panels, the
upper curves are the calibrated blazar magnitudes and lower ones
are  differential magnitudes of (Star1$-$Star2) in V, R , I and (V-R). }
\end{figure*}

\end{document}